\documentclass[twocolumn,rmp,showpacs,preprintnumbers,amsmath,amssymb]{revtex4}
\usepackage{times}

\def\mx{{\mathbf x}}
\def\mk{{\mathbf k}}

\def\pd#1#2{{\partial #1} \over {\partial #2}}
\def\p2d#1#2{{\partial^2 #1} \over {\partial #2^2}}

\usepackage{graphicx}
\usepackage{dcolumn}
\usepackage{bm}
\begin{document}


\title{Scaling Laws in the Distribution of Galaxies}

\author{Bernard J. T. Jones}
\email{jones@astro.rug.nl} \affiliation{Kapteyn Institute,
University of Groningen, P.O. Box 800, 9700 AV Groningen, The
Netherlands}
\author{Vicent J. Mart\'{\i}nez}%
\email{martinez@uv.es}
\affiliation{Observatori Astron\`omic de la Universitat de
Val\`encia, Edifici d'Instituts de Paterna, Apartat de Correus
22085, 46071 Val\`encia, Spain}
\author{Enn Saar}
\email{saar@aai.ee}
\affiliation{Tartu Observatory, T\~oravere, 61602 Estonia}
\author{Virginia Trimble}
\email{vtrimble@astro.umd.edu}
 \affiliation{Astronomy Department, University of Maryland, College Park
MD 20742, USA \\Physics
Department, University of California, Irvine CA 92697 USA
 Maryland, USA}
\date{\today}

\begin{abstract}
Research done during the previous century established our Standard
Cosmological Model.  There are many details still to be filled in,
but few would seriously doubt the basic premise. Past surveys have
revealed that the large-scale distribution of galaxies in the
Universe is far from random: it is highly structured over a vast
range of scales.  Surveys being currently undertaken and being
planned for the next decades will provide a wealth of information
about this structure.  The ultimate goal must be not only to
describe galaxy clustering as it is now, but also to explain how
this arose as a consequence of evolutionary processes acting on
the initial conditions that we see in the Cosmic Microwave
Background anisotropy data.

In order to achieve this we will want to describe cosmic structure
quantitatively: we need to build mathematically quantifiable
descriptions of structure. Identifying where scaling laws apply
and the nature of those scaling laws is an important part of
understanding which physical mechanisms have been responsible for
the organization of clusters, superclusters of galaxies and the
voids between them. Finding where these scaling laws are broken is
equally important since this indicates the transition to different
underlying physics.

In describing scaling laws we are helped by making analogies with
fractals: mathematical constructs that can possess a wide variety
of scaling properties. We must beware, however, of saying that the
Universe {\it is} a fractal on some range of scales: it merely
exhibits a specific kind of fractal-like behavior on those scales.
We exploit the richness of fractal scaling behavior merely as an
important supplement to the usual battery of statistical
descriptors.

We review the history of how we have learned about the structure
of the Universe and present the data and methodologies that are
relevant to the question of discovering and understanding any
scaling properties that structure may have. The ultimate goal is
to have a complete understanding of how that structure emerged. We
are getting close!

\end{abstract}

\maketitle
\tableofcontents

\section{\label{sec:phys} PHYSICAL COSMOLOGY}
With the discovery of the Cosmic Background Radiation by Penzias
and Wilson (1965), cosmology became a branch of physics: there was
a well defined framework within which to formulate models and
confront them with observational data.  Prior to that there had
been a few important observations and a few important solutions to
the Einstein Field Equations for General Relativity.  We suspected
that these were somehow connected: that the Friedman-Lemaitre
solutions of the Einstein field equations described the
cosmological redshift law discovered by Hubble.

With the discovery of the background radiation we were left in no
doubt that the Universe had a hot singular origin a finite time in
our past.  That important discovery also showed that our Universe,
in the large, was both homogeneous and isotropic, and it also
showed the appropriateness of the Friedman-Lemaitre solutions.

The establishment of the ``Big Bang" paradigm led to a search for
answers, in terms of known physical laws, to key questions: why
was the Universe so isotropic, how did the structure we observe
originate? and so on.  Cosmologists built models involving only
known physics and confronted them with the data. Cosmology became
a branch of physics with a slight difference: we cannot experiment
with the subject of our discussion, the Universe, we can only
observe it and model it.

With the current round of cosmic microwave background anisotropy
maps we are able to see directly the initial conditions for galaxy
formation and for the formation of large-scale structure.   That
observed structure is thought to reflect directly the fluctuations
in the gravitational potential that gave birth to cosmic structure
and it is a consequence of the physics of the early universe.  The
goal is to link those initial conditions with what we see today.

The aim of this article is to show how the ``homogeneous and
isotropic Universe with a hot singular origin" paradigm has
emerged, and to explain how, within this framework, we can
quantify and understand the growth of the large scale cosmic
structure.

\subsection{Cross-disciplinary physics}
Gravitation is the driving force of the cosmos and so Einstein's
General Theory of Relativity is the appropriate tool for modelling
the Universe.  However, that alone is not enough: other branches
of physics have played a key role in building what has emerged as
a ``Standard Model" for cosmology.

Nucleosynthesis played an early role in defining how the light
elements formed \cite{albega}: the abundances of Helium and
Deuterium play a vital part in confronting our models with
reality.  In following how the cosmic medium cooled sufficiently
to enable gravitational collapse to form galaxies and stars we
need to understand some exotic molecular chemistry.

Today, our understanding of high energy physics plays a key role:
some even defined a new discipline and refer to it as
``astro-particle physics".  We have strong evidence that there is
a substantial amount of dark matter in galaxies and clusters of
galaxies.  So far we have not been able to say what is the nature
of this dark matter.  There is also growing evidence that the
expansion of the Universe is accelerating: this would require an
all-pervading component of matter or energy that effectively has
negative pressure.  If this were true we would have to resurrect
Einstein's cosmological constant, or invoke some more politically
correct ``fifth force" concept such as quintessence.

\subsection{Statistical mechanics}
The statistical mechanics of a self-gravitating system is a
totally nontrivial subject.  Most of the difficulty arises from
the fact that gravitation is an always-attractive force of
infinite range: there is no analogue to the Debye shielding in
plasma physics. Perhaps the most outstanding success was the
discovery by Jeans in the 1920's of equilibrium solutions to the
Liouville equation for the distribution function of a collection
of stars (the Jeans Theorem).  This has led to a whole industry in
galaxy dynamics, but it has had little or no impact on cosmology
where we might like to view the expanding universe with galaxies
condensing out as a phase transition in action.

This has not deterred the brave from tackling the statistical
mechanics or thermodynamics of self-gravitating systems, but it is
perhaps fair to say that so far there have been few outstanding
successes. The discussion by \textcite{LBW} of the so-called
gravo-thermal collapse of a stellar system in a box is probably as
close as anyone has come. It was only in the 1970's that
cosmologists ``discovered" the two-point clustering correlation
function for the distribution of galaxies and it was not until the
late 1980's with the discovery by \textcite{lapparent} of remarkable
large scale cosmic structure that we even knew what it was we were
trying to describe.

The early work of \textcite{saslaw1,saslaw2} on
``Gravithermodynamics" predated the knowledge of the correlation
function.  Following the discovery of the correlation function we
saw the work of \textcite{fallsev}, \textcite{kandrup}, and
\textcite{fry84}, providing models for the evolution of the
correlation function in various approximations.

One major problem was how to describe this structure. By 1980, it
was known that the two-point correlation function looked like a
power law on scales\footnote{The natural unit of length to describe the large scale
structure is the megaparsec (Mpc): 1 Mpc = $10^6\,$pc $\simeq
3.086\times 10^{22}\,$m $\simeq 3.26 \times 10^6$ light years.
$h$ is the Hubble constant in
units of 100 Mpc$^{-1}$ km s$^{-1}$.} $< 10 h^{-1}$ Mpc.
It was also known that the
3-point function too had a power law behavior and that it was
directly related to sums of products of pairs of two-point
functions (rather like the Kirkwood approximation).   However,
$N$-point correlation functions were not really evocative of the
observed structure and were difficult to measure past $N=4$.

Two suggestions for describing large scale cosmic structure
emerged: void probability functions proposed by \textcite{white79}
and measured first by \textcite{vpf} and multifractal measures
\cite{Jones88}, the latter being largely motivated by the manifest
scaling behavior of the lower order correlation functions on
scales $< 10 h^{-1}$ Mpc. Both of these descriptors encapsulate
the behavior of high order correlation functions.

\subsection{Scaling laws in physics}

The discovery of scaling laws and symmetries in natural phenomena
is a fundamental part of the methodology of physics.  This is not
new: we can think of Galileo's observations of the oscillations of
a pendulum, Kepler's discovery of the equal area law for planetary
motion and Newton's inverse square law of gravitation. Some
authors claim that the actual discovery of the scaling laws is
attributable to Galileo in the context of the strength of
materials as discussed in his book {\it Two New Sciences}
\cite{Peterson}.

The establishment of a scaling relationship between physical
quantities reveals an underlying driving mechanism. It is the task
of Physics to understand and to provide a formalism for that
mechanism.

\begin{figure*}
\includegraphics[width=15. cm]{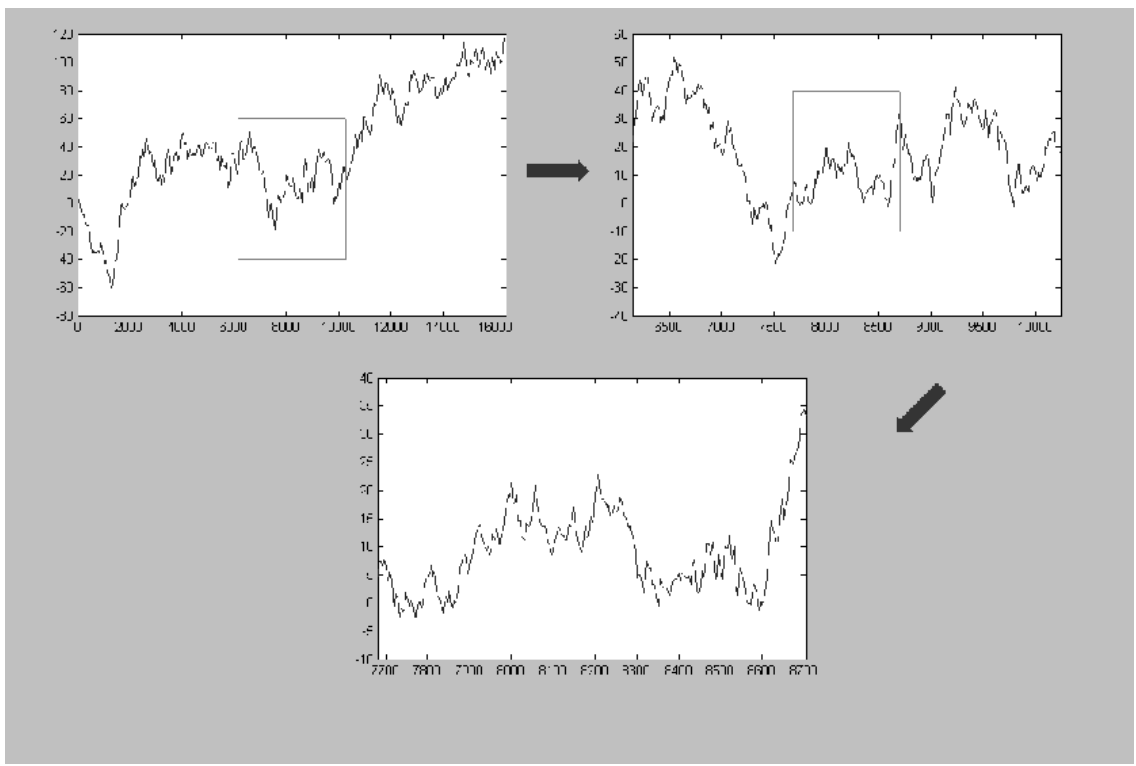}
\caption{\label{fig:brown} Scaling relations in one-dimensional
Brownian motion $x(t)$. In successive zooms the vertical
coordinate ($x$) is multiplied by 2, while the horizontal
coordinate (the time $t$) is multiplied by 4 to properly rescale
the curve.}
\end{figure*}

The self-affine Brownian motion is a good example for visual
illustration of a scaling process (see Fig.~\ref{fig:brown}). In
this case scaling is non-uniform, because different scaling
factors have to be applied to each coordinate to keep the same
visual appearance.

The breaking of symmetries and of scaling laws is equally
important and has played a key role in 20th century physics. Scale
invariance is typically broken when some new force or phenomenon
comes into play, and the result can look far more significant than
it really is.  \onlinecite{titius1, titius2} have suggested that
this may be the case for the Titius--Bode law (which is, of
course, not a law, and can be traced back before Titius and Bode
at least to David Gregory in 1702).  Their point is that, if the
primordial proto- planetary disk had a power-law distribution of
density and angular momentum then any process that forms planets
will give them something like the Titius--Bode distribution of
orbit sizes.  Thus the distribution cannot be used as a test for
any particular formation mechanism.

     Within cosmology, some of the examples of quantized redshifts reported
over the years \cite{Tifft76,BB67,Bur68}
may have been analogous cases, where the ``new phenomenon of
physics" was observational selection effects resulting when strong emission
lines passed into and out of the standard observed wavelength bands.

As we shall see, there are important scaling relationships in the
spatial distribution of galaxies. This scaling is almost certainly
a consequence of two factors: the nature of the initial conditions
for cosmic structure formation and the fact that the gravitational
force law is itself scale-free.

This scaling is observed to break down at very large distance.
This breakdown is a consequence of the large-scale homogeneity of
the Universe and of the fact that the Universe has a finite age:
gravitational agglomeration of matter has only been able to spread
over a limited domain of scales, leaving the largest scales
unaffected.

The scaling is also expected to break down for small objects where
non-gravitational forces have played a role: gas-dynamic processes
play an important role in the later stages of galaxy formation.
There are important scaling relationships among the properties of
galaxies which provide clues to the mechanisms of their formation.
We do not deal with these in detail here, although the main
scaling laws in the galaxy properties are summarized in 
Sect. VII.A.5.

\subsection{Some psychological issues}
Cosmology presents physics with a formidable challenge.  The
Universe is not a bounded and isolated system.  The Universe is
far from being in any form of dynamical equilibrium.  The
gravitational force is of infinite range and always attractive.
Nor can we experiment on the subject of interest, we are mere
observers.  Thus the usual concepts from statistical physics
cannot be simply imported, they have to be redefined to suit these
special circumstances.

This process of redefinition is apt to misdirect the struggle for
understanding the issues involved and is inevitably frustrating to
those who work in statistical physics or who seek to use
techniques from statistical physics. Indeed there have been
occasions where the notions of the standard model have been
abandoned simply in order to exploit standard concepts that would
otherwise be invalid (eg.: model universes having one spatial
dimension or model universes that have zero mean density in the
large).  Those papers may be interesting, but they have little or
nothing to do with the Universe as we know it.

\section{\label{sec:set} THE COSMIC SETTING}
The establishment of a definitive cosmological picture has been
one of the triumphs of 20th Century physics.  From Einstein's
first investigations into relativistic cosmological models,
through Hubble's discovery of the cosmic expansion, to the
discovery of the Cosmic Microwave Background Radiation in 1965,
most physicists would now agree on the basic ingredients of what
might as well be called ``the Standard Cosmological Model". The
astrophysics of the 21st century will consist largely of filling
in and understanding the details of this model: a nontrivial
process that will consume substantial human, technical and
financial resources.

While there are suggestions that the standard model may not be
complete, the data as a whole do not as yet demand any further
parametrization such as ``quintessence".  Of course, as our
understanding of fundamental physics deepens, the standard model
might be recast in a new wider, more profound, framework such as
that offered by brane cosmologies.

\subsection{Key factors}
There are several important factors to support our current view of
cosmic structure formation:
\begin{itemize}
\item
The discovery by Hubble in 1928 of the linear velocity-distance
relationship for galaxies \cite{hubble29}.  This relationship was
soon interpreted by \textcite{robert} as being due to the
expansion of the Universe in the manner described by the
Friedman-Lemaitre cosmological solutions of the Einstein Field
equations for gravitation.  These solutions described a
homogeneous and isotropic Universe emerging from a singular state
of infinite density: the Big Bang.  Later on, \textcite{bondi} and
\textcite{hoyle} provided an alternative homogeneous and isotropic
expanding model that avoided the initial singularity: the Steady
State Theory.
\item
The discovery in 1965 of the Cosmic Microwave Background Radiation
tells us the cosmological framework within which we have to work.
Our Universe is, in the large, homogeneous and isotropic; it was
initially hot enough to synthesize the element Helium. This is the
Hot Big Bang theory promoted early on by Gamow. This discovery
signaled the end of the Steady State Theory.

\item
The observation in 1992 by the COBE satellite of the large-scale
structure of the Universe at very early times provides us with
precise information about the initial conditions for structure
formation.  This is ongoing research that will lead to detailed
knowledge of the fundamental parameters of our Standard Model and
to detailed knowledge of the initial conditions in the Big Bang
that resulted in the currently observed structure.

\end{itemize}
We know a great deal about our Universe. Studies of cosmic
structure must fall within the precepts set by our Standard Model
or they will simply be dismissed at best as being academic
curiosities or at worst as being totally irrelevant.

\subsection{Some caveats}

The most important caveat in all of this is the fact that when
studying cosmic structure we observe only the luminous
constituents of the Universe.  It is true that we can observe
cosmic structure over an enormous range of the electromagnetic
spectrum, but nevertheless we face the prospect that about 85\% of what
there is out there may forever remain invisible except indirectly
though its gravitational influence.

Fortunately, we can directly study the gravitational influence of
the dark component in a number of ways.  If it is uniformly
distributed it has an influence on the overall cosmic expansion
and on the physics of the early Universe.  We can detect its
influence by studying the cosmic expansion law, or by studying the
nature of the spatial inhomogeneities seen in the cosmic microwave
background radiation. If it is not uniformly distributed it will
influence the dynamics of the large scale structure as seen in the
velocity maps for large samples of galaxies and it may reveal
itself through studies of gravitational lensing.

Our numerical simulations of the evolution of structure can in
principle take account of several forms of matter.  While this has
been a successful program, the lack of detailed knowledge about
the nature of the dark matter is nevertheless a serious
impediment. Some astrophysicists would turn the problem around and
argue that those simulations that best reproduce what is seen will
provide important information about the nature of the dark matter.

\section{\label{sec:early}EARLY IDEAS ABOUT THE GALAXY DISTRIBUTION}

\subsection{\label{subs:cosmogony} Cosmogony}
In the 4th. Century BCE, Epicurus taught that there are an infinite
number of worlds like (and unlike) ours, while Aristotle taught that
there is only one.  Neither hypothesis can currently be falsified,
and indeed we may see the continuation of this metaphysical battle
in the so-called inflationary cosmological models.

Philosophers since Anaximander \cite{anax} have long debated the
true nature of the Universe, presenting often remarkably prescient
ideas notwithstanding the lack of any real data. Given the lack of
data, the only basis for constructing a Universe was symmetry and
simplicity or some more profound cosmological principle.

The ancients saw nested crystalline spheres fitting neatly into one
another: this was a part of the then culture of thinking of
mathematics (i.e. geometry in those days) as being somehow a
fundamental part of nature \footnote{Einstein's great intellectual coup
was to geometrize the force of gravity: we are governed on large
scales by the geometry of space-time manifesting itself as the force
of gravity.}. Later thinkers such as Swedenborg, Kant and Descartes
envisioned hierarchies of nested whirls. While these ideas generally
exploited the scientific trends and notions of their time, none of
them were formulated in terms of physics.  Many are reviewed in
\textcite{jones76} where detailed references to the classical works
are given.

Perhaps the first detailed presentation of cosmogonic ideas in the
modern vein was due to Poincar\'e in his {\it Le\c{c}ons sur les
Hypoth\`eses Cosmogoniques} \cite{poincare}, some of which was to
be echoed by Jeans in his texts on Astronomy and Cosmogony
\cite{jeans}. Jeans' work is said to have had a profound effect on
Hubble's own thoughts about galaxy evolution and structure
formation \cite{chris}.

\subsection{Galaxies as ``Island Universes''}

Once upon a time there was a single galaxy.  William and Caroline
Herschel had drawn a map of the Galaxy \cite{herschel} on the basis
that the Sun was near the center of the Galaxy, and this image
persisted into the 20th Century with the ``Kapteyn Universe''
\cite{Kap} which depicted the the Milky Way as a relatively small
flattened ellipsoidal system with the Sun at its center, surrounded
by a halo of globular clusters. \textcite{Trump} recognized the role
played by interstellar absorption; he provided a far larger view of
the Galaxy and moved the Sun outwards from the center of the Galaxy
to a position some 30,000 light years from the Galactic Center.

Competing with this view was the hypothesis of Island Universes,
though at least some astronomers 100 years ago thought that had been
completely ruled out. Remember that 100 years ago it was not known
that the ``nebulae" were extragalactic systems: they were thought of
as whirlpools in the interstellar medium.

The controversy between the Great Galaxy and Island Universe views
culminated in the great debate between Curtis and Shapley in 1920
\cite{hoskin}. Shapley, who had earlier placed our Sun in the
outer reaches of the Greater Galaxy by observing the distribution
of globular clusters\footnote{We should recall that at about this
time \textcite{Lindblad} and  \textcite{Oort28} showed that the
stars in the Galaxy were orbiting about a distant center, thus
clearly placing the Sun elsewhere than at the center.}, defended
the Great Galaxy hypothesis and won the day for all the wrong
reasons.

However, it was left to Edwin P. Hubble to settle the issue in
favour of the Island Universes when he found Cepheid variables in
the galaxy NGC6822 and the Andromeda nebula
\cite{hubble25a,hubble25b}.

There was one anomaly that persisted into the early 1950's: our
Galaxy seemed to be the largest in the Universe. This was resolved
by Baade who recognized that there were in fact two populations of
Cepheid variables \cite{baade56}. This doubled the distances to the external
galaxies, thereby solving the problem.

For Hubble and most of his contemporaries what had been found were
``field galaxies" largely isolated from one another.  This was in
part due to the sorts of telescope and their fields of view that
Hubble was using \cite{hubble34, hubble36} and also in part due to
the lingering effects of the phrase ``Island universe" which
evoked images of isolation. Indeed, as late as the 1960's,
astronomers who should have known better said that galaxies were
the building blocks of the Universe (eg: \textcite{crea64} and
Abell in undergraduate lectures at UCLA 1961-1963).

In fact, most galaxies are clustered.  This is implicit in images
taken with smaller telescopes having larger fields (Shapley often
said that large telescopes were over-rated \cite{shapley32},
perhaps in part because he had deliberately cut himself off from
them by moving to Harvard) and explicit in the remarks of
\textcite{zwicky38, zwicky52} who had begun to look at the
Universe through Schmidt-coloured glasses. (The 18'' Schmidt
telescope on Palomar Mountain came into use a couple of years
before).

\subsection{Earliest impressions on galaxy clustering}
In the 19th century William Herschel and Charles Messier noted
that the amorphous objects they referred to as ``nebulae" were
more common in some parts of the sky than others and in particular
in the constellation of Virgo.

However, clusters of galaxies were not described in detail until
the work of \textcite{wolf24} who described the Virgo and Coma
clusters of galaxies.  It was not known at that time that the
nebulae, as they were then called, were in fact extragalactic
systems of stars comparable with our own Galaxy.

Hubble, using the largest telescopes, noted the remarkable overall
homogeneity and isotropy of the distribution of galaxies. The
first systematic surveys of the galaxy distribution were
undertaken by Shapley and his collaborators (often uncited and
under-acknowledged wealthy Bostonian women).  This lead to the
discovery of numerous galaxy clusters and even groups of galaxy
clusters.

\subsection{Hierarchical models}

The clustering together of stars, galaxies, and clusters of galaxies
in successively ordered assemblies is normally called a hierarchy,
in a slightly different sense of the dictionary meaning in which
there is a one-way power structure. The technically correct term for
the structured universes of Kant and Lambert is multilevel.  A
complete multilevel universe has three consequences.  One is the
removal of Olbers paradox (the motivation of John Herschel and
Richard Proctor in the 19th century).  The second, recognized by
Kant and Lambert, is that the universe retains a primary center and
is therefore nonuniform on the largest cosmic scales.  The third,
recognized by the Irish physicist Fournier d'Albe and the Swedish
astronomer Carl Charlier early in the 20th century is that the total
amount of matter is much less than in a uniform universe with the
same local density.  D'Albe put forward the curious additional
notion that the visible universe is only one of a series of
universes nested inside each other like Chinese boxes. This is not
the same as multiple 4-dimensional universes in higher dimensional
space and does not seem to be a forerunner of any modern picture.

\subsubsection{Charlier's Hierarchy}

The idea that there should be structure on all scales up to that
of the Universe as a whole goes back to \textcite{Lambert61} who
was trying to solve the puzzle of the dark night sky that is
commonly called ``Olber's paradox".  (It was not formulated by
Olbers and it is a riddle rather than a paradox \cite{Harri87}).
Simply put: if the Universe were infinite and uniformly populated
with stars, every line of sight from Earth would eventually meet
the surface of a star and the sky would therefore be bright.  The
idea probably originated with John Herschel in a review of
Humboldt's Kosmos where the clustering hierarchy is suggested as a
solution to Olber's Paradox as an alternative to dust absorption.

\begin{figure}
\includegraphics[width=8.5 cm]{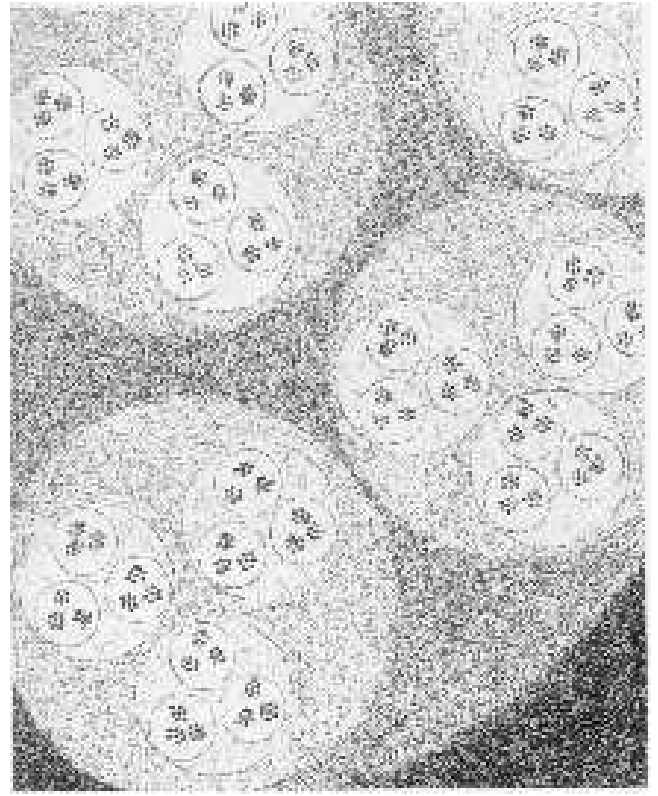}
\caption{\label{fig:hier} Hierarchical universes were very popular
at the end of the 19th century and the first half of the 20th
century. Reproduced from \textcite{harri},
Cosmology, Cambridge University Press.}
\end{figure}

At the start if the 20th century, The Swedish astronomer Carl
Charlier provided a cosmological model in which the galaxies were
distributed throughout the Universe in a clustering hierarchy
\cite{char08,char22}. His motivation was to provide a resolution
for Olber's Paradox.  Charlier showed that replacing the premise
of uniformity with a clustering hierarchy would solve the problem
provided the hierarchy had an infinite number of levels (see
Fig.~\ref{fig:hier}).

Charlier's idea was not new, though he was the first person to
provide a correct mathematical demonstration that Olber's Paradox
could indeed be resolved in this way.  It should be recalled that
he was working at a time before any galaxies had measured
redshifts and long before the cosmic expansion was known.

It is interesting that the Charlier model had de Vaucouleurs as
one of its long standing supporters \cite{vau70}.

More recently still there have been a number of attempts to
re-incarnate such a universal hierarchy in terms of fractal
models. Fractal models were first proposed by \textcite{four07}
and subsequently championed by \textcite{man82} and
\textcite{pietro87}. Several attempts have been made to construct
hierarchical cosmological models (a Newtonian solution was found
by \textcite{wertz71}, general-relativistic solutions were proposed by
\textcite{bonnor72,wesson78,ribeiro92}). All these solutions are,
naturally, inhomogeneous with preferred position(s) for the
observer(s), and thus unsatisfactory. So the present trend to
conciliate fractal models with cosmology is to use the measure of
last resort, and to assume that although the matter distribution
in the universe is homogeneous on large scales, the galaxy
distribution can be contrived to be fractal \cite{rib01}.
Numerical models of deep samples contradict this assumption.

\subsubsection{Carpenter's law}
\label{subsec:carp}

     Edwin F. Carpenter spent his early days at Steward Observatory
(of which he was director for more than 20 years, from 1938)
scanning zone plates to pick out extragalactic nebulae for later
study.  In 1931, he found a new cluster in the direction of Cancer
(independently discovered by Hubble at about the same time.)  He
measured its size on the sky, estimated its distance, and counted
the number of galaxies, $N$, he could recognize within its confines.
This gave him a sample of 7 clusters with similar data, all from Mt.
Wilson plates (5 in the Mt. Wilson director's report for 1929-30 and
one then just found by Lundmark).  He was inspired to graph
$\log(N)$ vs. the linear sizes of the clusters \cite{carp31} and
found a straight line relation, that is, a power law in
$N(\mbox{diameter})$, nowhere near as steep as $N \sim D^3$ or $N$
proportional to volume. The then known globular cluster system of
the Milky Way (with about 35 clusters within $10^5$ pc) also fit
right on his curve.

     Carpenter later considered a larger sample of clusters and found that a
similar curve then acted as an upper envelope to the data
\cite{carp38}.  If his numbers are transformed to the distance
scale with $H_0 = 100$ km s$^{-1}$ Mpc$^{-1}$, then the relations
are \cite{Vauc71}
\begin{equation}
\log N(\mbox{max}) = 2.5 + 1.5 \log R(\mbox{Mpc})
\end{equation}
or
\begin{equation}
\log N(\mbox{max}) = 2.19 + 0.5 \log V(\mbox{Mpc}^3)
\end{equation}
and the maximum number density in galaxies per Mpc$^3$ is also
proportional to $0.5 \log(V)$.  De Vaucouleurs called this
Carpenter's law, though the discoverer himself had been somewhat
more tentative, suggesting that this sort of distribution (which
we would call scale free, though he did not) might mean that there
was no fundamental difference among groups, clusters, and
superclusters of galaxies, but merely a non-random, non-uniform
distribution, which might contain some information about the
responsible process.
     It is, with hindsight, not surprising that the first few clusters that
\textcite{carp31} knew about were the densest sort, which define the upper envelope
of the larger set \cite{carp38}.  The ideas of a number of other proponents,
both observers and theorists,
on scale-free clustering and hierarchical structure
are presented (none too sympathetically) in Chapter 2 of \textcite{peeb80}.

\subsubsection{De Vaucouleurs hierarchical model}

\begin{figure}
\resizebox{0.4\textwidth}{!}{\includegraphics*{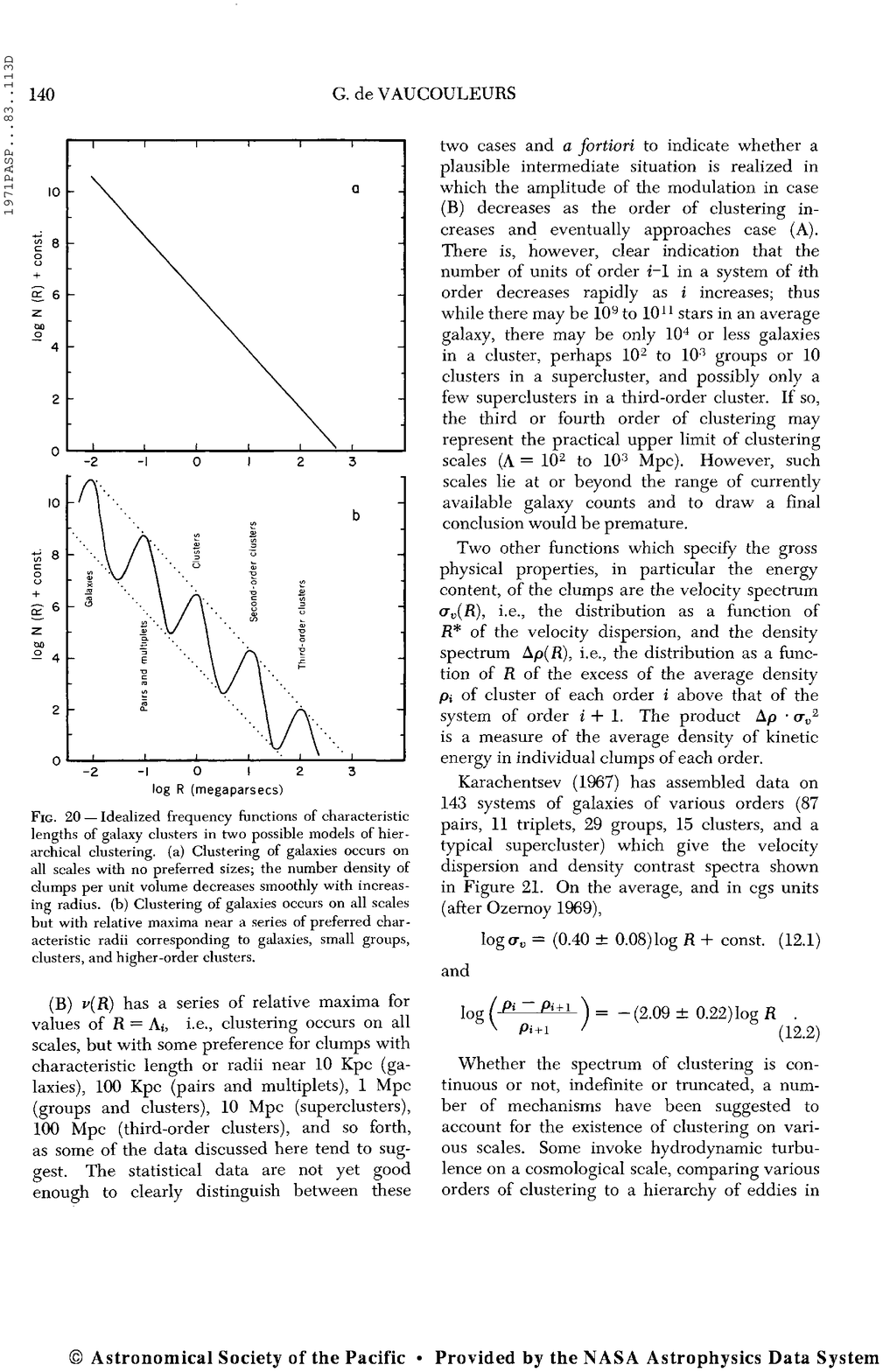}}
\caption{\label{fig:Vacu} In this idealized diagram de Vaucouleurs
shows two hierarchical frequency distributions of the number of
clumps per unit volume. In the top panel there are no
characteristic scales in the distribution. This is the model
proposed by \textcite{kiangsas}. The bottom panel shows a more
sophisticated alternative in which the overall decrease of the
number of clumps per unit volume does not behave monotonically
with the scale, but it displays a series of local maxima
corresponding to the characteristic scales of different cosmic
structures: galaxies, groups, clusters, superclusters, etc.
Reproduced from \textcite{Vauc71}, Astronomical
Society of the Pacific.}

\end{figure}

    De Vaucouleurs first appears on the cosmological stage doubting
what was then the only evidence for galaxy evolution with epoch,
the Stebbins-Whitford effect, which he attributed to observational
error \cite{Vauc48}. He was essentially right about this, but
widely ignored.  He was at other times a supporter of the
cosmological constant (when it was not popular) and a strong
exponent of a hierarchical universe, in which the largest
structures we see would always have a size comparable with the
reach of the deepest surveys \cite{Vauc60, vau70, Vauc71}.  He
pointed out that estimates of the age of the universe and of the
sizes of the largest objects in it had increased monotonically
(and perhaps as a sort of power law) with time since about 1600,
while the densities of various entities vs. size could all be
plotted as another power law,
\begin{equation}
\rho(r) \sim r^{-x}, \mbox{with $x$ between 1.5 and 1.9}.
\end{equation}
   By putting ``Carpenter's Law" into modern units, de Vaucouleurs
showed that it described this same sort of scale-free universe.  A
slightly more complex law, with oscillations around a mean,
falling line in a plot of density vs. size (see
Fig.~\ref{fig:Vacu}), could have galaxies, binaries, groups,
clusters, and superclusters as distinct physical entities, without
violating his main point that what you see is what you are able to
see.

De Vaucouleurs said that it would be quite remarkable if, just at
the moment he was writing, centuries of change in the best
estimate for the age and density of the universe should stop their
precipitous respective rise and fall and suddenly level off at
correct, cosmic values.  Thus he seemed to be predicting that
evidence for a universe older than 10-20 Gyr and for structures
larger than 100 Mpc should soon appear.  (He held firmly to a
value of $H_0$ near 100 km s$^{-1}$ Mpc$^{-1}$ for most of his
later career, except for the 1960 paper where it was 75, but
thought of local measurements of $H_0$ as being relevant only
locally).

Remarkable, but apparently true.  Instead of taking off again,
estimates of the age of the universe made since 1970 from
radioactive decay of unstable nuclides, from the evolution of the
oldest stars, and from the value of the Hubble constant,
increasingly concur.  And galaxy surveys have now penetrated a
factor 10 deeper in space than the Shane-Wirtanen and Harvard
counts in which de Vaucouleurs saw his superclusters.

\subsection{The cosmological principle}
The notion that the Earth is not at the center of the Universe is
generally referred to as the ``Copernican Principle'', though it
traces its origins back to Aristarchus who thought that the Sun and
the stars were in fact fixed, with the stars being at great
distances.

The modern notion that the Universe on the very largest scales
should be homogeneous and isotropic appears to have originated with
\textcite{ein17}. At that time there could have been no
observational basis for this assumption. However, homogeneity is a
consequence of the notion that we are not in a special place in the
Universe and the assumptions of homogeneity and isotropy provide for
easy solutions of the Einstein field equations. The first
cosmological models of Einstein and of de Sitter were based on this
principle. Robertson and Walker derived their famous solution of the
Einstein equations using only that principle.

It was frequently stated in the years that followed that the
Universe in the large looked homogeneous and isotropic.  The first
systematic study was \textcite{hubble26} who used a sample of 400
galaxies with magnitudes, the sample was thought to be complete to
magnitude 12.5.  He found his counts fitted the relationship 
\begin{equation}
\log N(<m) =
0.6 m + \mbox{constant}
\end{equation}
and concluded, importantly, that ``The agreement
between observed and computed $\log N$ over a range of more than 8
mag. is consistent with the double assumption of uniform luminosity
and uniform distribution or, more generally, indicates that the
density function is independent of the distance.''  He goes on to
look at systematics in the residuals in this plot and concludes that
they may be due to ``... clustering of nebulae in the vicinity of
the galactic system.  The cluster in Virgo alone accounts for an
appreciable part.''

Hubble only had data to magnitude 12. Anyone looking at the
considerably fainter Shane and Wirtanen's isoplethic maps of galaxy
counts based on the Lick Sky Survey (\textcite{shane67}), or the
more recent Center for Astrophysics 
(CfA-II) slices data \cite{gelhu} might be forgiven for
questioning the homogeneity conjecture!

The first demonstration of homogeneity in the galaxy distribution
was probably the observation by Peebles that the (projected)
two-point correlation function estimated from diverse catalogs
probing the galaxy distribution to different depths followed a
scaling law that was consistent with homogeneity.  The advent of
automated plate-measuring machines provided deeper and more reliable
samples with which to confirm the uniform distribution
number-magnitude relationship.  However, at the faintest magnitude
levels, these counts show significant systematic deviations from
what is expected from a uniform distribution: these deviations are
due to the effects of galaxy evolution at early times and their
interpretation depends on models for the evolution of stellar
populations in galaxies.  Recent, very deep studies
(\textcite{metcalfe}) show convincingly ``... that space density of
galaxies may not have changed much between $z=0$ and $z=3$''.

The first incontrovertible proof of cosmic isotropy came only as
recently as early 1990s from the COBE satellite all-sky map of the
cosmic microwave background radiation \cite{cobe}. This map is
isotropic to a high degree, with relative intensity fluctuations
only at the level of $10^{-5}$. With this observation, and with the
reasonable hypothesis that the Universe looks the same to all
observers (the Copernican Principle) we can deduce that the Universe
must be locally Friedman-Robertson Walker, ie: homogeneous as well
as isotropic \cite{egs}.

\section{DISCOVERING COSMIC STRUCTURE}

\subsection{Early catalog builders}

Observational cosmology, like most other physical sciences, is
technology driven. With each new generation of telescope and with
each improvement in the photographic process, astronomers probed
further into the Universe, cataloguing its contents.

Early on, Edward Fath used the Mount Wilson 60" telescope to
photograph Kapteyn's selected areas. That survey showed
inhomogeneities that were later analyzed by \textcite{Bok} and
\textcite{mow} who demonstrated statistically, using counts in
cells, that the galaxy distribution was nonuniform.  About this
time, \textcite{carp38} noticed that small objects tend to be
dense while vast objects tend to be tenuous.  He plotted a
remarkable relationship between scale and density ranging all the
way from the Universe, through galaxies and stellar systems to
planets and rock, as it has been explained in
Sect.~\ref{subsec:carp}. This was perhaps the first example of a
scaling relationship in cosmology.

By 1930, the Shapley/Ames catalog of galaxies revealed the Virgo
cluster as the dominant feature in the distribution of bright
galaxies. It was already clear from that catalog that the Virgo
Cluster was part of an extended and rather flattened supercluster.
This notion was hardly discussed except by de Vaucouleurs who
thought that this was indeed a coherent structure whose flattening
was due to rotation.

The Lick Survey of the sky provided extensive plate material that
was later to prove one of the key data sets for studies of galaxy
clustering.  The early isoplethic maps drawn by \textcite{shane54}
provided the first cartographic view of cosmic structure.  Their
counts of galaxies in cells was to provide \textcite{Rubin54} and
\textcite{Limber54} with the stimulus to introduce the two point
clustering function as a descriptor of cosmic structure.

\begin{figure}
   \includegraphics[width=8.5 cm]{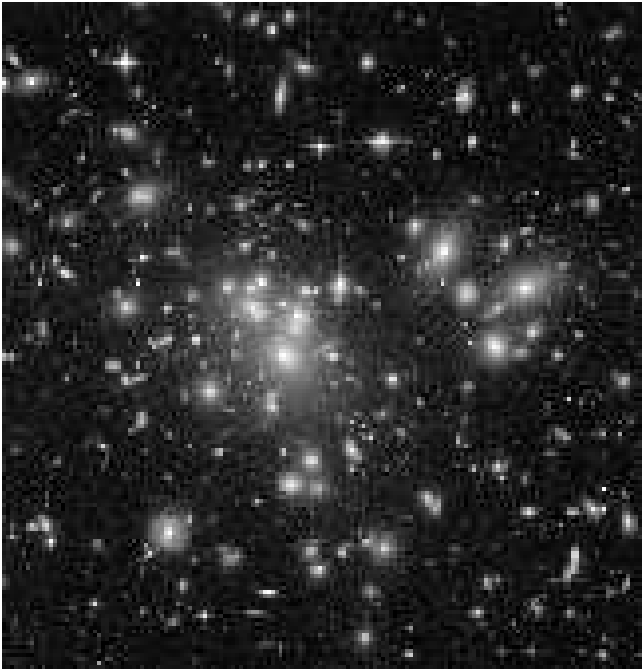}
\caption{\label{fig:a1689} The cluster of galaxies Abell 1689 at
redshift $z=0.18$ seen by the HST with its recently installed
Advanced Camera for Surveys (ACS). The arcs observed amongst
hundreds of galaxies conforming the cluster are multiple images of
far-away individual galaxies whose light has been amplified and
distorted by the total cluster mass (visible and dark) acting
as a huge gravitational lens, (image courtesy of NASA, N. Benitez
(JHU), T. Broadhurst (The Hebrew University), H. Ford (JHU), M.
Clampin (STScI), G. Hartig (STScI), G. Illingworth (UCO/Lick
Observatory), and the ACS Science Team, and ESA).}
\end{figure}

But it was the Palomar Sky Survey using the new 48'' Schmidt
telescope that was to provide the key impetus in understanding the
clustering of galaxies.  Zwicky and his collaborators at Caltech
systematically cataloged the position and brightness of thousands of
brighter galaxies on these plates, creating what has become known as
the ``Zwicky Catalog".  \textcite{abell58} made a systematic survey
for rich clusters of galaxies and drew up a catalog listing
thousands of clusters. This has become simply known as the ``Abell
catalog". Fig.~\ref{fig:a1689} shows a modern image of the cluster
Abell 1689 obtained by the ACS camera aboard of the Hubble Space
Telescope (HST). A catalog of galaxy redshifts noting the clusters
to which galaxies belonged was published in 1956 by
\textcite{Humason56}.

\subsubsection{The Lick survey}
The first map of the sky revealing widespread clustering and
super-clustering of galaxies came from the Lick survey of galaxies
undertaken by \textcite{shane67} using large field plates from the
Lick Observatory.  This was, or anyhow should have been, the
definitive database.  It was the subject of statistical analysis
by \textcite{neyman53}, which was a major starting point for what
have subsequently become known  as Neyman--Scott processes in the
statistics literature. Ironically, although these processes have
become a discipline in their own right, they have since that time
played only a minor role in astronomy.

Scott in the IAU Symposium 15 \cite{scott62}  mentions that there
are clearly larger structures to be seen in these counts, as
\textcite{shane54} had already noted.  They spoke of ``larger
aggregations" or ``clouds" as being rather general features. The
Lick survey was later to play an important role in Peebles'
systematic assault on the problem of galaxy clustering. Peebles
obtained from Shane the notes containing the original counts in
10'x10' cells and computerized them for his analysis. The counts
in 1 degree cells had been used first by Vera Cooper-Rubin (as
Vera Rubin was then known) to study galaxy clustering in terms of
correlation functions, a task set by her adviser George Gamow.
Rubin did this at a time when there were no computers.   It was
\textcite{totsuji} who first did this on a computer and
published the first two-point correlation function as we now know
it with the power law that has dominated much of cosmology for the
past three decades and more\footnote{BJ ``discovered" this paper
at the time of writing his Review of Modern Physics article
(Jones, 1976) while perusing the Publications of the Astronomical
Society of Japan in the Institute of Theoretical Astronomy Library
in Cambridge. There do not appear to be any citations prior to
that time.}.

\subsubsection{Palomar Observatory sky survey}
The two main catalogs of clusters derived from the Palomar
Observatory Sky Survey (POSS) were that of \textcite{abell58} and
that of Zwicky and his collaborators \cite{zwicky6168}.

Abell went on immediately to say that there was significant higher
order clustering in his data, giving, in 1958, a scale for
superclustering of 24 $(H_0/180)^{-1}$ Mpc. In 1961 at a meeting
held in connection with the Berkeley IAU Abell published
\cite{abell61} a list of these ``super-clusters", dropped the
Hubble constant to 75 km s$^{-1}$ Mpc$^{-1}$  and estimated masses
of $10^{16}-10^{17}$ $M_\odot$ with velocity dispersions in the
range 1000-3000 km s$^{-1}$.  At about the same time,
\textcite{Bergh61} remarks that Abell's most distant clusters
(distance class 6 having redshifts typically around 50,000 km
s$^{-1}$) show structure on the sky on a scale of some $20^\circ$,
corresponding to 100 Mpc, for his $H_0=180$ km s$^{-1}$
Mpc$^{-1}$, or about 300 Mpc using current values.

Zwicky explicitly and repeatedly denied the existence of higher
order structure \cite{Zwicky63, Zwicky65, Zwicky66a, Zwicky66b}.
Some of his ``clusters" were on the order of 80 Mpc across (for
$H_0$ less than 100), had significant substructure, and would to
any other person have looked like superclusters! Herzog, one of
Zwicky's collaborators in the cluster catalog, found large
aggregates of clusters in the catalog and had the temerity to
say so publicly in a Caltech astronomy colloquium. He was offered
``political asylum" at UCLA by George Abell. \textcite{Kara66}
also reported finding large aggregates in the Zwicky catalog.

\subsubsection{Analysis of POSS clusters}
Up until about 1960 most of those involved seemed to envisage a
definite hierarchy of structures: galaxies (perhaps binaries and
small groups), clusters and superclusters.  Kiang  remarked that
the existing data were best described by continuous,
``indefinite", clustering: quite different from the clustering
hierarchy as understood at the time \cite{kiang, kiangsas}. Kiang,
incidentally, bridged a critical era in data processing, using
``computers" (i.e., poorly paid non-PhD labour, mostly women after
the style of Shapley) and later on real computers (Atlas).
\textcite{Flin74} came independently to the same conclusion, and
in his presentation at IAU Symposium 63 was scolded by Kiang for
not having read the literature.

The later investigation by \textcite{peha} using the power
spectrum of the cluster distribution showed superclustering quite
conclusively: clusters of galaxies are not randomly distributed
and as they are correlated they are themselves clustered.  Later
analyses revealed a variation of cluster clustering with cluster
richness.

Nevertheless, there still remained mysteries to be cleared up: the
level measured for clustering of clusters was far in excess of
what would be expected on the basis of the measured clustering of
the galaxies from which they are built.  Many solutions have been
proposed to explain this anomaly, including the argument that the
Abell catalog is too subjective and biased.  However, the
phenomenon still persists in cluster catalogs constructed by
machine scans of photographic plates.

\subsection{Redshift Surveys}

\subsubsection{Why do this?}

Those early catalogs simply listed objects as they appeared
projected onto the celestial sphere.  The only indication of depth
or distance came from brightness and/or size.  These catalogs
were, moreover, subject to human selection effects and these might
vary depending on which human did the work, or even what time of
the day it was.

What characterizes more recent surveys is the ability to scan
photographic plates digitally (eg: the Cambridge Automatic
Plate Machine, APM), or to create the survey in digital format
(eg: IRAS, Sloan Survey and so on). Moreover, it is now far easier
to obtain radial velocities (redshifts) for large numbers of
objects in these catalogs.

Having said that, it should be noted that handling the data from
these super-catalogs requires teams of dozens of astronomers
doing little else. Automation of the data gathering does little to
help with the data analysis!

Galaxy redshift surveys occupy a major part of the total effort
and resources spent in cosmology research.  Giving away hundreds
of nights of telescope time for a survey, or even constructing
purpose built telescopes is no light endeavour. We have to know
beforehand why we are doing this, how we are going to handle and
analyze the data and, most importantly, what we want to get out of
it.  The early work, modest as it was by comparison with the giant
surveys being currently undertaken, has served to define the
methods and goals for the future, and in particular have served to
highlight potential problems in the data analysis.

We have come a long way from using surveys just to determine a
two-point correlation function and wonder at what a fantastic
straight line it is. What is probably not appreciated by those who
say we have got it all wrong (eg: \textcite{labini98}) is how much
effort has gone into getting and understanding these results by a
large army of people.  This effort has come under intense scrutiny
from other groups: that is the importance of making public the
data and the techniques by which they were analyzed.  The analysis
of redshift data is now a highly sophisticated process leaving
little room for uncertainty in the methodology: we do not simply
count pairs of galaxies in some volume, normalise and plot a
graph!

The prime goals of redshift surveys are to map the Universe in
both physical and velocity space (particularly the deviation from
uniform Hubble expansion) with a view to understanding the
clustering and the dynamics. From this we can infer things about
the distribution of gravitating matter and the luminosity, and we
can say how they are related. This is also important when
determining the global cosmological density parameters from galaxy
dynamics: we are now able to measure directly the biases that
arise from the fact that mass and light do not have the same
distribution.

Mapping the universe in this way will provide information about
how structured the Universe is now and at relatively modest
redshifts. Through the cosmic microwave background radiation we
have a direct view of the initial conditions that led to this
structure, initial conditions that can serve as the starting point
for $N$-body simulations.  If we can put the two together we will
have a pretty complete picture of our Universe and how it came to
be the way it is.

Note, however, that this approach is purely experimental.  We
measure the properties of a large sample of galaxies, we
understand the way to analyse this through $N$-body models, and on
that basis we extract the data we want.  The purist might say that
there is no understanding that has grown out of this. This brings
to mind the comment made by the mathematician Russell Graham in
relation to computer proofs of mathematical theorems: he might ask
the all-knowing computer whether the Riemann hypothesis (the last
great unsolved problem of mathematics) is true.  It would be
immensely discouraging if the computer were to answer ``Yes, it is
true, but you will not be able to understand the proof".  We would
know that something is true without benefiting from the experience
gained from proving it.  This is to be compared with Andrew Wiles'
proof of the Fermat Conjecture \cite{wiles95} which was merely a
corollary of some far more important issues he had discovered on
his way: through proving the fundamental Taniyama-Shimura
conjecture we can now relate elliptic curves and modular forms
\cite{horgan93}.

We may feel the same way about running parameter-adjusted computer
models of the Universe.  Ultimately, we need to understand why
these parameters take on the particular values assigned to them.
This inevitably requires analytic or semi-analytic understanding
of the underlying processes.  Anything less is unsatisfactory.

\subsubsection{Redshift distortions}
\begin{figure}
\includegraphics[width=8.5 cm]{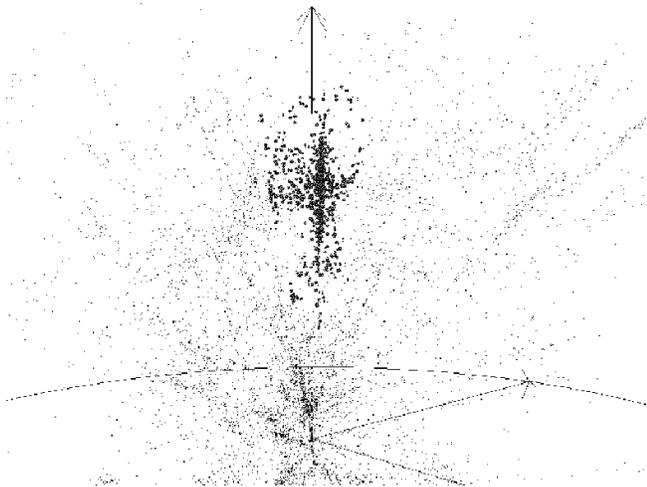}
\caption{\label{fig:finger} A view of the three dimensional
distribution of galaxies in which the members of the Coma cluster
have been highlighted to show the characteristic ``finger-of-God"
pattern, from \textcite{lars}.}
\end{figure}

Viewed in redshift space, which is the only three-dimensional view
we have, the universe looks anisotropic: the distribution of
galaxies is elongated in what have been called ``fingers-of-god"
pointing toward us (a phrase probably attributable to Jim
Peebles).  These fingers-of-god appear strongest where the galaxy
density is largest (see Fig.~\ref{fig:finger}), and are
attributable to the extra ``peculiar" (ie: non-Hubble) component
of velocity in the galaxy clusters. This manifests itself as
density-correlated radial noise in the radial velocity map.

Since we know that the real 3-dimensional map should be
statistically isotropic, this finger-of-god effect can be filtered
out. There are several techniques for doing that: it has become
particularly important in the analysis of the vast 2dF (2 degree 
Field) and SDSS (Sloan Digital Sky Survey)
surveys \cite{TegHam}. The earliest discussion of this was probably
\textcite{Davis83}.

There is another important macroscopic effect to deal with
resulting from large scale flows induced by the large scale
structure so clearly seen in the CfA-II Slice \cite{lapparent}.
Matter is systemically flowing out of voids and into filaments;
this superposes a density-dependent pattern on the redshift
distribution that is not random noise as in the finger-of-god
phenomenon.  This distorts the map \cite{Sargent77, Kaiser87,
ham98}. As this distortion enhances the visual intensity of galaxy
walls, which are perpendicular to the line-of-sight, it is called
``the bull's-eye effect'' \cite{bullseye}.

\subsubsection{Flux-limited surveys and selection functions}
Whenever we see a cone diagram of a redshift survey (see
Fig.~\ref{fig:cone}), we clearly notice a gradient in the number
of galaxies with redshift (or distance). This  artefact is
consequence of the fact that redshift surveys are flux-limited.
Such surveys include all galaxies in a given region of the sky
exceeding an apparent magnitude cutoff. The apparent magnitude
depends logarithmically on the observed radiation flux. Thus only
a small fraction of intrinsically very high luminosity galaxies
are bright enough to be detected at large distances.

For the statistical analyses of these surveys there are two
possible approaches:
\begin{enumerate}
\item {\it Extracting volume-limited samples.}
Given a distance limit, one can calculate, for a particular
cosmological model, the minimum luminosity of a galaxy that still
can be observed at that distance, considering the flux limit of
the sample. Galaxies in the whole volume fainter that this
luminosity will be discarded. The remaining galaxies form a
homogeneous sample, but the price paid ---ignoring much of the
hard-earned amount of redshift information--- is too high.

\item {\it Using selection functions.}
For some statistical purposes, such as measuring the two-point
correlation function, it is possible to use all galaxies from the
flux-limited survey provided that we are able to assign a weight
to each galaxy inversely proportional to the probability that a
galaxy at a given distance $r$ is included in the sample: this is
dubbed {\it the selection function} $\varphi(r)$. This quantity is
usually derived from the luminosity function, which is the number
density of galaxies within a given range of luminosities. A
standard fit to the observed luminosity function is provided by
the Schechter function \cite{Schechter}
\begin{equation}
\phi(L) dL = \phi_{\ast} \left ({L \over L_{\ast}} \right )^{\alpha}
\exp \left (- {L \over L_{\ast}} \right ) d \left ( {L \over
L_{\ast}} \right ) ,
\label{Eq:lum}
\end{equation}
where $\phi_{\ast}$ is related to the total number of galaxies and
the fitting parameters are $L_{\ast}$, a characteristic
luminosity, and the scaling exponent $\alpha$ of the  power-law
dominating the behavior of Eq.~\ref{Eq:lum} at the faint end.

The problem with that approach is that the luminosity function has
been found to depend on local galaxy density and morphology. This
is a recent discovery and has not been modelled yet.
\end{enumerate}

\subsubsection{Corrections to redshifts and magnitudes}

The redshift distortions described earlier can be accounted for
only statistically \cite{TegHam}; there is no way to improve
individual redshifts. However, individual measured redshifts are
usually corrected for our own motion in the rest frame determined
by the cosmic background radiation. This motion consists of
several components (the motion of the solar system in the Galaxy,
the motion of the Galaxy in the Local Group (of galaxies), and the
motion of the Local Group with respect to the CMB rest frame). It
is usually lumped together under the label ``LG peculiar
velocity'' and its value is $v^{LG}=627\pm22$ km s$^{-1}$ toward
an apex in the constellation of Hydra, with galactic latitude
$b=30^\circ\pm3^\circ$ and longitude $l=276^\circ\pm3^\circ$ (see,
e.g., \textcite{ham98}). If not corrected for, this velocity
causes a so-called ``rffect'' \cite{Kaiser87}, an apparent
dipole density enhancement in redshift space. Application of this
correction has several subtleties: see \textcite{ham98}.

Most corrections to measured galaxy magnitudes are usually made
during construction of a catalog, and are specific to a catalog.
There is, however, one universal correction: galaxy magnitudes are
obtained by measuring the flux from the galaxy in a finite width
bandpass. The spectrum of a far-away galaxy is redshifted, and the
flux responsible for its measured magnitude comes from different
wavelengths. This correction is called the ``K-correction''
\cite{Humason56}; the main problem in calculating it is
insufficient knowledge of spectra of far-away (and younger)
galaxies. In addition, directional corrections to magnitudes have to be
considered due to the fact that the sky is not equally transparent in
all directions. Part of the light coming from extragalactic objects is
absorbed by the dust of the Milky Way. Due to the flat shape of our galaxy,
the more obscured regions correspond to those of low galactic latitude,
the so-called zone of avoidance, 
although the best way to account for this effect is to use the
extinction maps elaborated from the observations \cite{dust}.

\subsection{The first generation of redshift surveys}

\subsubsection{CfA surveys}
The first CfA redshift survey was undertaken by \textcite{Huchra83} 
who mapped some
2400 galaxies down to $m \simeq 14.5$ taken from the Zwicky
catalog. This survey was too sparse to show definite structure.

\begin{figure*}
   \includegraphics[height=7.404cm]{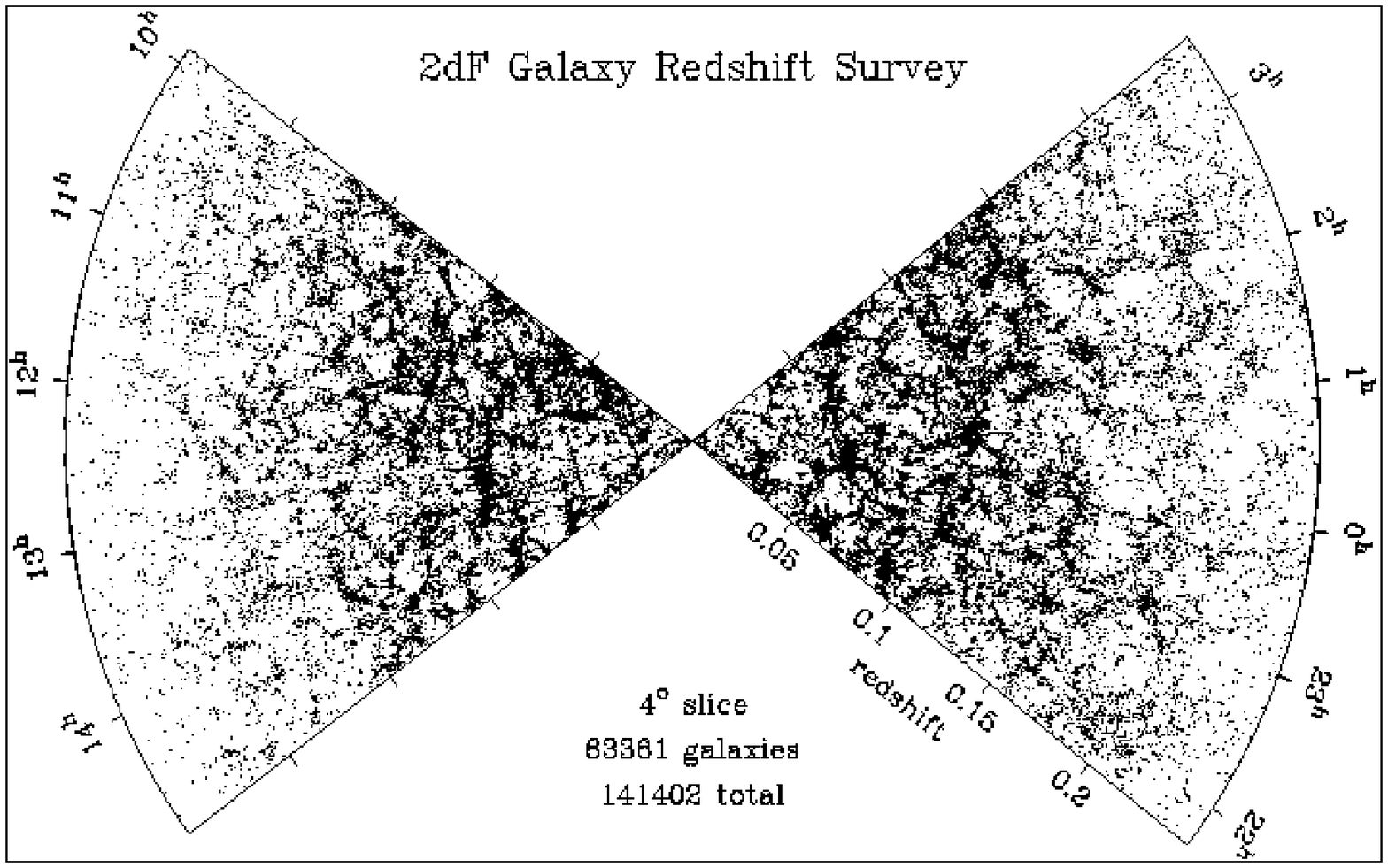}
   \includegraphics[height=1.254cm]{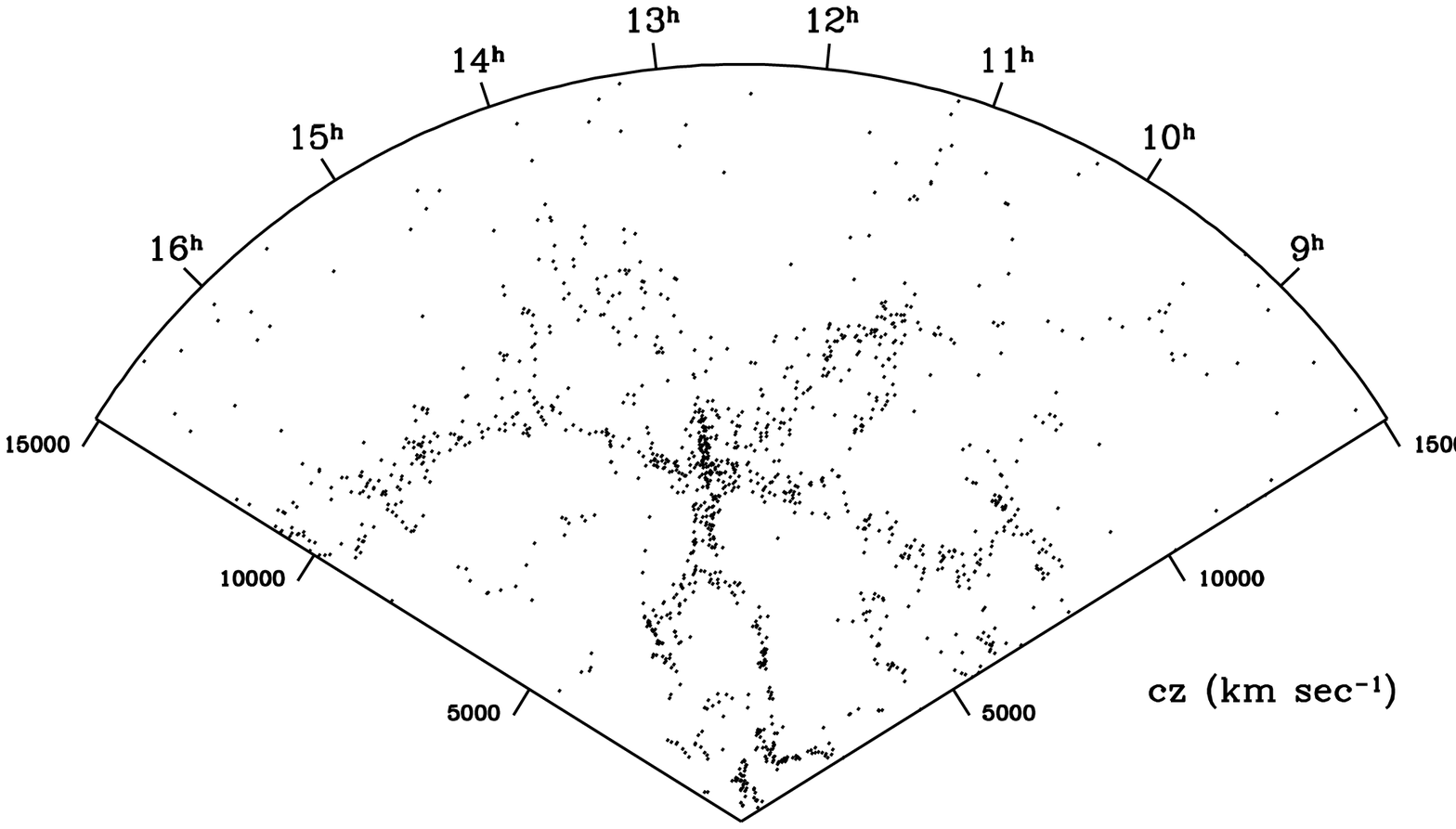} \\
   \includegraphics[height=10.92cm]{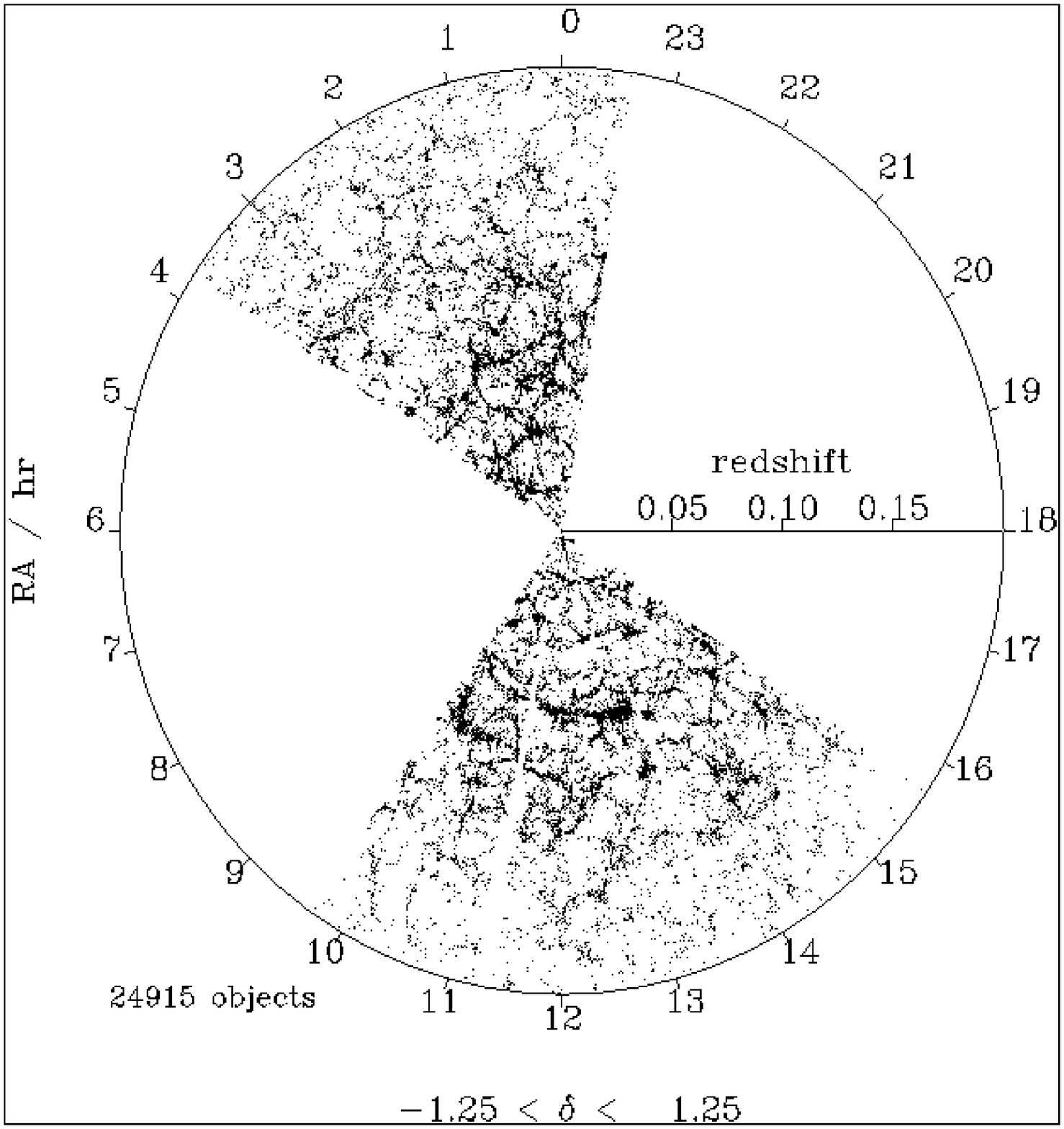}
\caption{\label{fig:cone} The top diagram shows two slices of
$4^\circ$ width and depth $z=0.25$ from the 2dF galaxy redshift
survey, from \textcite{PeaNat01}. The circular diagram at the
bottom has a radius corresponding to redshift $z=0.2$ and shows
24,915 galaxies from the SDSS survey, from \cite{loveday}). As an
inset on the right, the first CfA-II slice from
\textcite{lapparent} is shown to scale.}
\end{figure*}

The first survey to truly reflect the cosmic structure was the
first CfA-II slice of \textcite{lapparent}, the ``Slice of the
Universe'' (the smallest wedge in Fig.~\ref{fig:cone}). The slice
showed very clearly the ``bubbly'' nature of the large-scale
structure, as the authors defined it.  This important discovery
generated a lot of publicity: cartoons appeared in newspapers
depicting females with their arms in a sink full of soap bubbles,
and the {\it Encyclopaedia Britannica} was updated to include a
picture of the slice.

Prior to that there had been smaller surveys, such as the
Perseus-Pisces region survey of \textcite{gioha} and the
Coma-A1367 survey of \textcite{chinca}. These surveys had revealed
rich structures in the distribution of galaxies, similar to
Zel'dovich's predicted pancakes and voids.  But since they were
restricted to a volume around a major cluster of galaxies they
could not be thought of as being representative of the universe as
a whole.

At first glance it may seem that similar critique applies also to
the CfA surveys, since the first CfA slice \cite{lapparent} was
indeed centered on the Coma cluster. However, the breadth of the
slice (some 120 degrees on the sky) samples a far greater volume,
and it was very deep for that time, extending to about
150$h^{-1}$Mpc. The slice also contains an unusual number of rich
galaxy clusters. Subsequent surveys, the following CfA slices and
the ESO Southern survey \cite{costa91} amply confirmed the
impression given by the CfA slice.

The main source for redshifts during those years was 'Zcat', a
heterogeneous compilation of galaxy redshifts by J. Huchra. But it
took many years before the data from the CfA slices entered the
public domain. This was unfortunate since many other groups would
have liked to try their own analysis techniques on such a well
defined sample.  By the time that the data became available there
existed already more substantial surveys with publicly available
data and much of the impetus of the CfA slices, apart from the
fine work done by the Harvard group itself, was lost.

The work to improve and extend the CfA surveys has continued. The
Century Survey \cite{Geller97} covers the central $1^\circ$ region
of the famous CfA-II slice, but is much deeper, extending to $R =
16.1$ in the apparent magnitude and to 450$h^{-1}$Mpc in space.
The final CfA catalog is the Updated Zwicky Catalog \cite{falco}
that includes uniform measurements of almost all (about 19,000)
galaxies of the Zwicky catalog (with the magnitude limit of
$m_{Zw} \approx 15.5$) in the northern sky. 
Nowadays catalogs are made public as soon as possible; the CfA
redshift catalogs can be obtained from the web-page of the
Smithsonian Astronomical Observatory Telescope Data Center
(http://tdc-www.harvard.edu/).

\subsubsection{SSRS and ORS}
The Southern Sky Redshift Survey \cite{costa91} was meant to
complement the original CfA survey, mapping galaxies in the
southern sky. It includes almost 2000 redshifts; the followup
survey, the extended SSRS \cite{costa98} with about 5400 redshifts
mirrored the Second CfA survey for the southern sky. These
catalogs were mostly used for comparison with the CfA survey
results; they were made public at once and produced many useful
results.  Presently they are available from the Vizier database
(http://vizier.u-strasbg.fr).

The Optical Redshift Survey \cite{Santiago}, had a depth of
80$h^{-1}$Mpc, similar to the first CfA survey, but attempted a
complete coverage of the sky (except for the dusty avoidance zone
around the galactic equator). They measured about 1300 new
redshifts, including about 8500 redshifts in total. This survey
was heavily exploited to describe the nearby density fields, to
estimate the luminosity functions, galaxy correlations, velocity
dispersions etc. The catalog and the publications can be found in
http://www.astro.princeton.edu/$\sim$strauss/ors/.

\subsubsection{Stromlo-APM and Durham/UKST redshift surveys}
The Stromlo-APM redshift survey \cite{Loveday96} is a sparse
survey (1 in 20) of some 1800 optically selected galaxies brighter
than the apparent  magnitude limit $B \approx 17$ taken from the
APM survey of the Southern sky. As the APM survey \cite{maddox90}
itself, the Stromlo-APM survey was an important data source and
generated several important results on correlation functions in
real and redshift space, power spectra, redshift distortions,
cosmological parameters, bias and so on.  It was eventually put
into the public domain, although rather too late to be of much use
to any third party investigators.

The APM survey was also used to generate a galaxy cluster
catalog. The APM cluster redshift catalog \cite{Dalton97} was
the first objectively defined cluster catalog.  It not only
provided important data on the distribution of clusters, it also
provided an assessment of the reliability of the only cluster
source available before that, the Abell cluster catalog.

The Durham/UKST redshift survey \cite{Ratcliffe} measured
redshifts for about 2500 galaxies around the South Galactic Pole.
The depth of the survey was similar to that of the Stromlo-APM
survey, and it was also a diluted survey sampling 1 galaxy in 3.

These catalogs can be found now at the Vizier site (see above).

\subsubsection{IRAS redshift samples: PSCz} 

The story of the IRAS (Infrared Astronomical Satellite) 
redshift catalogs stresses the importance
of having a good base photometric catalog before starting to
measure redshifts. As galactic absorption in infrared is much
smaller than in the optical bands, the IRAS Point Source Catalog (PSC)
covers uniformly almost all of the sky. This catalog was used to
select galaxies for redshift programs, which extended down to
successively smaller flux limits: the 2 Jy survey of
\textcite{Strauss92} with 2658 galaxies; the 1.2 Jy survey of
\textcite{Fisher95} added 2663 galaxies; and the 0.6 Jy
sparse-sampled (1 in 6) QDOT  survey of \textcite{Lawrence99} with
2387 galaxies. This culminated in the PSCz survey of some 15000
galaxies by \textcite{Saunders00}, which includes practically all
IRAS galaxies within the 0.6 Jy flux limit.

The IRAS redshift catalogs have been used for the usual battery
of large-scale studies, but their main advantage is their full-sky
coverage (about 84\%). This allows using the Wiener-type
reconstruction methods to derive the true density and velocity
fields, and to get an independent estimate of the biasing
parameter. The first fields to be studied were taken from the 2 Jy
survey by \textcite{Yahil91}, the last fields came from the PSCz
survey by \textcite{branchini99} and \textcite{schmoldt99}.

The PSCz survey has also been used for fractal studies. Although
the IRAS samples are not too deep (PSCz extends to about
200$h^{-1}$Mpc), \textcite{pancoles00} found that multifractal
analysis shows a definite crossover to homogeneity already before
this scale.

\subsubsection{ESO Deep Slice and the Las Campanas redshift survey}
The ESO Deep Slice \cite{ESO} measured redshifts of 3300 galaxies
down to the blue magnitude $b_J = 19.4$ in the $B_J,R,I$
photometric system \cite{gullixson}. The surveyed region is a
$1^\circ\times22^\circ$ strip of depth about 600$h^{-1}$Mpc. The
most interesting discussion that this data caused was about the
fractal nature of the large-scale galaxy distributions. While
\textcite{Scara98} found the correlation dimension $D=3$,
\textcite{Joyce99} showed that a more reasonable choice of the
K-correction (redshift-dependent apparent dimming of galaxies)
gave a clearly fractal $D=2$ correlation dimension.

The Las Campanas Redshift Survey \cite{lcrs} had a similar
geometry, six thin parallel slices ($1.5^\circ \times 90^\circ$)
with the depth about 750$h^{-1}$Mpc ($z\approx0.25$). The survey
team measured redshifts of about 24000 galaxies in these slices.
This was the first deep survey of sufficient volume that it could
be used to test if our knowledge of the nearby Universe was
sufficient to describe more distant regions. The usual tests
included the luminosity functions (these were found to depend on
galaxy density and morphology), second- and third-order
correlation functions, power spectra, and fractal properties. A
catalog of groups of galaxies was generated. The survey results
were quickly made public: the general interest in the data was
high and close to a hundred papers have been published using these
data.

\subsection{Recent and on-going Surveys}

\subsubsection{2dF galaxy redshift survey}
The 2dF multi-fiber spectrograph on the 3.9m Anglo-Australian
Telescope is capable of observing up to 400 objects simultaneously
over a field of view some 2 degrees in diameter, hence the name of
the survey.  The sample of galaxies targeted for having their
redshifts measured consists of some 250,000 galaxies located in
extended regions around the north and south Galactic poles. The
source catalog is a revised APM survey. The galaxies in the
survey go down to the magnitude $b_J = 19.45$.  The median
redshift of the sample is $z = 0.11$ and redshifts extend to about
$z \simeq 0.3$.  In mid-2001 the survey team released the data on
the first 100,000 galaxies, and published also an interim report
on the analysis of some 140,000 galaxies: \textcite{PeaNat01} and
\textcite{Percival01}.

The survey is already complete, and the resulting correlation
functions, redshift distortions and pairwise velocity dispersions
\cite{hawkins02} demonstrate the quality of the data set. The 2dFGRS
currently provides us with the best estimates for a large number of
cosmological parameters describing the population of galaxies. Not
only can we determine clustering properties of the sample as a
whole, but the sample can be broken down by galaxy absolute
brightness or by morphological type \cite{percival04}. 
The surveys's web
page is http://www.mso.anu.edu.au/2dFGRS/.

\subsubsection{Sloan digital sky survey}

Hot on the heels of the 2dF survey is an even larger survey: the
Sloan Digital Sky Survey (SDSS). The survey team has close to two
hundred members from 13 institutions in U.S., Europe, and Japan,
and uses a dedicated 2.5 m telescope. The initial photometric
program is measuring the positions and luminosities of about
$10^8$ objects in $\pi$ sterradians of the Northern sky, and the
follow-up spectroscopy is planned to give redshifts of about
$10^6$ galaxies and $10^5$ quasars. Good descriptions of the
survey can be found in \textcite{loveday} and on the surveys's web
page (http://www.sdss.org/).

The first official data release was done in 2003, but the
astronomical community had already have the chance to see and use the
data from a preliminary Early Data Release \cite{stoughton02}. These
data and the data from the commissioning phase of the project have
served as a basis for more than one hundred papers on such diverse
subjects as the study of asteroids, brown dwarf stars in the
vicinity of the Sun, remnants of destroyed satellites of our Galaxy,
star formation rates in galaxies, galaxy luminosity functions, and,
of course, on the statistics of the galaxy distribution.

The main difference between the 2dF and the SDSS surveys, apart of
their data volume and sky coverage, is the fact that they are
based on different selection rules. While the 2dF survey is a
blue-magnitude limited survey with $b_{\lim} =19.45$, 
the limiting magnitude of the SDSS
survey is red $r_{\lim} =17.77$. 
This causes considerable differences in galaxy
morphologies of the two surveys. Also, while the depths of the
main surveys are similar ($z\approx0.25$), a part of the SDSS
survey, including about $10^5$ luminous red galaxies, will reach
redshifts $z\approx0.5$.

\subsubsection{2MASS and 6dF}
 
\begin{figure*}
\includegraphics[width=15 cm]{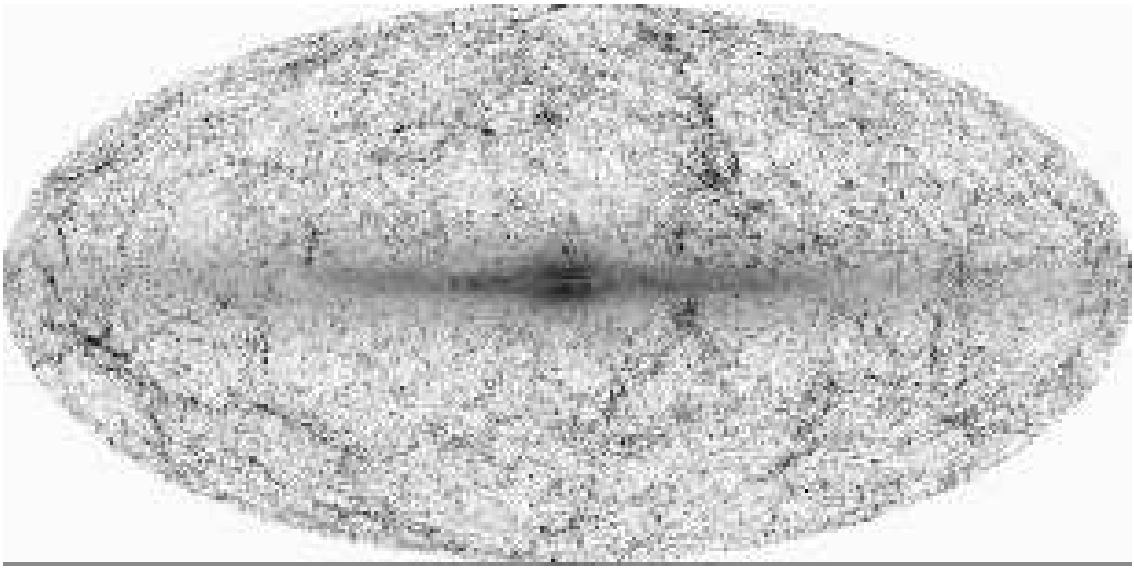}
\caption{\label{fig:2mass} The near-infrared view of the local
universe provided by the 2MASS survey. Beyond the Milky Way lying
at the Galactic equator, more than 1.5 million galaxies are depicted
using a grey-scale code based on their photometrically
deduced redshift, from \textcite{jarret04}.}
\end{figure*}

The Two Micron All Sky Survey (2MASS) has scanned the whole sky in
three different near-infrared bands. The Extended Source Catalog
(XSC) is the 2MASS galaxy catalog \cite{jarret04} and contains more
then 1.5 million galaxies, mapping rather well the zone of avoidance, 
The view of our local universe provided by
2MASS is shown in Fig.~\ref{fig:2mass}.

The 6dF galaxy survey \cite{jones04} targeted on the 2MASS galaxy
catalog (XSC) will encompass twice the volume of the PSCz and will
contain ten times more galaxies, allowing combined knowledge of
galaxy masses and redshift. It will be the best sample for studies
of the peculiar velocity field, allowing a better understanding of
the relation of galaxy clustering with mass, and hence providing
important clues to understand how bias depends on the scale.

\subsubsection{Deep spectroscopic and photometric surveys}

Deep spectroscopic surveys such as the Canadian Network for
Observational Cosmology (CNOC2) \cite{cnoc2}, DEEP2 \cite{deep2}, and
the Visible Imaging Multi-Object Spectrograph (VIRMOS-VLT) survey 
\cite{virmos} have allowed the study of the evolution of
clustering with redshift and with various morphological properties
of galaxies \cite{carlberg, coil}. Nevertheless, it is extremely
difficult to measure redshifts of very faint objects. The
present limit reached making use of the largest 
ground-based telescopes is about $I \simeq 24$. An alternative 
to spectroscopy, is the poor man $z$- machine \cite{koo}, 
provided by multi-wavelength imaging. 

Following the pioneering work of \textcite{baum} and \textcite{koo},
\textcite{fsoto} have shown that it is possible to reliably 
estimate redshifts using CCD images at different wavebands 
---the so called photometric redshifts---. 
This technique is particularly useful when mapping the
very distant universe because galaxies in deep surveys could not be
spectroscopically observable. Bayesian techniques have been
introduced to improve the accuracy of the photometric redshift
estimation \cite{benitez00}.

Different surveys reaching extremely large depths are providing us
with the possibility of analyzing the evolution of clustering with
cosmic time. We can mention the COMBO17 survey (Classifying Objects
by Medium-Band Observations) which lists photometry in 17 passbands
\cite{wolf04}, the Calar Alto Deep Imaging Survey (CADIS), used by
\textcite{phleps} to show how the clustering strength grows from
$z=1$ to the present epoch and its dependence on morphological type, 
and the recently released  Great Observatories Origins Deep Survey
(GOODS) described in \textcite{goods}. 
The SDSS provides also photometric information in five bands
allowing the measurement of photometric redshifts for a
volume-limited sample containing more than 
2 million galaxies within the
range $0.1 < z <0.3$. Analyzing the
angular two-point correlation function of this survey, 
\textcite{budavari} have
found an interesting bimodal behavior between {\it red}
elliptical-like galaxies and {\it blue} galaxies. 

The recent project named the ALHAMBRA-survey (Advanced Large, 
Homogeneous Area Medium Band Redshift Astronomical survey) is being 
carried out by Moles and collaborators using the 3.5m Calar 
Alto telescope. The photometric survey will cover an area of
eight square degrees. Imaging will be performed using 20 optical
filters plus three standard bands in the near infrared. It is 
expected to collect about 600,000 photometric galaxy redshifts
with an accuracy of $\Delta z < 0.015 (1+z)$. This 
photometric survey, midway between the wide-angle 
spectroscopic surveys and the narrow imaging surveys, is 
deep enough and wide enough to be extremely useful for 
all kind of studies involving cosmic evolution.

\subsection{The radio, X-ray and $\gamma$-ray skies}

The 1950's was a great era for cataloguing radio sources, much of
the work being done at Cambridge in England (with the 2C, 3C, etc.
surveys) and at Parkes in Australia. The surveys were done at
considerably different frequencies and gave disparate views of the
source counts.  This had a strong influence on the Steady State
versus Big Bang debate, each survey being used to support a
different cosmological hypothesis.

The sources in early surveys were randomly distributed over the
sky (for instance, \onlinecite{Holden66} on the Third Cambridge
Catalog and \onlinecite{Payne67} on the southern counterpart).
This remained true for later surveys at low frequencies, which
found, for the most part, intrinsically very bright sources at
somewhat larger distances (for instance, \onlinecite{Webster76}
analyzing the Fourth Cambridge and Greenbank surveys, and
\onlinecite{Masson79} on the Sixth Cambridge Catalog).  Indeed
it remains true down to the present day  \cite{Trimble01}, for the
low-frequency surveys that pick out large, bright, steep-spectrum,
extended double sources: \textcite{Artyukh00, Venturi00} reported
that they did not even identify the Shapley concentration).  What
this means is that, on average, there is only one of these sources
in each of the largest-scale structures to be found in the local
universe.  The absence of clustering is, therefore, in some sense
evidence for the existence of ``largest structures," though
Artyukh and Venturi {\it et al.} note that mergers of small groups
into large clusters and superclusters may well turn off fainter
radio sources that would otherwise reveal intermediate structure.

In contrast, higher frequency surveys that yield intrinsically
fainter radio galaxies find that they are clustered very much like
radio-quiet galaxies of the same Hubble types
(\onlinecite{Cress96} on the Faint Images of the Radio Sky at 
Twenty-cm (FIRST) survey from the Very Large Array (VLA), and
\onlinecite{Maglioc} a further analysis of FIRST, showing that the
distribution of those radio sources in space is consistent with
their having grown by gravitational instabilities from Gaussian
initial conditions). Returning to the Shapley concentration,
\textcite{Venturi02} found no fewer than 124 radio sources there.

Distant radio sources (of which quasars are an important sort) are
rather sparsely distributed throughout the Universe and are
consequently not good indicators of large scale structure.  It is
therefore not surprising that radio source catalogs provide
little evidence for the large scale clustering.

Galaxy clusters are prominent  features of the X-ray sky that can
provide a good measure of the large scale clustering.  X-ray
selected samples of clusters are less prone to bias than
catalogs for clusters selected from maps of the galaxy
distribution.  One problem, however, is that the selection
criteria for galaxy clusters selected from X-ray surveys
\cite{borguz} are quite different from the selection criteria for
clusters selected from optically scanned photographic plates
\cite{Dalton97} and it is not so easy to relate studies based on
the two sources of data.

The REFLEX (ROSAT-ESO Flux Limited X-ray) cluster survey contains
449 clusters, covering an area of $4.24$ steradians in the
southern hemisphere ($\delta < 2.5^{\circ}$). It is complete at $
\geq 90$\%, down to a nominal flux limit of $3 \times 10^{-12}$
erg s$^{-1}$ cm$^{-2}$ in the $0.1 - 2.4$ keV band. REFLEX, as
other cluster samples, shows unambiguously very large-scale
inhomogeneities that appear when the clustering power is measured
and compared with that of galaxies at the same scales
\cite{guzzorev}.

\subsection{Distribution of quasars and Ly-$\alpha$ clouds}

The spectra of quasars are populated by narrow absorption lines
from intervening gas clouds along the line of sight (the
Ly-$\alpha$ forest). Owing to the great redshift of most quasars
these absorption clouds provide an important probe of clustering
at large distances and at times long in our past.

\textcite{wurees} used the large-scale uniformity of the
Ly-$\alpha$ forest to argue against fractal distribution of
matter. Recently, \textcite{croft02} showed that it is possible to
estimate the full 3-D power spectrum of density fluctuations
$P(k)$ from the (one-dimensional) Ly-$\alpha$ flux power spectrum.
This is extremely important, as it allows us to check for
theoretical predictions at large redshifts ($z\approx$2--4). It
also allows us to recover the linear (post-recombination) power
spectrum for small scales, which have turned nonlinear by now.

Lines of sight to quasar pairs, be they optical pairs or pairs
that are a consequence of gravitational lensing, provide
additional clues to the clustering transverse to the line of sight
\cite{wurees}.

The statistical analysis of the distribution of quasars and
Ly-$\alpha$ clouds has provided additional evidence for
the large scale homogeneity in the universe \cite{andreani91,carbone}.

\subsection{The cosmic microwave background}
The importance of the CMB anisotropy measurements cannot be
over-emphasized and would warrant an entire review by itself. From
the point of view of this article we are concerned with knowing
the initial conditions for galaxy formation and the parameters of
the cosmological framework within which galaxy formation takes
place.  Given that data, the task is to derive the currently
observed clustering properties of galaxies in the Universe.

\subsubsection{Structure before our eyes}
Arguably the most important observation in the study of clustering
is the recent measurement of the structure in the cosmic microwave
background radiation at the time of recombination.  This structure
was predicted independently by \textcite{silk67} and by
\textcite{sacwolf}, although the phenomenon is generally referred
to as the ``Sachs-Wolfe" effect. Understanding the details of how
the structure in the microwave background arises in any of a vast
number number of cosmological models has been a cosmic
folk-industry spanning some 30 years.  The results are
encapsulated in a run-it-yourself computer program of Zaldarriaga
and Seljak, 2000 (see
http://physics.nyu.edu/matiasz/CMBFAST/cmbfast.html).

\begin{figure}
\centering

\resizebox{9.2cm}{!}{\rotatebox{90}{\includegraphics*{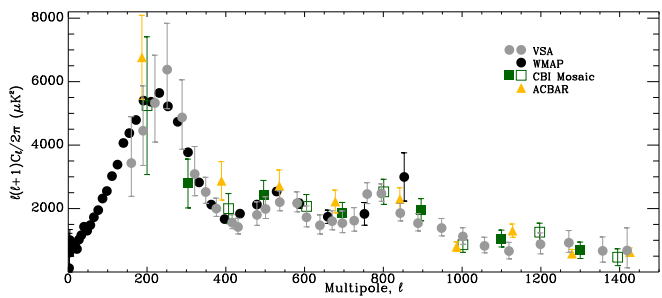}}}
\caption{\label{fig:rebolo} The agreement between the estimated
power spectrum of the CMB anisotropies from four different experiments
with similar sensitivity, from \textcite{Dickinson04}.}
\end{figure}

The structure was first seen at about $7^\circ$ in angular
resolution in the data of the COBE satellite DMR experiment (Bennett
et al, 1996). Smaller structure has been detected in recent high
angular resolution experiments with names like DASI \cite{dasihal,
dasipryke}, MAXIMA-1 \cite{MAXhan, MAXlee, MAXbalbi} and
BOOMERANG-98 \cite{BOOM00, BOOM01, BOOM02}, and in the WMAP
first-year full-sky data \cite{Bennett03}. An analysis of the
cosmological conclusions to be drawn from the combination of these
is given by \textcite{CMBall} and by \textcite{Sperg03}; an example
of present data sets and the curves fitted to them is shown in
Fig.~\ref{fig:rebolo} where,
in addition to the WMAP power-spectrum, several other
recent experiments are shown (VSA analyzed by \textcite{Dickinson04},
CBI \cite{mason03} and ACBAR \cite{kuo04}), having
similar sensitivity, but being different in the frequency range and
observing techniques.

Here we observe unambiguously the structure in the gravitational
potential that will lead to the birth and clustering of galaxies
and clusters of galaxies as we see them today. We also observe
structure on scales far larger than can be traced by galaxies.

The units in Fig.~\ref{fig:rebolo} could use a little bit of
explanation. As the sky we see can be thought of as a surface of a
sphere, the distribution of temperature on the sky is analysed
into scales using Legendre polynomials $Y_l^m(\theta, \phi)$.  A
polynomial of order $l$ picks out structure on an angular scale
that is roughly, in degrees,
\begin{equation}
\theta^\circ \approx {180^\circ \over l}
\end{equation}
This corresponds to structure on a linear scale today of
\begin{equation}
 L=\frac{2\pi c}{H_0 \Omega_m^{0.4}\ell}
    \approx\frac{19000}{\ell} \Omega_m^{-0.4} h^{-1}\mbox{\rm Mpc}.
 \end{equation}
for a flat universe with $\Omega_m+\Omega_\Lambda=1$ \cite{vitosilk}.
The range of $l$-values covered by current experiments range over
about two decades:
\begin{equation}
10 < l < 1500
\end{equation}
with the limit of higher $l$-values being pushed upward all the
time. The low resolution end is from the COBE and WMAP data
\cite{cobe96,Bennett03}
and reveals inhomogeneities on scales in excess of $100 h^{-1}$
Mpc.

Notice that the highest resolution data still only cover linear
scales in excess of around $30h^{-1}$ Mpc and so we do not yet see
the initial condition for the scales over which the two-point
galaxy clustering correlation function is significantly greater
than zero.  We are just seeing the scales where rich cluster
clustering may be significant.  The prominent peak in the spectrum
at $l \sim 250$ corresponding to scales of around $50 h^{-1}$ Mpc
is intriguing.  We must not forget, however, that this is a peak
in a normalized spectrum; in the real matter $P(k)$ these peaks
are much less pronounced. There is evidence of oscillations in the
observed power spectra of clusters and galaxies, but current
surveys are not able yet to detect such structure with confidence
\cite{miller02,elgaroy}.

\subsubsection{Defining the standard model}
The presence of significant peaks in the angular distribution of
the cosmic microwave background strongly constrains the global
parameters that describe our Universe.  If these data are combined
with data from other sources, such as local determinations of the
Hubble constant and observations of very distant supernovae
\cite{snriess,snperl}, we
arrive at the so-called {\it concordance model} \cite{tegmark01}.
We hasten to add that this is not a term we invented: it might
have been OK to use the term {\it standard model}, but the high
energy physicists got there first.  The actual values of the
parameters in the concordance model depends on whose paper we
read: there is a little disaccord here, though it would seem to be
relatively minor.  It all depends on what prior knowledge is
assumed when making fitting the model to the data.  The error bars
are impressively small.

\subsubsection{Initial conditions for galaxy formation}
One of the best determined parameters is the slope $n$ of the
power spectrum of the pre-recombination inhomogeneities.  It was
suggested by Harrison and by Zel'dovich that $n=1$ on the grounds
that (a) the spectrum had to be a power law (what else could it
be?) and that (b) this value of the slope was the value that did
the minimal violence to the geometry of space-time on either the
large or small scales.  Following on Guth's brilliant notion of
inflationary cosmology \cite{guth}, many subsequent revisions of
the inflationary model and theories for the origin of cosmic
fluctuations gave physical reasons why we should have $n=1$ (e.g.:
\textcite{guthpi}, \textcite{star82}, \textcite{linde82,linde83,linde94}).

The DASI experiment \cite{dasipryke} gives
\begin{equation}
n = 1.01 ^{+0.08}_{-0.06}
\end{equation}
where the error bars are $68$\% confidence limits.  This result
comes from fitting the DASI data alone, making typical prior
assumptions about such things as the Hubble constant. The recent
WMAP data gives a value
\begin{equation}
n=0.99 \pm 0.04
\end{equation}
\cite{Sperg03}. (This latter value comes from the WMAP data alone,
no other data is taken into account.) Other similar numbers come
from \textcite{wang} and \textcite{miller01}.

It is perhaps appropriate to point out that this fit comes from
data on scales bigger than the scale of significant galaxy
clustering and that it is a matter of belief that the primordial
power law continued in the same manner to smaller scales.
In fact, more complex inflationary models predict a slowly
varying exponent (spectral index) (see, e.g., \textcite{Turner95});
this is in accordance with the WMAP data.
The scales which are relevant to the clustering of galaxies are just
those scales where the effects of the recombination process on the
fluctuation spectrum are the greatest. We believe we understand
that process fully \cite{hu97, hu01} and so we have no hesitation
in saying what are the consequences of having an initial $n=1$
power spectrum.  That, and the success of the $N$-body experiments,
provide a good basis for the belief that $n\approx1$ on galaxy
clustering scales.
Anyway, it is probable that the
Sunyaev-Zel'dovich effect \cite{SZ80} will dominate on the scales
we are interested in so we may never see the recombination-damped
primordial fluctuations on such scales.

We therefore have a classical initial value problem: the
difficulty lies mainly in knowing what physics, subsequent to
recombination, our solution will need as input and knowing how to
compare the results of the consequent numerical simulations with
observation.
CMB measurements can also give us valuable clues for these later
epochs in the evolution of the universe. A good example is the
discovery of significant large-scale CMB polarization by the WMAP
team \cite{Kogut03} that pushes the secondary re-ionization
(formation of the first generation of stars) back
to redshifts $z\approx20$.

\section{MEASUREMENTS OF CLUSTERING}

\subsection{The discovery of power-law clustering}

The pioneering work of Rubin and Limber has already been
mentioned. These early authors were limited by the nature of the
catalogs that existed at the time and the means to analyze the
data -- there were no computers!

It was \textcite{totsuji} and, independently, \textcite{peeb74}
who were first to present a computer-based analysis of a complete catalog
of galaxies. Totsuji and Kihara used the published Lick counts in
cells from \textcite{shane67}, while Peebles and coworkers
analysed a number of catalogs: the Reference catalog of Bright
Galaxies, the Zwicky catalog, the Lick catalog and later on
the very deep Jagellonian field \cite{peha, peebles75a,
peebles75b}. All this work was done on the projected distribution
of galaxies since little or no redshift information was available.

The central discovery was that the two-point correlation function
describing the deviation of the galaxy distribution from
homogeneity scales like a simple power law over a substantial
range of distances.  This result has stood firm through numerous
analyses of diverse catalogs over the subsequent decades.

The amplitudes of the correlation functions calculated from the
different catalogs were found to scale in accordance with the
nominal depth of the catalog.  This was one of the first direct
proofs that the Universe is homogeneous. Before that we knew about
the isotropy of the galaxy distribution at different depths and
could only infer homogeneity by arguing that we were not at the
center of the Universe.

\subsection{The correlation function: galaxies}

\subsubsection{Definitions and scaling}

The definition of the correlation function used in cosmology
differs slightly from the definition used in other fields.  In
cosmology we have a nonzero mean field (the mean density of the
Universe) superposed on which are the fluctuations that correspond
to the galaxies and galaxy clusters. Since the Universe is
homogeneous on the largest scales, the correlations tend to zero
on these scales.

On occasion, people have tried to use the standard definition and
in doing so have come up with anomalous conclusions.

The right definition is: In cosmology, the 2-point galaxy
correlation function is defined as a measure of the excess
probability, relative to a Poisson distribution, of finding two
galaxies at the volume elements $dV_1$ and $dV_2$ separated by a
vector distance $\mathbf{r}$:
\begin{equation}
dP_{12} = n^2 [1 + \xi(\mathbf{r})] dV_1 dV_2,
\label{eq:xidef1}
\end{equation}
where $n$ is the mean number density over the whole sample volume.
When homogeneity\footnote{This property is called stationarity in
point field statistics.} and isotropy are assumed
$\xi(\mathbf{r})$ depends only on the distance $r=|\mathbf{r}|$.
From Eq.~(\ref{eq:xidef1}), it is straightforward to derive the
expression for the conditional probability that a galaxy lies at
$dV$ at distance $r$ given that there is a galaxy at the origin of
$\mathbf{r}$.
\begin{equation}
dP = n [1 + \xi(r)] dV.
\label{eq:xidef2}
\end{equation}
Therefore, $\xi(r)$ measures the clustering in excess ($\xi(r) >
0$) or in defect ($\xi(r) < 0$) compared with a random Poisson
point distribution, for which $\xi(r)=0$. It is worth to mention
that in statistical mechanics the correlation function normally
used is $g(r)=1+\xi(r)$ which is called the \textit{radial
distribution function} \cite{statmec}. Statisticians call this
quantity  the \textit{pair correlation function} \cite{stoyan2}.
The number of galaxies, on average, lying at a distance between
$r$ and $r+dr$ from a given one is $ng(r)4\pi r^2$.

A similar quantity can be defined for projected catalogs: surveys
compiling the angular positions of the galaxies on the celestial
sphere. The angular two-point correlation function, $w(\theta)$,
can be defined by means of the conditional probability of finding
a galaxy within the solid angle $d\Omega$ lying at an angular
distance $\theta$ from a given galaxy (arbitrarily chosen):
\begin{equation}
dP = {\cal N}  [1+ w(\theta)]d\Omega , \label{angcor}
\end{equation}
Now,  ${\cal N}$ is the mean number density of galaxies per unit
area in the projected catalog. Since the first available
catalogs were two-dimensional, with no redshift information,
$w(\theta)$ was measured before any direct measurement of $\xi(r)$
was possible. Nevertheless, $\xi(r)$ can be inferred from its
angular counterpart $w(\theta)$  by means of the Limber equation
\cite{Rubin54,Limber54} which provides an integral relation
between the angular and the spatial correlation function for small
angles,
\begin{equation}
\label{limber}
w(\theta)=\int_0^\infty y^4\phi^2(y)\,dy
    \int_0^\infty\xi\left(\sqrt{x^2+y^2\theta^2}\right)\,dx.
\end{equation}
Here $y$ is the comoving distance and $\phi(y)$ is the radial
selection function normalized such that $\int\phi(y)y^2dy=1$. If
$\xi(r)$ follows a power law $\xi(r)=(r/r_0)^{-\gamma}$, it is
straightforward to see that the angular correlation function is
also a power law,  $w(\theta) = A  \theta^{1-\gamma}$
\cite{peeb80}. \textcite{totsuji} were the first to derive a
power-law model for $\xi(r)$ on the basis of the angular data.
Their canonical value for the scaling exponent $\gamma =1.8$ has
remained unaltered for more than 30 years. Eq.~\ref{limber}
provides the basis for an important scaling relation.
\textcite{peeb80} has shown that, in a homogeneous universe,
$w(\theta)$ must scale with the sample depth $D_{\ast}$ as
\begin{equation}
w(\theta) = \frac{1}{D_\ast} W(\theta D_\ast) \label{angscl}
\end{equation}
where the function $W$ is an intrinsic angular correlation
function which does not depend on the apparent limmiting
magnitude of the sample.
The characteristic depth $D_\ast$ is the distance at which a
galaxy with intrinsic luminosity $L_\ast$ is seen at the limiting
flux density $f$, which is in the Euclidean geometry (neglecting
expansion and curvature),
\begin{equation}
D_\ast = \sqrt{L_\ast \over 4 \pi f},
\end{equation}
or, in terms of magnitudes,
\begin{equation}
D_\ast = 10^{0.2(m_0-M_\ast) -5}h^{-1}\mbox{Mpc},
\end{equation}
where $m_0$ is the apparent limiting magnitude of the sample. The
scaling relation in Eq.~(\ref{angscl}) can be deduced from the Limber
equation (\ref{limber}) assuming that distribution of galaxies is
homogeneous on average and therefore ${\cal N} \propto D_\ast^3$.
\textcite{peebles} has shown that the analysis of the deep
catalogs of galaxies on the basis of the scaling law
(Eq.~\ref{angscl}) argues strongly against an unbounded
self-similar fractal distribution of galaxies. In the 1970's and
early 1980's a number of catalogs going to a variety of
magnitude limits were available and analysed by Peebles and his
collaborators.  Because of the way the galaxy luminosity function
works, most of the galaxies in a catalog fall within a
relatively narrow range of distance that depends on the limiting
magnitude of the catalog: catalogs reaching to fainter
magnitudes are probing the Universe at greater distances.

As the distance increases, the angular scale subtended by a given
physical distance decreases.  Hence, if the Universe is
homogeneous, the two-point angular correlation function of one
catalog should look like a rescaled version of the two point
angular correlation function of a deeper catalog (see
Eq.~\ref{angscl}), i.e., for catalogs with varying characteristic
distance $w \propto D_\ast^{-1}$ at a given angular separation $\theta
D_\ast$; or, in other words, if we calculate the angular correlation
function on two samples, with characteristic depths $D_\ast$ and
$D_\ast'$, Eq.~\ref{angscl} implies that
$w'((D_\ast /D_\ast')\theta)=(D_\ast/D_\ast')w(\theta)$.
The scaling relationship can be predicted
precisely, though for catalogs that probe to very great depths
it is necessary to be careful of K-corrections and geometric
effects due to the cosmological model \cite{bonometto}.

The earliest catalogs available were the de Vaucouleurs catalog
of Bright Galaxies, the Zwicky catalog, the Shane--Wirtanen
catalog and the Jagellonian Field.  Matching their correlation
functions provided the first direct evidence for large scale
cosmic homogeneity \cite{gp77,gp86}. The scaling relation
has been confirmed with more recent catalogs, in particular, the APM
galaxy survey has povided one of the strongest observational
evidences supporting this law \cite{madcorr,baugh,maddox96}.

\begin{figure}
\includegraphics[width=8.5 cm]{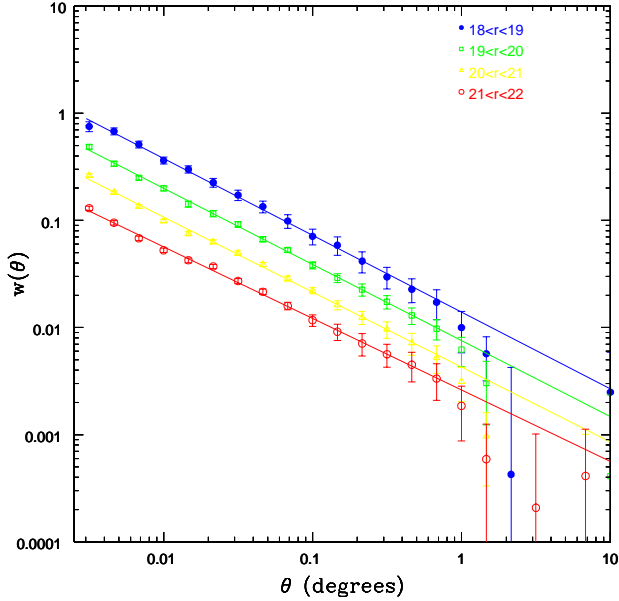}
\caption{\label{fig:wn} The angular correlation function from the
SDSS as a function of magnitude from \textcite{connolly02}. The
correlation function is determined for the magnitude intervals
$18<r^*<19$, $19<r^*<20$, $20<r^*<21$ and $21<r^*<22$. The fits to
these data, over angular scales of 1' to 30', are shown by the
solid lines.}
\end{figure}

Now we can do much better since we have bigger and better
catalogs with partial or complete  redshift information.  Such
catalogs can be divided into magnitude slices and the same test
performed on the two point angular correlation function of the
slices.  The result \cite{connolly02} reproduced in
Fig.~\ref{fig:wn} is as good a vindication of the homogeneity of
the Universe as one could wish for.  More data will be forthcoming
from the 2dF and SDSS surveys.

The scaling properties of the correlation function are usually
shown in the form of the correlation integral. For a point
distribution the integral expresses the number of neighbors, on
average, that an object has within a sphere of radius $r$. It is
given by
\begin{equation}
N(<r) =n \int_0^r 4 \pi s^2 (1+\xi(s)) ds
\label{corrint}
\end{equation}

The distribution is said to follow fractal scaling if within a
large range of scales the behavior of $N(<r)$ can be well fitted
to a power law
\begin{equation}
N(<r)\propto r^{D_2},
\label{corintsc1}
\end{equation}
or, alternatively
\begin{equation}
1+\xi(r)\propto r^{D_2-3}.
\label{corintsc2}
\end{equation}
where $D_2$ is the so-called correlation dimension. The scaling
range has to be long enough to talk about fractal behavior.
However, the term has been used very often for describing scaling
behaviors within rather limited scale ranges \cite{avnir98}.  In
Sect.~\ref{sec:d2} we show recent determinations of $D_2$ for
several galaxy samples at different scale ranges.

\subsubsection{Estimators}
The two-point correlation function $\xi(r)$ can be estimated in
several ways from a given galaxy sample. For a discussion of them
see, for example, \textcite{marsaa,pons,kerm}. At small distances,
nearly all the estimators provide very similar performance,
however at large distances, their performance is not equivalent
any more and some of them could be biased. Considering the galaxy
distribution as a point process, the two-point correlation
function at a given distance $r$ is estimated by counting and
averaging the number of neighbors each galaxy has at a given
scale. It is clear that the boundaries of the sample have to be
considered, because as no galaxies are observed beyond the
boundaries, the number of neighbors is systematically
underestimated at larger distances. If we do not make any
assumption regarding the kind of point process that we are dealing
with, the only solution is to use the so-called minus--estimators,
the kind of estimators favored by Piertonero and co-workers
\cite{labini98}: The averages of the number of neighbors at a
given distance are taken omitting those galaxies lying closer to
the border than $r$. At large scales only a small fraction of the
galaxies in the sample enters in the estimation, increasing the
variance. To make full use of the surveyed galaxies, the estimator
has to incorporate an edge-correction. The most widely used
estimators in cosmology are the Davis and Peebles estimator
\cite{Davis83}, the Hamilton estimator \cite{Hamilton93}, and the
Landy and Szalay estimator \cite{Landy93}. Here we provide their
formulae when applied to a complete galaxy sample in a given
volume with $N$ objects. A Poisson catalog, a binomial process
with $N_{\rm{rd}}$ points, has to be generated within the same
boundaries.
\begin{equation}
\widehat\xi_{\rm{DP}} (r) =
    \frac{N_{\rm{rd}}}{N}
    \frac{DD(r)}{DR(r)}
    -1,
\label{dpest}
\end{equation}
\begin{equation}
\widehat{\xi}_{\rm{HAM}} (r) = {\frac{DD(r)\cdot RR(r)} {[DR(r)]^2}} - 1,
\label{eham}
\end{equation}
\begin{equation}
\widehat{\xi}_{\rm{LS}} (r) = 1 +
    \left(\frac{N_{\rm{rd}}}{N}\right)^2
    \frac{DD(r)}{RR(r)}
    - 2 \frac{N_{\rm{rd}}}{N} \frac{DR(r)}{RR(r)}
\label{els}.
\end{equation}
where $DD(r)$ is the number of pairs of galaxies with separation
within the interval $[r-dr/2, r+dr/2$, $DR(r)$ is the number of
pairs between a galaxy and a point of the Poisson catalog, and
$RR(r)$ is the number of pairs with separation in the same
interval in the Poisson catalog. At large scales the performance
of the Hamilton and Landy and Szalay estimators has been proved to
be better \cite{pons,kerm}.

\begin{figure}
\includegraphics[width=8.5cm]{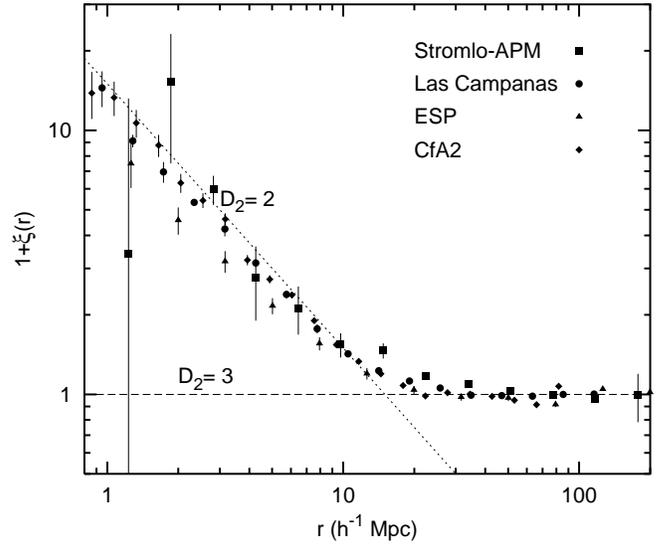}
\caption{\label{fig:corrfun} The correlation function $1+\xi(r)$
for different samples calculated with different estimators. We can
see that the small scale fractal regime is followed by a gradual
transition to homogeneity.}
\end{figure}

\subsubsection{Recent determinations of the correlation function}

Earlier estimates of the pairwise galaxy correlation function were
obtained from shallow samples, and one could suspect that they
were not finding the true correlation function. The first sample
deep enough to get close to solving that problem was the Las
Campanas Redshift Survey (LCRS). The two-point correlation
function for LCRS was determined by \textcite{Tucker97} and by
\textcite{Jing98} (see
Fig.~\ref{fig:corrfun}). Jing {\it el al.} get slightly smaller values
for the correlation length ($r_0=5.1h^{-1}$Mpc) than Tucker {\it
et al.} ($r_0=6.3h^{-1}$Mpc). When making comparisons, it is
necessary to take care that the length scales have been
interpreted in the same underlying cosmological model.  Older
papers tend to set $\Lambda = 0$ whereas more recent papers are
often phrased in terms of a flat-$\Lambda$ plus cold dark matter
cosmology.

\begin{figure}
\includegraphics*[width=8.5 cm]{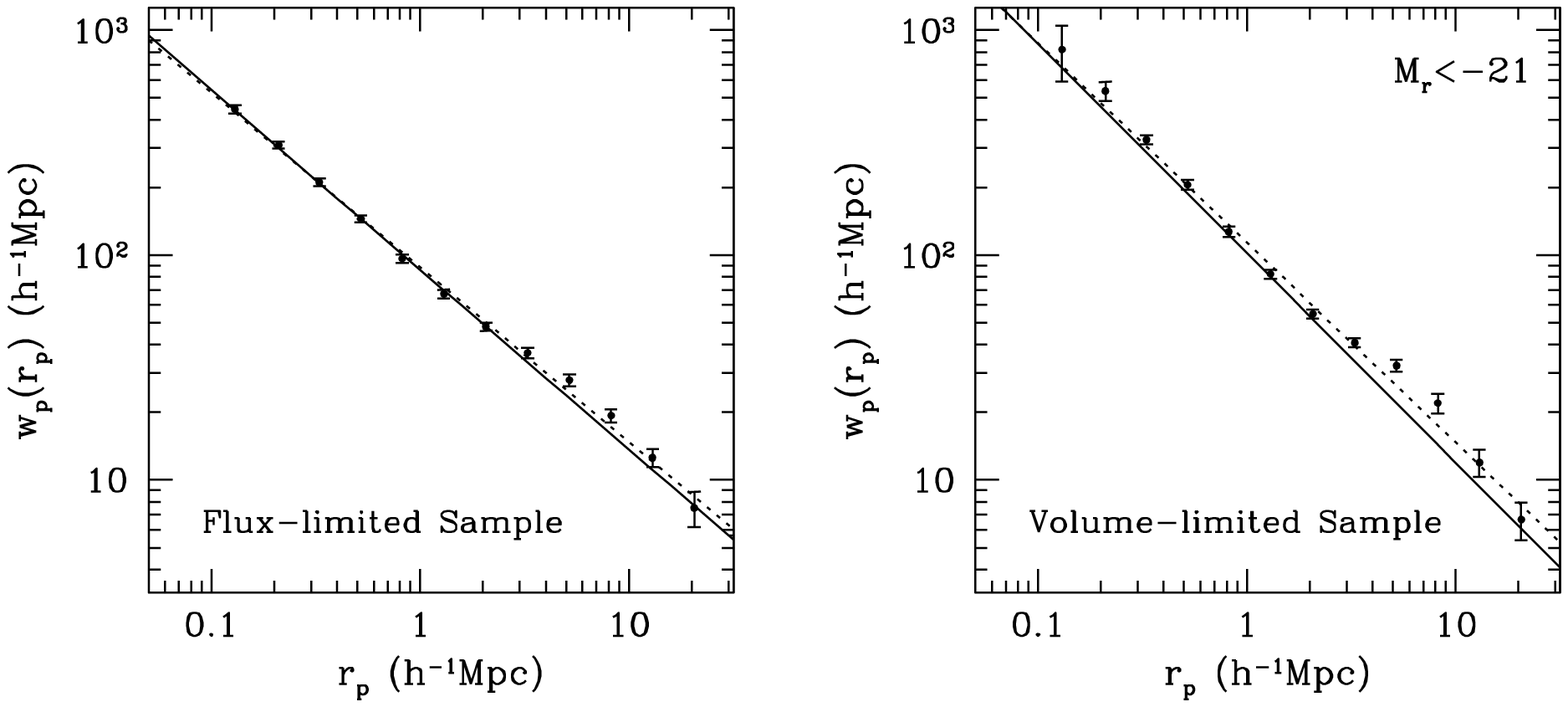}
\caption{\label{fig:wprp} The (projected)
real space two point-correlation function
of the SSDS data from \textcite{Zehavi03}. The two straight lines
show different fits corresponding to different weighting schemes}
\end{figure}

Analyzing data from the first batch of the SSDS,
\textcite{Zehavi02} analyse 29300 galaxies covering a 690 square
degree region of sky, made up of a number of long narrow segments
(2.5 - 5 degrees).  They arrive at an average real-space
correlation function of
\begin{equation}
\xi(r) = \left ( {r} \over{6.1 \pm 0.2 h^{-1} \hbox{Mpc}}
\right )^{-1.75 \pm 0.03}
\end{equation}
for 0.1$h^{-1}$ Mpc $< r <$ 16 $h^{-1}$ Mpc. This comes
close to the LCRS result of
\textcite{Tucker97}. More recently, the
same group \cite{Zehavi03} has updated the result,
using a more complete
sample with 118,149 galaxies (see
Fig.~\ref{fig:wprp}), and the best power-law fit is
\begin{equation}
\xi(r) = \left ( {r} \over{5.77 h^{-1} \hbox{Mpc}}
\right )^{-1.80}
\end{equation}

This is a remarkable scaling law covering some 3 orders of
magnitude in distance. The smallest scale measured ($100 h^{-1}$
kpc) is barely larger than a typical galaxy. Interestingly, this
lower scale is set, in the \textcite{Zehavi02} analysis, by the
requirement that, at the outer limit of the survey (corresponding
to a radial velocity of 39,000 km s$^{-1}$), pairs of galaxies
should be no closer than can be reached by two neighboring fibers
on the multifiber system. There would be some interest in looking
at nearer galaxies and tracing the correlation function to even
smaller scales to see whether the old and remarkable extrapolation
of \textcite{Gott79}\footnote{Gott and Turner estimated the
small-scale end of the correlation function down to a scale of
30$h^{-1}$ kpc from the distribution of projected distances
between isolated galaxy pairs (double galaxies). As strange as it
may seem, this correlation function fitted neatly the general
galaxy correlation function.} is valid in this newer data set (see
also \textcite{SDSSpairs}).  The largest distance ($16 h^{-1}$
Mpc) is larger than the size of a great cluster.  It should be
emphasized that this is a real space correlation function: the
finger-of-god effects have been filtered.

There is a substantial luminosity effect seen in the scale length.
The absolute magnitude $M_\star$ of the ``knee" of the Schechter
galaxy luminosity function \cite{Schechter} is taken as a
reference point (being a ``typical" galaxy luminosity, whatever
that means).  For galaxies with absolute magnitudes centered on
$M_\star - 1.5$ the scale length is $r_0 \approx 7.4 h^{-1}$ Mpc.
For samples centered on $M_\star$ the scale length is $r_0 \approx
6.3 h^{-1}$ Mpc. And for samples centered on $M_\star + 1.5$ the
scale length is $r_0 \approx 4.7 h^{-1}$ Mpc.  The slope for these
samples is essentially the same. A similar strong dependence of
the correlation function on the color, morphology, and redshift of
galaxies was found before, in the Canadian Network for
Observational Cosmology Field Galaxy Redshift Survey (CNOC2) by
\textcite{cnoc-corr}.

The angular correlation function for the SDSS  \cite{connolly02} is
independent of redshift distortions and agrees well with the value
inferred from the redshift survey.  This encourages one to believe
that the redshift corrections are being handled effectively.

However, the latest careful analysis of the (almost) full 2dF
survey \cite{hawkins02} gives the correlation length
$r_0=5.05h^{-1}$Mpc, substantially smaller than the SDSS result.
\textcite{hawkins02} ascribe this to the different galaxy content
of the two surveys: the SDSS is a red-magnitude selected survey
and the 2dFGRS is a blue magnitude selected survey.

\subsubsection{Correlation dimension}
\label{sec:d2} Recently, many authors have measured the
correlation dimension of the galaxy distribution at different
scales using all available redshift catalogs. \textcite{wurees}
and \textcite{Kuro01} summarized these results in a table. A more
completed and updated version of a similar table, including more
references and new catalogs is presented here (see Table
\ref{tab:table1}). The estimates of the correlation dimension have
been performed using different methods depending on the authors'
preferences. It is worthwhile to mention the elegant technique
introduced by \textcite{Amendola99} based on radial cells that
maximizes the scale at which the minus-estimator can be applied.
The table shows unambiguously that the correlation dimension is a
scale dependent quantity, increasing gradually from values $D_2
\simeq 2$ for scales less than $20-30 h^{-1}$ Mpc (and even larger
values of $D_2$ in IRAS based redshift surveys) to values
approaching $D_2 \simeq 3$ for larger scales.

\begin{table*}
\caption{\label{tab:table1}The correlation dimension estimated on different
redshift surveys at different scale ranges.}
\begin{ruledtabular}
\begin{tabular}{lccc}
Reference & Sample & Range of scales ($h^{-1}$ Mpc)& $D_2$\\
\hline
\onlinecite{marjones}& CfA-I & 3-10 &$1.15 - 1.40$\\
\onlinecite{Lemson91}& CfA-I & $1-30$ &2\\
\onlinecite{Dom94}& CfA-I & $1.5-25$ &2\\
\onlinecite{Kuro99}& CfA-II & $7-27$ &$1.89 \pm 0.06$\\
\onlinecite{Guzzo91}& Perseus-Pisces & $1-3.5$ &$1.25 \pm 0.10$\\
& Perseus-Pisces & $3.5-27$ &$2.21 \pm 0.06$\\
& Perseus-Pisces & $27-70$ &$\simeq 3$\\
\onlinecite{Martinez98}& Perseus-Pisces & $1-20$ &$1.8-2.3$\\

\onlinecite{martcoles94}& QDOT & $1-10$ &2.25\\
& QDOT & $10-50$ &2.77\\
\onlinecite{Martinez98}& Stromlo-APM & $30-60$ &$2.7-2.9$\\
\onlinecite{Hatton99}& Stromlo-APM & $12-55$ &2.76\\
\onlinecite{Amendola99}& Las Campanas & $ \le 20-30$ &2\\
 & Las Campanas & $ > 30$ & $\to 3$\\
\onlinecite{Kuro01}& Las Campanas & $ 5 -32$ &$1.96 \pm 0.05$\\
& Las Campanas & $ 32 - 63$ & $\simeq 3$\\
\onlinecite{pancoles00}& PSCz & $ < 10 $ &$2.16$\\
& PSCz & $ 10-30 $ &$2.71$\\
& PSCz & $ 30-400 $ &$2.99$\\

\end{tabular}
\end{ruledtabular}
\end{table*}

\subsubsection{Correlation length as a function of sample depth}

The first indication that correlation length might depend on the
sample depth was found in the CfA-I data \cite{eks86}. The
correlation length increased, when deeper samples were chosen.
Although the authors explained the effect by the specific geometry
of the mass distribution in shallow samples, this paper motivated
the early campaign to explain the galaxy distribution as fractals
\cite{ruffini1,pietro87}, because for a fractal $r_0$ increases
proportionally with the sample depth \cite{coleman,guzna}. The Ruffini group realized from the
beginning that fractal scaling cannot extend to large scales and
started to look for crossover to homogeneity \cite{ruffini3}, but
the Pietronero group has continued the fractal war until now,
fighting for an all-fractal universe. Their stand is summarized in
\textcite{labini98}.

\begin{figure}
\includegraphics[width=8.5 cm]{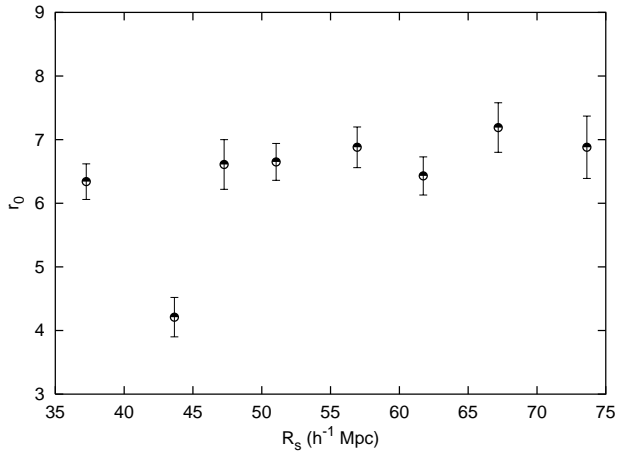}
\caption{\label{fig:r0flat} The correlation length as a function
of the sample depth for the CfA-II catalog, from \textcite{mart01}. The
observed {\it plateau} argues against the fractal interpretation
of the galaxy distribution.}
\end{figure}

The deep samples now at our disposal have solved this problem once
and for all --- the galaxy correlation functions may depend on
their intrinsic properties (luminosity, morphology, etc.), but not
on the sample size \cite{mart01,Kerseg}.
As an example, Fig.~\ref{fig:r0flat} shows the
results of a recent study.

\subsection{\label{subsec:cluscorr}Galaxy-galaxy and
cluster-cluster correlations}

Having re-discovered the power of the two-point correlation
function as a tool for measuring clustering, it was evident that
the Princeton group would go on to analyze every available
catalog of extragalactic objects they could lay their hands on.
One of these catalogs was the Abell catalog of rich galaxy
clusters identified on the Palomar Sky Survey \cite{hape}. The
technique used was power spectrum analysis since it was felt this
would give a better method of dealing with the incomplete sky
coverage.

It came as somewhat of a surprise to discover (a) that these Abell
clusters were themselves clustered and (b) that, on a given scale,
they were more clustered than the galaxies.  The former was a
surprise because serious doubts had previously been expressed
about the reality of superclustering.  Here was direct evidence
that clusters were likely to be found in pairs and even in groups.
The latter was a surprise because it had been (naively) expected
that clusters identified from a set of points would necessarily
have the same correlation function as the set itself.  The galaxy
clusters were themselves clustered on scales where the
galaxy-galaxy correlation was so small as to be immeasurable.

Both the galaxy and cluster correlation functions are
approximately power laws $\xi(r)=(r/r_0)^{-\gamma}$ with the same
exponent $\gamma\approx 1.8$, but the correlation amplitudes for
clusters are much larger than those for galaxies.

There is a simple reason why the cluster-cluster correlation
function might have an amplitude exceeding the galaxy-galaxy
correlation function amplitude: it arises because of the way
clusters are identified as regions where groups of points have a
substantially higher than average density.  Such regions contain
most of the close pairs that go into defining the value of the
galaxy-galaxy correlation function.  Moreover, eliminating the
points which are not in such clusters biases the expected number
of pairs that would have been found had this been a Poisson
distribution containing the same number of points.  The boost in
the value of the correlation function achieved from such
censorship depends directly on the volume of space occupied by
these clusters.

This entirely obvious point was made in a preprint by
\textcite{Jones2}: the paper was never published.  As with many
useful ideas, it became common knowledge and moved into the realm
of folklore.

There remained some important questions:

\begin{itemize}
\item[a:]
Does the Abell catalog provide a sufficiently good sample for this
purpose: is it free from systematic biases that may prejudice the
result?  Abell identified clusters by eye, a procedure which would
lack the objectivity of an automatic plate scanning machine.
\item[b:]
If in the cluster sample we reject the least impressive ones,
would this change the correlation function? This corresponds to
selection by cluster richness.
\item[c:]
How would changing the selection threshold affect the correlation
function? This is not quite the same as selecting by cluster
richness: less rich clusters are still included, though they would
appear as smaller objects on increasing the discrimination
threshold.
\item[d:]
If clusters were selected other than by virtue of their contrast
with the background, eg: from identifying clusters in an X-Ray
survey, would we still see enhanced clustering?
\item[e:]
What does the galaxy-cluster cross correlation tell us?
\end{itemize}

It was well known that there were systematic biases in the Abell
Catalog.  The subsample of low richness clusters was incomplete,
and the more distant clusters were systematically richer than than
nearby counterparts.  This was not in itself enough to remove the
``discrepancy" between the galaxy-galaxy correlation function and
the cluster-cluster correlation function, but it might prejudice
conclusion about richness dependence of the discrepancy.

It was not until 1992 that a sufficiently good alternative to the
Abell Catalog became available: this was the APM cluster catalog
\cite{Dalton92, Dalton97} derived from the Cambridge APM Galaxy
Survey (Automatic Plate Measuring Machine) of UK Schmidt Telescope
plates.  Now we await results from the large 2dF and SDSS redshift
catalogs which have already provided detailed information about
the galaxy-galaxy correlation function.

\subsubsection{Analysis of recent catalogs}
Currently the best data on galaxy cluster clustering comes from
redshift surveys of clusters identified in machine generated
galaxy catalogs and of clusters observed in X-Ray surveys.  The
2dF and SDSS surveys will undoubtedly settle this matter once and
for all since they contain a large number of clusters that can be
selected on the basis of redshift.  However, it is already
apparent (as in the \textcite{cnoc-corr} study of the CNOC2
sample, for \textcite{Zehavi02} study of the Early SDSS Data,
and for \textcite{Norberg01,Madgwick03} correlation analysis of the 2dFGRS)
that talking about {\it the} galaxy-galaxy correlation function is
somewhat of an over-simplification in the first place: the
galaxy-galaxy correlation depends strongly on the absolute
magnitude, galaxy colour and galaxy spectral type. Galaxies are
clearly not unbiased tracers of the underlying mass distribution.

In automated cluster searching, clusters are generally discovered
via a nearest-neighbour, friends-of-friends, type of analysis.
They are discovered by virtue of their central concentration and
so catalogs contain clusters that are defined in terms of a
``distance to your nearest neighbour" threshold length.  If the
threshold length is increased the catalog contains more
clusters: the number of poorer, less centrally dense, clusters
increases.  It is not a priori obvious how the mean density of
galaxies within a cluster so found relates to its central density:
there will clearly be a correlation.  It might well be that
selecting clusters by virtue of their mean galaxy density rather
than their peak density would yield different catalogs and lead
to different conclusions about the systematics of cluster-cluster
clustering.

\subsubsection{Theoretical expectations}
It is easier to build theoretical (analytic) models based on
selection by mean cluster density, ie: clusters selected via a
density threshold, than it is to build models based on clusters
selected by peak density.  The latter requires an understanding of
how the cluster dynamics works to produce the density profile of
the galaxy distribution.  This may contribute to some of the
confusion that exists when looking for trends in the clustering of
clusters.

The earlier theoretical models \cite{Jones2, Kaiser84,
Bahcallwest} for the clustering of clusters were based on
threshold selection. The same is true of more recent hierarchical
models based on multi-fractal models for the distribution of
galaxies \cite{martpara, Paredes}. Most of the conclusions about
superclustering in which the clusters are defined via the peak
density excursion comes from $N$-body simulations of various sizes
and sophistication \cite{Bahcallcen, Croft94, virgo}.

Since clusters found in X-ray surveys are found by virtue of their
gas temperature, that is total potential, these surveys should
agree rather well with the conclusions based on $N$-body
experiments.

\subsubsection{Richness dependence of the correlation length}
The seminal paper on the effect of cluster richness on the
cluster--cluster correlation function was that of
\textcite{Szalay85}. They suggested that the scaling length for
clustering should itself depend on the cluster density.  Which
cluster density, peak or mean, was never stated.

The formula for the cluster two-point correlation function
$\xi(r;\nu)$ is usually written as \cite{Kaiser84}:
\begin{equation}
\label{xinu}
\xi(r;\nu)=\frac{\nu^2}{\sigma^2}\xi(r),
\
\end{equation}
where $\nu$ is the height of the peaks in units of the rms error
$\sigma$ of the galaxy density field, and $\xi(r)$ is the
correlation function of the galaxy field\footnote{As the
correlation functions and $\sigma$ are defined for the density
contrast $\delta=(\rho-\bar{\rho})/\bar{\rho}$, all quantities in
(\ref{xinu}) are dimensionless; there is no dimensionality
conflict.}.

\begin{figure}
\includegraphics[width=8.5 cm]{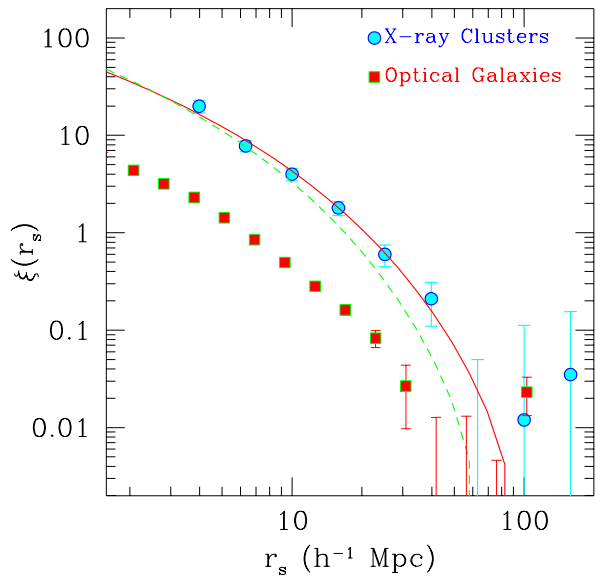}
\caption{\label{fig:guzzo} The two-point correlation function
for the X-ray selected clusters from the REFLEX survey (circles) and
for the Las Campanas galaxy redshift survey (squares). The solid and
dashed lines are the expected results for
an X-ray similar survey in a LCDM model
with different values for the cosmological parameters,
from \textcite{borguz}.}
\end{figure}

\begin{figure}
\includegraphics[width=8.5 cm]{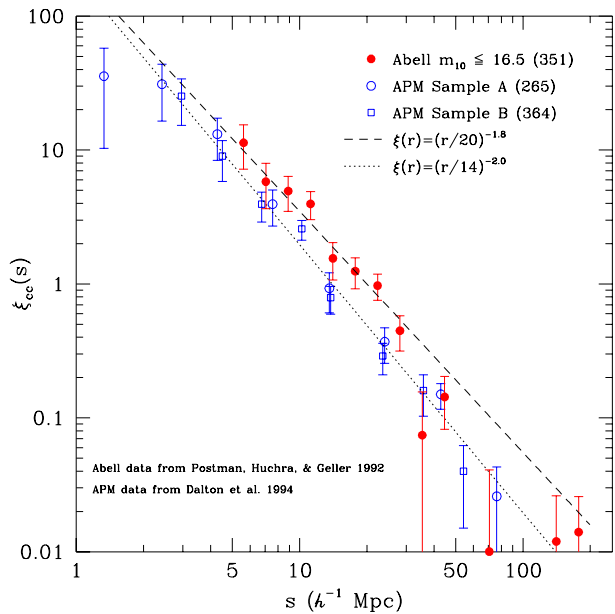}
\caption{\label{fig:clclfun} The two-point correlation functions
for the Abell clusters and two subsamples of the APM survey. The
best power-law fits are shown in the plot, from \textcite{Postman99}.}
\end{figure}

The empirical determination of the the cluster--cluster
correlation function, $\xi_{cc}(r)$, is much more uncertain than
the galaxy--galaxy correlation function, $\xi_{gg}(r)$. The
selection effects associated with the cluster identification
method \cite{eke96} are the major source for this uncertainty.
The possible dependence of clustering properties on cluster
richness makes the issue still more difficult. Nevertheless
$\xi_{cc}(r)$ is usually fitted to a power law
\begin{equation}
\label{ccxi} \xi_{cc} = \left ( \frac{r}{r_c} \right )^{\gamma_c}.
\
\end{equation}
Eq. \ref{xinu} holds if $\gamma_c=\gamma$, where $\gamma$ is the
exponent of the power-law galaxy--galaxy correlation function. As
already mentioned,  this seems to be the case, see for example in
Fig.~\ref{fig:guzzo},
the remarkable agreement between the slopes of the correlation
function of the REFLEX cluster catalog and the Las Campanas 
galaxy redshift survey \cite{borguz, guzzorev}. Nevertheless,
depending on the analyzed cluster sample and cluster
identification procedure, the scatter of the reported values for
the slope of the correlation function is very high with $\gamma_c
= 1.6$ to $2.5$. For the correlation length the values go from
$13h^{-1}$ Mpc to $40h^{-1}$ Mpc \cite{Bahcallwest, Dalton94,
Nichol92, Postman92, borguz}. Fig.~\ref{fig:clclfun} illustrates this
variability displaying the differences between the correlation
function of the Abell and APM cluster samples.

\begin{figure}
\includegraphics*[width=8.5 cm]{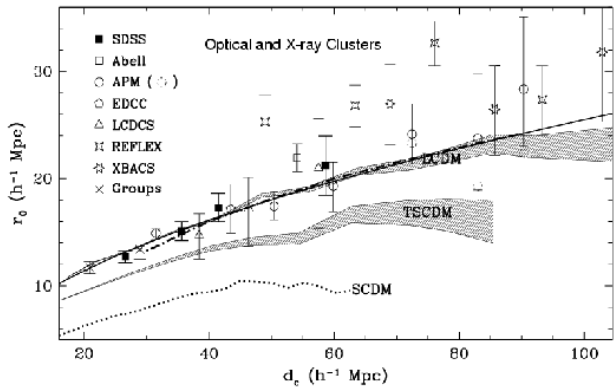}
\caption{\label{fig:r0dc}The correlation length of different
cluster samples as a function of the intercluster distance. The
solid line shows the relation $r_c=2.6 \sqrt{d_c}$ that fits
well the observations and the LCDM model, from \textcite{bahcallnew}.}
\end{figure}

Rich clusters have many members and are rare, therefore the
distance between then $d_c = n_c^{-1/3}$ is larger.
\textcite{Bahcallwest} derived a linear relation between the
cluster correlation length $r_c$ and the mean intercluster
separation $d_c$, $r_c =0.4 d_c$ from power-law fits (constrained
to have a fixed value of $\gamma_c=1.8$) to correlation functions
calculated on cluster samples with different richness.
Fig.~\ref{fig:r0dc} shows that this relation is not confirmed by
the new data. In fact, at large values of $d_c$ the relation must
level off, and a weaker dependence of $r_c$ versus $d_c$ agrees
better with the observations, for example $r_c = 2.6 \sqrt{d_c}$
as shown in the figure \cite{bahcallnew}.

Since $r_c$ and $\gamma_c$ are not independent, the slope is
usually constrained to a fixed value $\gamma_c=1.8$.  Dependence
of $\gamma_c$ on cluster richness has been proposed
\cite{martsci}, although this dependence is better parametrized by
the correlation dimension --- the exponent of the power law
fitting the correlation integral $N(r) = Ar^{D_2}$ (see Eq. 16).
Multiscaling is the term used for scaling laws in which $D_2$
displays a slowly varying behavior with the density threshold that
characterizes the richness of clusters. The higher the threshold,
the richer the clusters, and the smaller the value of $D_2$.
Within the multiscaling framework, the relation $r_0$ versus $d_c$
gets a more complicated form flattening for large values of $d_c$
as the observations confirm \cite{martsci}.

\subsection{The pairwise velocity dispersion}

The pairwise velocity dispersion of galaxies is a measure of the
temperature of the ``gas" of galaxies. By energy conservation, the
kinetic energy of this gas has to be balanced by its gravitational
energy, which depends mainly on the mean mass density of the
Universe. Thus, measuring the pairwise velocity dispersion gives
us a handle on the density.  This is, however, more easily said
than done since we measure only the radial component of the
velocity, and that is biased by larger density inhomogeneities
than a linear theory can handle.

The following short argument shows how the velocity dispersion
relates to the fluctuations in the density field. The non-Hubble
component of a galaxy velocity through the Universe, (its {\it
peculiar} velocity), is due to the acceleration caused by clumps
in the matter distribution.  This is easy to estimate during the
phase of linear evolution of cosmic structure since linear
perturbation theory applies.

A particle that has experienced a peculiar acceleration $g_p$ for
a time $t$ would have acquired a peculiar velocity $v_p \sim
g_pt$. If this acceleration is due to a mass fluctuation $\delta
M$ at distance $r$, we have
\begin{equation}
g_p = G \delta M / r^2 = (4\pi / 3) G \delta \rho r = 0.5 \Omega_0 H_0 v_H
\end{equation}
which leads to
\begin{equation}
v_p/v_H\simeq (1/3)  f(\Omega) \delta, \quad f(\Omega)=(3/2)H_0t
\simeq \Omega^{0.6}.
\end{equation}
For a more general approximation including the cosmological
constant see \textcite{lahav91}.
As one can see the ratio of the peculiar to Hubble velocity is the
quantity that gives a direct measure of the amplitude of
primordial density fluctuations on a given scale for a given value
of $\Omega$. If we have a scaling law for the density fluctuations
we should also see a scaling law in the peculiar velocity field.

A more detailed calculation, still using linear theory, gives a
direct relation between the {\it rms} amplitude of the peculiar
velocity and the power spectrum of primordial density
fluctuations \cite{strauss95}:
\begin{equation}
\langle {v_p(R)^2} \rangle = \frac{H_0^2f^2}{2\pi^2}\int
{\cal{P}}(k)\widetilde{W}^2(kR)dk
\end{equation}
where $\widetilde{W}(kR)$ is the Fourier transform of a spherical
window function of radius $R$, $W(R)$. This equation also works quite
well for rather high $\delta$, well beyond the linear regime.  The
main problem then becomes dealing with the redshift distortion of
the observed velocity field.

This equation, however, contains information only about the {\it rms}
magnitude of $v_p$ on a given scale.  More information about
peculiar motions in different cosmological scenarios can be
obtained from other types of the velocity correlation functions
that can be estimated from data sets.

As direct data on peculiar velocities of galaxies are hard to
obtain, the pairwise galaxy velocity dispersion is measured from
ordinary redshift surveys by modelling its effect on the redshift
space correlation function.  This modelling is not very certain,
as it depends on the choice both of the adopted mean streaming
velocity model and of the model for the pairwise velocity
distribution itself. The latter is usually modelled as an
exponential distribution \cite{peeb80}.

The first determination of the pairwise velocity dispersion
$\sigma_{12}^2$ was made by \textcite{Davis83}, who found
$\sigma_{12}\approx340\,\mbox{km\,s}^{-1}$. Subsequent
determination from the IRAS data \cite{Fisher94, fisherkb} gave a
similar value ($\sigma_{12}\approx 317\,\mbox{km\,s}^{-1}$). These
values were much lower than those predicted for the Standard Cold
Dark Matter (SCDM) model
($\sigma_{12}\approx1000\,\mbox{km\,s}^{-1}$), and served as an
argument for discarding the model.

Later determinations have given larger values for this dispersion:
the estimates of \textcite{Jing98, Marzke95, Zehavi02} all
converge at the value
$\sigma_{12}\approx\mbox{550--600\,km\,s}^{-1}$; not enough for
the SCDM model, but in concordance with the present standard, the
$\Lambda$CDM model.

In the stable clustering model the pairwise velocity dispersion
should scale with pair distance as $r^{0.2}$; this scaling has not
been observed. Also, it is well known that the value of
$\sigma_{12}$ is sensitive to the presence of rich clusters in the
sample. \textcite{Davis97} and \textcite{Landy98} propose
alternative schemes for estimating the pairwise velocity
dispersion, which again lead to small values of $\sigma_{12}$.

The galaxy velocity field is also rather inhomogeneous; a
well-known fact is the extreme coldness of the flow in our
neighbourhood, out to $5h^{-1}$Mpc, where
$\sigma_{12}=60\,\mbox{km\,s}^{-1}$ \cite{coldflow}.

\subsection{Light does not trace mass}
It has long been realized that there is a difference between the
distribution of light in the Universe and the distribution of
mass. The first clues came with the apparent systematic increase
of mass-to-light ratios with scale determined from galaxies,
binary galaxies, groups and clusters of galaxies: this was later
made more explicit by \textcite{nateinsaar}, \textcite{joeveer} and
\textcite{Ostriker74}. It was also known that galaxy morphology is
related to the clustering environment \cite{Hubble36b, Zwicky37,
abell58, Davis76, Dressler80, Einasto80, GuzzoStr97}.

The recognition that clustering depends on galaxy luminosity is
more recent \cite{Ham88, dom89, White88, mart93, Loveday95,
Willmer98, Benoist99, Kerseg}.

It is not difficult to understand why this should be so. We may be
even surprised that the results were in any way surprising!  There
was early work of \textcite{Bahcall83, bbks, melott86}. However,
it has not been easy to model these luminosity---  and
type-dependent phenomena since we have only the barest
understanding of the galaxy formation process and it is probably
fair to say that our knowledge of what causes galaxies to have
vastly different morphologies is still rather incomplete.

The recent advances in augmenting $N$-body simulations with
semi-analytic models and  computational hydrodynamics is
promising, though at a relatively early stage  \cite{Katz92,
Cen92, Benson00, Blanton99, Colin99, Kauffmann99, Pearce99,
White01, Yoshikawa}. Modelling the formation of individual
galaxies shows just how many physical processes must be taken into
account, quite apart from trying to fold in our ignorance of the
star formation process (and that is what gives rise to the
luminosity).  A brave attempt is exemplified by the paper of
\textcite{Sommer02}.

\subsubsection{Mass distribution and galaxy distribution: biasing}

The concept of biasing was introduced by \textcite{Kaiser84} in
order to explain the observed relation between the correlation
functions of galaxies and galaxy clusters. Using the high-peak
approximation to a Gaussian density field, he obtained a formula
(\ref{xinu}) showing that the two correlation functions were
proportional.

The same idea was later applied to galaxy distributions: as
different types of galaxies have different clustering properties,
they cannot all follow directly the overall density field. Thus we
normalize the correlations by writing
\begin{equation}
\label{linbias}
\sigma^2_{gal}=b^2\sigma^2_{total},
\end{equation}
(note that $\sigma^2=\xi(0)$), and call $b$ the bias factor. As
baryonic matter comprises about four per cent of the total matter
plus energy content of the universe, we can also say that the
above relation connects the galaxy and dark matter distributions.

Bias cannot be measured directly, and indirect observational
determinations of bias values have not yet converged to a single
value for a given type of galaxies. Moreover, \textcite{lahav99}
showed that bias is, in general, nonlinear and stochastic. And
later determinations have found that bias is also scale-dependent
\cite{padma00}. Such bias can easily destroy scaling relations
that could be inherent in the matter distribution.

\subsubsection{Mass and light fluctuations}
An alternative measure of the scale dependence of clustering  is
to plot the variance of the mass or light density fluctuations on
a variety of scales. This is little more than what Carpenter had
done in the 1920's, and was first formalized by
\textcite{Peebles65} in his remarkable  paper on galaxy formation.
\footnote{Several things are remarkable about Peebles' 1965 paper.
It was Peebles' first paper on galaxy formation and its submission
to the Astrophysical Journal preceded the the announcement of the
discovery of the microwave background. In that paper we see the
entire roadmap for the following decades of galaxy formation
theory, albeit in terms of initial isothermal fluctuations.}  It
is relatively easy to calculate a density fluctuation spectrum:
sample the density field in windows of different sizes, for each
window size calculate the mean and variance of the contents of the
window and plot the result.  This works equally well in two or
three dimensions.  Some important technical questions arise: what
to do at the boundaries and what the shape and profile of the
window should be.  By the profile it is meant what weight is
attached to an object falling at a given location in the window.
The ``top hat" profile counts a weight of one if the object is in
the window and zero outside: this is the simplest choice, though
not a particularly good one. Fuzzy edged windows are to be
preferred since they reduce the effects of shot noise.

This process is analogous to two other methods of analyzing a
density field: counts in cells and wavelet analysis.  Counts in
cells statistics do precisely what has just been described, using
various coverings of the data set, and are most often hard-edged.
The wavelet analysis does the same, but the choice of analyzing
wavelet determines how ``hard" the sampling volume is. Simple Haar
wavelets are a bad choice since they too are hard-edged, but there
are many fine alternatives.  This an an area which requires more
research since wavelets are particularly good at sniffing out
scaling relationships.

The density fluctuation spectrum is in some sense a half-way house
towards the power spectrum: the variance of the mass fluctuations
are referred to a physical variable, mass scale, rather than the
$k$-space wavenumber (which is itself an inverse length scale).
The problem with the mass spectrum is that its amplitudes are
correlated and depend on the adopted mass profile filter; the
conventional power spectrum (spectral density) has independent
amplitudes as it will be explained in
Sect. VI.C.

\section{FURTHER CLUSTERING MEASURES}

\subsection{Higher order correlation functions}
The two-point correlation function is not a unique descriptor of
clustering, it is merely the first of an infinite hierarchy of
such descriptors describing the galaxy distribution of galaxies
taken $N$ at a time. Two quite different distributions can have
the same two-point correlation function.  In particular, the fact
that a point distribution generated by any random walk (e.g., as a
L\'evy flight as proposed by \textcite{mand75} has the correct
two-point correlation function does not mean much unless other
statistical measures of clustering are tested.

The present day galaxy distribution is manifestly not a Gaussian
random process: there is, for example, no symmetry about the mean
density.  This fact alone tells us that there is more to galaxy
clustering than the two-point correlation function.

So what kind of descriptors should we look for? Generalizations of
the two-point functions to 3-, 4- and higher order functions are
certainly possible, but they are difficult to calculate and not
particularly edifying. However, they do the job of providing some
of the needed extra information and through such constructs as the
BBGKY hierarchy\footnote{The BBGKY hierarchy, (after Bogolyubov,
Born, Green, Kirkwood and Yvon),
is an infinite chain of
equations adapted from plasma physics \cite{BBGKYbook}
to describe self-gravitating non-linear clustering (see for example
\textcite{fallsev}, \textcite{peeb80}, and \textcite{saslaw}.)}
they do relate
to the underlying physics of the
clustering process. We shall describe the observed scaling of the
3-point correlation function below.

One alternative is to go for different clustering models: anything
but correlation functions.  These may have the virtue of providing
immediate gratification in terms of visualization of the process,
but they are often difficult to relate to any kind of dynamical
process.

If we knew all higher order correlation functions we would have a
complete description of the galaxy clustering process.  However,
calculating an estimate of a two point function from a sample of
$N$ galaxies requires taking all pairs from the sample of $N$,
while calculating a three point functions requires taking all
triples from $N$.  The amount of computation escalates rapidly and
restrictions have to be imposed on what is actually being
calculated.

Nevertheless, calculating restricted $N$-point functions may be
useful: these functions may be related to one another and have
interesting scale dependence. \textcite{gazta92} has calculated
restricted $N$-point functions and showed that these have power
law behavior over the range of scales where they can be
determined.

\subsection{Three-point correlation functions}

The simplest high-order correlation function is the 3-point
correlation function $\zeta(\mx_1,\mx_2,\mx_3)$. It appears to be
simply related to the two-point function through a Kirkwood-like
relationship (see \textcite{peeb80}):
\begin{eqnarray}
\label{hier}
&&\zeta(\mx_1,\mx_2,\mx_3)=\zeta(r_{12},r_{23},r_{31})=\\
&&\quad=Q\left[\xi(r_{12}\xi(r_{23})+\xi(r_{23})\xi(r_{31})
+\xi(r_{31})\xi(r_{12})\right]\nonumber,
\end{eqnarray}
where $Q\approx 1$ is a constant, and the first equality is due to
the usual assumption of homogeneity and isotropy.
This scaling law is called ``the
hierarchical model'' in cosmology, and it agrees rather well with
observations. The full Kirkwood law\cite{BBGKYbook}
would require an additional
term on the right-hand side of Eq.~(\ref{hier}), proportional to
$\xi(r_{12})\xi(r_{23})\xi(r_{31})$.

As observations show \cite{peeb80,peebles,meiksin}, there is
no intrinsic 3-point
term, either Kirkwood type or more general. If this term were present
the 3-point function would be enormous at small scales. Therefore it
makes no contribution. The
absence of such a 3-point term is probably a consequence of the
fact that gravity is a two-body interaction and is the only force
that plays a role in the clustering process.

\subsection{The power spectrum}

The power spectrum $P(k)$ is the description of clustering in
terms of wavenumbers $k$ that separates the effects of different
scales. If $F(\mk)$ is the Fourier transform of a random field,
then
\begin{equation}
{\cal P}(\mk) = {\bf \rm E}\left[F(\mk) \overline {F(\mk)}\right]
\label{powerspect}
\end{equation}
where ${\bf \rm E}$ denotes the statistical expectation value.

The Fourier modes of a Gaussian random field (our basic model for
the matter distribution in the universe at early times) are
independent, and the only function that defines the field is the
power spectrum. As the initial fluctuations from the inflation
period are described naturally in terms of Fourier modes, the
power spectrum is the best descriptor of the matter distribution
for these times.

Inflationary models predict a power-law power spectrum, $P(k)\sim
k^{n}$ (see \textcite{peeb02} for a recent review), with the most
popular exponent $n=1$. This simple scaling is, however, broken,
once the wavelength of a mode gets smaller than the horizon;
interactions between matter, radiation and gravity deform the
power spectrum in a computable, but complex manner
\cite{ehu98,ehu99}.

Nevertheless, if we restrict ourselves to a smaller scale interval
(say, two orders of magnitude), the power spectrum remains close
to a power law. For the scales of the observed structure the
exponent of this power law is negative, ranging from $n=-1$ for
larger scales to $n\ge-3$ for galaxy scales.

If we combine a scale-free power spectrum $P\sim k^n$ with a
scale-free expansion law $a(t)\sim t^\alpha$ we should get a
perfect scaling regime for evolution of structure. Unfortunately,
this is not true, as there are two completely different regimes of
evolution of gravitating structures: the linear regime, when every
Fourier mode grows at the same rate, and the nonlinear regime,
when we can assume that objects are virialized and their physical
structure does not change. The latter assumption is called
``stable clustering'' \cite{peeb74}.

The linear regime is characterized by small density amplitudes and
large scales (small wavenumbers), the stable clustering regime has
large density amplitudes and occurs at small scales (large
wavenumbers). The scaling solution for the correlation function in
the stable clustering regime was found by \textcite{peeb74}:
$\xi(r)\sim r^{-\gamma}$, where $\gamma=(9+3n)/(5+n)$. The first
attempt to get a solution that would interpolate between the two
regimes was made by \textcite{hklm}. For that they rescaled the
distances $r$, assuming no shell crossing during evolution of
objects, and found an empirical relation between the nonlinear and
linear correlation functions, using $N$-body models. This is known
as the HKLM scaling solution. \textcite{dodds} found a similar
relation for power spectra. These results have been used
frequently for likelihood search in large volumes of cosmological
parameter space, which could not be covered by time-consuming
$N$-body modelling.

However, nowadays it seems that the stable clustering hypothesis
does not describe well either the observed structure, or
present-day numerical simulations, mostly because of merging of
objects in the later stages of evolution of structure. A scaling
solution in terms of a nonlinearity wavenumber that does not
assume stable clustering is described by \textcite{smith02}. Let
us define the nonlinearity wavenumber $k_{NL}$  by
\[
\sigma^2(k_{NL},a)\sim\int_0^{k_{NL}}P(k,a)\,k^2dk=1;
\]
it separates the linear regime $k<k_{NL}$ from the nonlinear
regime $k>k_{NL}$. One then expects the scaling solution to have
the form
\[
P(k,a)=F(k/k_{NL}).
\]
As an example, for the Einstein-de Sitter cosmological model
$a(t)\sim t^{2/3}$, the scale-free power spectrum can be written
as $P(k,a)=a^2k^n$, and the nonlinearity wavenumber $k_{NL}\sim
a^{-2/(n+3)}$. Numerical experiments confirm that scaling
solutions exist.

The latest real-space power spectrum of the SDSS survey 
\cite{tegmark04} shows clearly curvature,
departing from a single
power-law, providing, as the authors say, ``another nail into the 
coffin of the fractal universe hypothesis".
 
\subsection{The bispectrum}

The power spectrum (Eq.~\ref{powerspect}) is a quadratic
descriptor of a random field: it contains information about the
amplitudes of the Fourier components, but not about any phase
relationships that might have evolved through nonlinear processes.
The power spectrum characterizes fully a Gaussian field.  Since
the present-day high-amplitude fluctuating density field is not
Gaussian (there cannot be any region with negative density), the
power spectrum by itself is provides only a partial description.
There are several ways of providing further information in Fourier
space, one of which is to look at higher order correlations among
Fourier components.

The next order descriptors are cubic, the three-point correlation
function and its Fourier counterpart, the bispectrum. The
bispectrum is the third moment of the Fourier amplitudes of a
random field, depending on three wavenumbers. If we denote the
Fourier amplitudes of a random field by $F(\mk)$, the bispectrum
of the field is defined as
\[
{\cal B}(\mk_1,\mk_2,\mk_3) = {\bf \rm E}\left[F(\mk_1)F(\mk_2)F(\mk_3)\right],
\]
where ${\bf \rm E}$ denotes the statistical expectation value. For
homogeneous random fields the bispectrum is non-zero only for
closed triangles of vectors $\mk_1,\mk_2,\mk_3$ (see, e.g.,
\textcite{marsaa}). Consequently, for real-valued homogeneous
random fields the bispectrum can be calculated by
\[
{\cal B}(\mk_1,\mk_2) = {\bf \rm E}\left[F(\mk_1)F(\mk_2)
\overline{F(\mk_1+\mk_2)}\right],
\]
where the overline denotes conjugation. In the signal processing
world the bispectrum is known as the bicoherence spectrum and it
is used to measure the phase coherence among triples of spectral
components that arises as a consequence of nonlinear wave
coupling.

The hierarchical ansatz that we wrote for the three-point
correlation function can be written also for the bi-spectrum:
\begin{eqnarray*}
{\cal B}(\mk_1,\mk_2,\mk_3)&=&Q\left[P(\mk_1)P(\mk_2)+
    P(\mk_1)P(\mk_3)+\right.\\
&+&\left.P(\mk_2)P(\mk_3)\right].
\end{eqnarray*}

A similar expression is predicted by perturbation theory
\cite{fry84}, but with different coefficients for every term.

It is not easy to determine the bi-spectrum from observations, as
its argument space is large (the set of all triangles), and it is
strongly modified by galaxy bias. The estimates so far have
confirmed that the bispectrum follows approximately the predictions of the
perturbation theory \cite{bcgs}.
 As it depends on the bias
parameters, it can be used to estimate galaxy bias. An example
is provided by a recent study \cite{verde02} that found that
the bi-spectrum of the 2dFGRS galaxies is compatible with no bias;
these galaxies seem to faithfully trace the total matter distribution.

\subsection{Fractal descriptors of clustering}

None of the previous descriptors is motivated by the requirement
that the galaxy distribution should, in some sense, be scale free,
which might be expected on the grounds that the gravitational
force which drives the clustering is scale free. What one would
like to do is to generate a set of scaling indices that describes,
say, the scaling of the moments of the galaxy counts distribution
with cell size.

This was in a sense achieved by \textcite{gazta92,Gaztanaga94}
when he determined the scaling laws of restricted $N$-point
correlation functions.  However, one might argue that the scaling
of some high order correlation function has less immediate
intuitive appeal than the scaling of the moments of cell counts.

There is a formalism for describing moments of cell counts that is
commonly used when describing fractal point sets that was adopted
as a clustering descriptor by \textcite{martpara}.  If it is
possible to determine a set of such scaling indices we can turn
the argument around and say that, over the range of scales where
scaling is observed, the galaxy clustering can be represented by a
fractal of a given type.

One should be aware that having a power law correlation function
is not necessarily an indication of scale invariance!  Conversely,
the fractal description implies no particular underlying physical
process: it is merely a statement of how moments of counts in
cells behave as a function of cell size.

It is an interesting question of physics to formulate the physical
process that might generate this distribution of scaling indices.
This has been attempted by \textcite{Jones99} for a simple
nonlinear gravitational clustering model.

\subsubsection{A cautionary word}

There is a considerable difference between using the concept of
fractal measure to describe a statistical process in some
particular regime and saying ``this distribution is such-and-such
a fractal". There has been a set of papers observing scaling of a
low order correlation function and jumping to the conclusions that
(a) this scaling law holds at all scales \cite{labini98} and (b)
this scaling law must be a consequence of some exotic phenomenon
\cite{Bak}.

In the first case scaling laws can only be expected to hold over
scales where nonlinear gravitational clustering has been at work.
In the linear regime we merely see a reflection of the initial
conditions: these have been revealed to us by the COBE experiment
and by other microwave background anisotropy measurements. Indeed,
it is a prediction of gravitational clustering theory that there
should be a break in the scaling laws that reflect the transition
between the linear and nonlinear regimes.  We expect to see this
as the transition to homogeneity that must occur on large scales.

There is no way out of this: the COBE results tell us that there
will be large scales where the Universe is almost homogeneous.

In the second case there is absolutely no indication that anything
more exotic than the force of gravitation is involved in the
growth of clustering. On the contrary, the manifest successes of
gravitational $N$-body experiments testify to the adequacy of
gravity. We are not observing a critical phenomenon, nor are we on
the verge of some marginal instability.

\subsubsection{Structure from counts in cells}

The first analyses of galaxy sky maps were done by dividing the
sky into cells and counting the cell occupancy.  As mentioned
earlier, \textcite{Bok} and \textcite{mow} established the
non-uniformity of the galaxy distribution by counting galaxies in
cells, and later \textcite{Rubin54}, \textcite{Limber54}, and
\textcite{totsuji} used the Lick catalog published as cell
counts in $1^\circ$ cells.  Peebles used the unpublished higher
resolution data from the original notes of Shane and Wirtanen.
Today, cell counts still provide an important mechanism for
analysing point distributions since they are easier to deal with
than the raw, unbinned, data.

\subsubsection{Scaling properties of counts in cells}

Whether we evolve a model numerically or make some analytic
approximation it is necessary to characterize the clustering that
develops in a quantitative manner.  Conventionally, this is done
by presenting the two-point correlation function $\xi(r)$ for the
mass distribution. However, by itself this does not fully
characterize the distribution of points.  An important alternative
is to look at the distribution of counts in cells as a function of
cell size.

The relationship between the probability $P_N(V)$ of finding $N$
galaxies in a sample volume $V$ and the correlation functions of
all orders was given by \textcite{white79}.  The expression is not
of any real use unless all correlation functions are known, or if
there is a known relationship between them.  \textcite{Frycount}
and \textcite{Balian1} computed the properties of the
counts-in-cells distribution $P_N(V)$ on the hypothesis that the
correlation functions of all orders form a particular scaling
hierarchy in which the $q$th order correlation function
$\xi^{(q)}$ based on a $q$-agon of points $r_i$ scales as
\begin{equation}
\xi^{(q)}(r_1,  \dots, r_q) = \lambda^{\gamma{(q-1)}}
\xi^{(q)}(\lambda r_1, \dots, \lambda r_q).
\end{equation}
The hierarchy is described by a single scaling index $\gamma$.
The data available at the time, the CfA survey, appeared
to support both the form of $P_N(V)$ and this scaling hypothesis.

The special case of $P_0(V)$ is the ``Void Probability Function''
(VPF), that is the probability of a volume $V$ containing zero
galaxies. One can construct the probability distribution for
having a void of a given size $V$ in a distribution of galaxies
with given correlation properties \cite{Fall76}.  It is given by
\textcite{white79}
\begin{equation}
P_0(n_0V) =e^{-n_0Va}
\end{equation}
with
\begin{equation}
a = 1+\sum_{i = 2}^{\infty} (-n_0)^{i-1} \int w_i dV_1 \ldots dV_{i-1}.
\end{equation}
Here $n_0$ is the mean space density of galaxies (or clusters),
and $w_i$ is the $i$-point correlation function of $(i-1)$
coordinates and is determined on linear scales by (among other
things) the power spectrum of the primordial density fluctuations.
For purely Gaussian fluctuations the sum in $a$ is cut off beyond
the second term. However, gravitational evolution destroys the
Gaussian character of fluctuations and we are thus forced to make
an {\it ansatz} regarding the relationship between second and
higher order correlation functions either through BBGKY
hierarchies \cite{fry84} or by pure guess.

\textcite{white79} shows the relation between $P_0(V)$ and the
cell count probabilities $P_N(V)$. Different clustering models
have been proposed based on particular choices for the counts in
cells \cite{coljones, saslaw, Borgani93}. A particular ---and
rather popular--- way of analyzing the statistical properties of
point sets is through the possible scaling of the moments of the
counts in cells as it is explained in next section. Alternatively,
one can consider the scaling of moments of counts of neighbors
\cite{martcoles94}.

\subsubsection{Quantifying structure using multifractals}

Given a model for the development of galaxy clustering we might like to
predict the resulting distribution of cell counts since this provides a
straightforward way of confronting the model with data.

Denote by $p(X;L)$ the probability that some quantity $x$ takes on
the value $X$ when measured in a cell of size $L$.  The
distribution $p$ can be characterized by its moments:
\begin{equation}
m_q(L) = \sum_{cells} p(X;L) X^q
\label{moments}
\end{equation}
If for some monotonic function $D(q)$ the moments scale with cell
size $L$ as
\begin{equation}
\sum_{cells} p(X;L) X^q \propto L^{(q-1)D(q)} \label{scaling}
\end{equation}
the point distribution is said to have scaling properties
characterized by dimensions $D(q)$.  The exponent is written in
this way since the case $q=1$ corresponds to the total number of
particles in the sample volume, which is obviously independent of
the cell size.  The case $q=2$ is related to the variance of the
cell counts and to the two-point correlation function.

Eq.~\ref{scaling} does not describe arbitrary point distributions,
but it does describe a large and important set of such
distributions that have the property of multifractal
scaling\cite{borganipr}. It
has been argued that the observed galaxy distribution and the
distribution of particles in an evolved $N$-Body simulation
exhibit multifractal scaling.

There is a slightly different way of getting at the scaling
exponents $D(q)$: via the {\it partition function} $Z(q,r)$.
$Z(q,r)$ is related to the $q$th statistical moment of the
distribution of points as viewed in cells of size $r$.  Suppose
the sample is drawn from a probability distribution $p(n;r)$ for
finding $n$ galaxies in a randomly chosen cell of scale $r$. The
$q$th moment of the cell occupancy is defined as
\begin{equation}
m_q = \sum_{n=0}^\infty p(n;r)n^q
\label{moments1}
\end{equation}
The partition function is then defined as
\begin{equation}
Z(q,r) = {{\cal N}_r \over {N^q}} m_q.
\label{part}
\end{equation}
If $n_i(r)$ denotes the occupancy of the $i$th cell in a partition
of the sample space into ${\cal N}_r$ cells of scale $r$,  the
sample estimate for the partition function is
\begin{equation}
\bar{Z}(q,r) = \sum_{i=1}^{{\cal N}_r} \left[{{n_i(r)} \over N }  \right]^q
\label{Zestimate}
\end{equation}
where $N$ is the total number of points ($\sum n_i(r)$).  Note
that the ordering of the cells is not important and so the
information on the relationship between neighbouring cells appears
through the $r$-dependence of $Z(q,r)$.

The situation of interest is where, for all values of $q$,
$Z(q,r)$ is found to scale as a power law in $r$:
\begin{equation}
Z(q,r) \propto r^{(q-1)D(q) }\propto r^{\tau(q)},
\label{Zscale}
\end{equation}
where $\tau(q)$ is the scaling index of the partition function;
see, e.g., \textcite{martpara}. The function $D(q)$ defined in
this way is a measure of some generalised dimension of order $q$
for the distribution.   This is simply a restatement of
Eq.~\ref{scaling}. Since Eq.~\ref{part} tell us $Z \propto m_q$,
Eq.~\ref{moments} and Eq.~\ref{moments1} are essentially the same.

$D(q)$ is the logarithmic slope of the moment generating function
and of the partition function:
\begin{eqnarray}
D(q) &&= {1 \over (q-1)} { {d \log m_q(r)} \over {d \log r}} \\
     &&=  {1 \over (q-1)} { {d \log Z(q,r)} \over {d \log r}},
     \quad q \neq 1
\end{eqnarray}
In computing $D(q)$ for a sample we would therefore expect to be
able to see a reasonably straight line of data points in plot of
either $m_q$ or $Z(q,r)$ against $r$.  Several aspects of
finite-size data sample mitigate against this.

It should be noted that, technically, Eq.~(\ref{Zscale}) needs be valid
only in the limit $r \rightarrow 0$.  This limit is impossible to
take in the case of a discrete sample which is dominated by shot
noise at distances much smaller than the mean particle separation.
We can only ask for scaling over some well observed range.
Likewise, we are unable to reliably compute (\ref{Zestimate}) for
large $q$ since at large values of $q$ the sum is dominated by
whatever happens to be the single largest cluster of points in the
sample.

\subsubsection{Intermittency}

An important feature of many statistical distributions is the
phenomenon known as {\it intermittency}.  Mathematically this
describes a situation where the higher moments of the {\it
spatial} distribution of some quantity dominate over the lower
moments in a special way: there is an anomalous ratio between
successive statistical moments as compared with a Gaussian
process.  The physical manifestation of this is that the quantity
becomes spatially localised.

It is important to realize that, although we traditionally
characterize the galaxy distribution via its two-point and
three-point correlation functions, these have little or nothing to
do with the visual appearance of the clustering pattern: voids,
walls and filaments.  These macroscopic features are
manifestations of the fact that the higher order moments of the
density distribution are dominant: the statistical distribution of
galaxies is intermittent.

Intermittency can be quantified through a simple non-dimensional
function involving higher order statistical moments of the
distribution.  Consider some random function of position
$\psi({\bf x})$ having a non-zero mean and a statistical
distribution whose moments $\left< \psi^q \right>$ are known . The
{\it intermittency exponent} $\mu_q$ is defined in terms of the
scaling properties of the moments by

\begin{equation}
{{\left< \psi^q \right>} \over {\left< \psi \right>^q}} \sim \left({{L}
\over {l}} \right)^{\mu_q},
\label{intermit}
\end{equation}
where $l$ is some fiducial length scale. The {\it spatial}
intermittency pattern is characterized by the $q$ dependence of
this ratio of moments. It is well known that a quadratic $q-$
dependence of $\mu_q$ corresponds to a lognormal
distribution of $\psi$ (eg. Jones et al., 1993).

Notice that $\left< \psi^q \right>$ is simply the moment
generating function for the process $\psi({\bf x})$, and so the
property of intermittency is a feature of the underlying
statistics.

The assumption that the individual moments scale as per
Eq.~(\ref{scaling}) guarantees the existence of $\mu_q$ and in this
case we have
\begin{equation}
\mu_q =  (q - 1) D(q).
\label{sindex}
\end{equation}

Since the quantity $\left< \psi^q \right>$ for $q = 1$ has no
scale dependence (it is the mean value for the field), Eqs.
(\ref{intermit}) and (\ref{sindex}) provide the scaling law of the
moments in the case of multifractal scaling:
\begin{equation}
\left< \psi^q \right> \propto l^{(q-1)D(q)},
\end{equation}
$\mu_q$ is the standard notation for the intermittency exponent.
$\mu_q$ is also called $\tau(q)$ in the multifractal literature;
as in Eq.~(\ref{Zscale}).

\begin{figure*}
\includegraphics[width=12 cm]{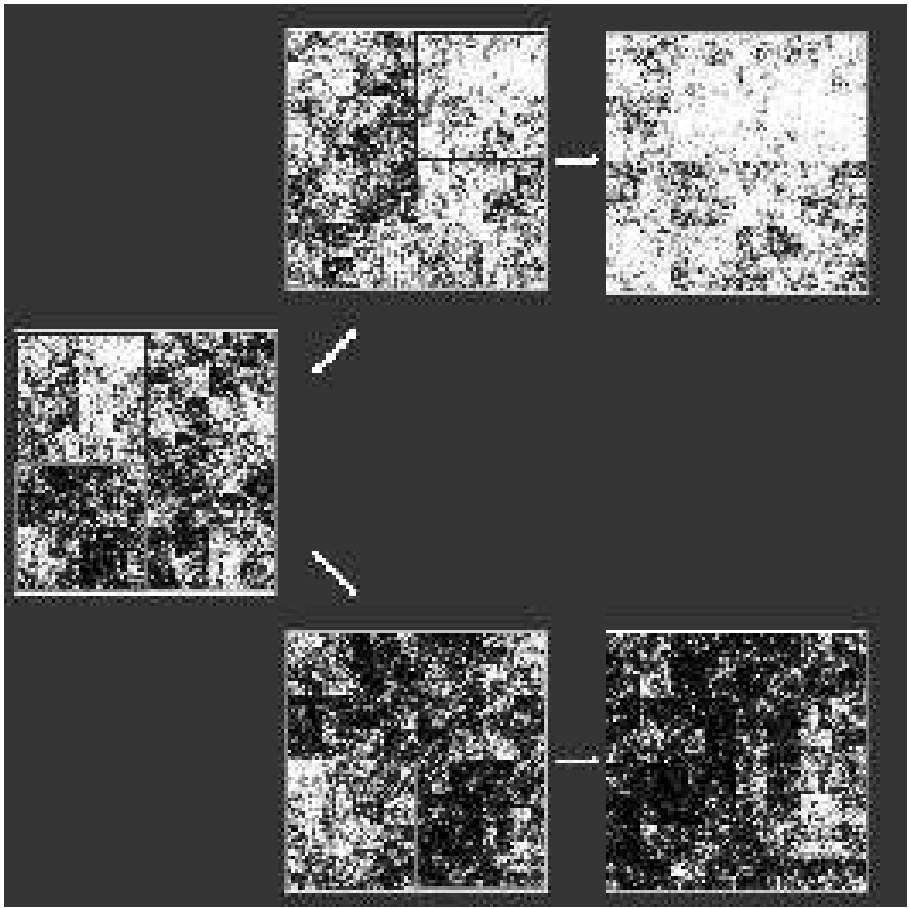}
\caption{\label{fig:multif} A multifractal mass distribution over a
square of side $L_0$. The distribution has been generated following
a multiplicative
cascade process \cite{martpara}. The gray scale represents the quantity
of mass ($X$) in each pixel. Successive enlargements of two
different regions of the original plot illustrate the inhomogeneity
of the mass distribution.}
\end{figure*}

\subsubsection{Multifractality}
People are generally familiar with the notion of simple scaling in
which a function of one variable is independent of the scale of
the variable.  A power law is the prototypical example: if $n(r)
\propto r^\alpha$ then rescaling $r\rightarrow s = \lambda r$
recovers the same power law behavior, $n(r) \propto s^\alpha$.
Only the amplitude and scale of the function have changed, the
shape is the same.

This kind of scaling can be expressed mathematically in a way that
is particularly relevant to the current discussion.  Suppose that
$p(X,L)$ is the probability of measuring a value $X$ for some
property of a system when the sample volume has been binned into
cells of size $L$. Then the property $X$ is said to exhibit
``simple or finite scaling'' when for some constants $\beta$ and
$\nu$
\begin{equation}
p(X,L) = L^{-\beta} g \left({X \over {L^\nu}} \right).
\label{simscaling}
\end{equation}
for some function $g(x)$.  In the jargon of fractals we say that the
quantity $X$ is distributed on a fractal with a single scaling index.

Following \textcite{Kadanoff} we can define a more complicated
kind of scaling, {\it multifractal scaling}, in which we have
\begin{equation}
{{\log p(X,L)} \over \displaystyle{\log {L \over L_0}}} = -f
\left({\displaystyle{\log {X \over X_0}}} \over
{\displaystyle{\log {L \over L_0}}} \right) \label{mscaling}
\end{equation}
Here $X_0$ and $L_0$ can be thought of as physical units in which the
quantities $X$ and $L$ are to be measured.

It is, at first glance, not easy to comprehend what this equation
is telling us about how the the distribution of $X$ looks!. Define
a {\it local scaling index} $\alpha$ by the equation
\begin{equation}
\alpha = {\log \displaystyle{X \over X_0} \over \log
\displaystyle{L \over L_0}}.
\end{equation}
Since $\alpha$ depends on the realization of the value of $X$ in a
cell of scale $L$, $\alpha$ is a possibly random, function of
position.  This is why it is referred to as a {\it local} scaling
index. With this, the probability of finding a value $X$ in a cell
of size $L$ is just
\begin{equation}
p(X,L) = p(X_0,L_0) \left( {L \over L_0} \right)^{-f(\alpha)}.
\label{falpha}
\end{equation}
We have power law scaling with cell size, but the scaling index
$\alpha$ is an arbitrary function of the quantity $X$ and the cell
size $L$.

These two forms Eq.~(\ref{simscaling}) and Eq.~(\ref{mscaling})
of scaling agree
only when $g(x)$ is a power law and $f(x)$ is linear.


If we look only at points such that $\alpha$ in Eq.~(\ref{falpha}) has
some specific value, the distribution $p(X,L)$ has the form
(\ref{simscaling}): the set of points with $\alpha$ = constant is
a simple fractal of dimension $f(\alpha)$.  Since the set consists
of a range of values of $\alpha$ it can be called a {\it
multifractal}, a set of intertwined simple fractals having
different dimensions (see Fig. \ref{fig:multif}).

Note, however, this cautionary tale. A set of points distributed
in power-law clusters is not  necessarily a multifractal.  It is
only a multifractal if the scaling indices $\alpha$ are themselves
constant on homogeneous fractal set.  Thus not all point
distributions are multifractals, even if they are distributed in
power law clusters. A modified version of the scaling indices
formalism, the ``weighted scaling indices", has been recently introduced
\cite{scaind}. This method allows us to statistically
quantify the local morphological
properties of the galaxy distribution.

It can be shown that the descriptions of a point set via its
statistical moments (\ref{scaling}) or via the distribution of its
scaling indices (\ref{falpha}) are totally equivalent.  The
functions $f(\alpha)$ and $\tau(q)$ are related to one another via a
Legendre transform \cite{jcm}:
\[
\tau(q)=\alpha q - f(\alpha), \qquad \quad \alpha(q)=\frac{d\tau}{dq}.
\]

\section{CLUSTERING MODELS}

\subsection{Cosmological simulations}

\subsubsection{Aarseth}

The simplest way to explain the observed clustering is to do
nonlinear numerical simulations of the galaxy clustering process.
Although such simulations provide no deep explanations for what is
going on, the ability to reproduce cosmic clustering simply by
using a distribution of particles moving under their mutual
gravitational interactions is quite striking.

$N$-body models have served to disprove several popular hypotheses
on the evolution of large-scale structure, and motivated to
introduce new assumptions. The downfall of the ``Standard Cold
Dark Matter Model'' (SCDM) started with $N$-body models that gave
top-heavy large-scale structure and too large pairwise velocity
dispersion compared to the observations. Another example is the
present controversy over cuspy centers of dark halos, which were
found in high-resolution $N$-body simulations, but which are not
observed. This motivated intensive study of Warm Dark Matter
models.

The origin of $N$-body experiments as we know them today is the
work of Sverre Aarseth at Cambridge England \cite{aa1}.  Aarseth
was a student of Fred Hoyle whose visionary insight foresaw as
long ago as 1965 the role that computers would play in
astronomical research. Aarseth not only developed series of
$N$-body codes tailor-made for different problems, he made these
codes available to all and never even asked to be named as a
collaborator.

The particle-particle codes developed by Aarseth were originally
aimed at simulating problems in stellar dynamics. The particles
were point masses and integrating of tight binaries was through
two-body regularization. This was adapted to the cosmological
problem by making the particles soft rather than point-like, and
so dropping the need for the time-consuming calculation of binary
encounters.  The first papers using this modified code
\cite{Aarseth1, Aarseth2} used a mere 1000 equal mass particles
and simple Poisson initial conditions. Yet they were able to
reproduce a power-law correlation function for the clustering of
these points.

\subsubsection{Subsequent developments}

During the 1970's the application of $N$-body codes to the problem
of gravitational clustering mushroomed.  Faster computers and
improved numerical techniques drove particle numbers up. Following
on from that work there has been a gradual growth in the number of
particles used in simulations: 30,000 by the 1980s
\cite{Efstathiou85}, 1,000,000 by \textcite{Bertschinger91} in the
1990s (see also the review \textcite{bert98}) and
\textcite{couch}, and now more than 1000,000,000 by the ``Virgo
Consortium'' \cite{virgo02}.

\begin{figure}
\centering
\parbox{4mm}{\rotatebox{90}{{\large redshift}}}
\parbox{10mm}{\resizebox{!}{.90\textheight}{\includegraphics*{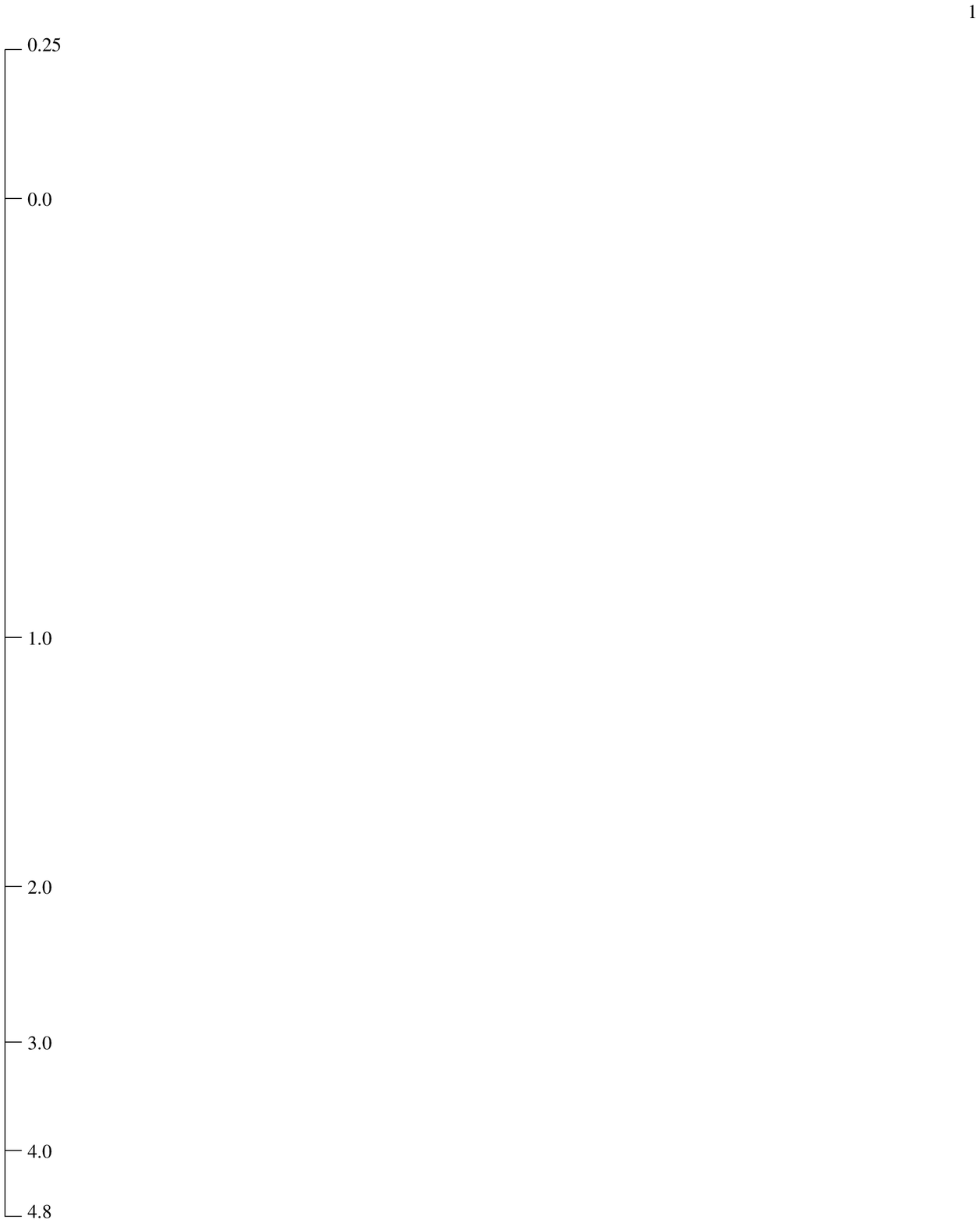}}}
\parbox{40mm}{\rotatebox{180}{\resizebox{!}{.93\textheight}{\includegraphics*{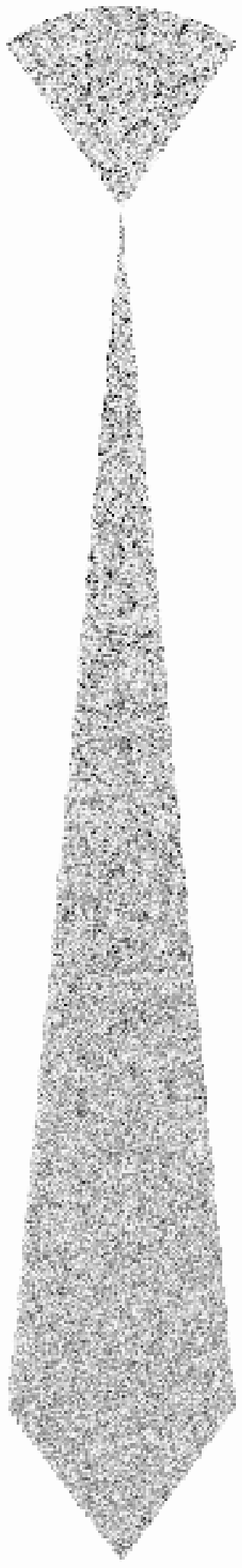}}}}
\caption{\label{fig:wedge}A deep simulated wedge of the Universe.
Figure by Gus Evrard and Andrzej Kudlicki, courtesy of the ``Virgo
Consortium''; details in text.}
\end{figure}

The $N$-body models cover a wide range of cosmic parameters and
have enough particles to be used in trying to discriminate the
clustering properties of the different models. We show in
Fig.~\ref{fig:wedge} a recent $10^9$-point lightcone simulation of
the ``Virgo Consortium'', a deep wedge 40$h^{-1}$Mpc thick and
3.5$h^{-1}$Gpc deep, extending to $z=4.8$ (the universe was then
about one eleventh of its present age). The upper sector of the
``tie'' shows a picture that we hope to get from the SDSS survey,
a wider wedge reaching $z=0.25$. Progressing in time from the
largest redshift until present, we see how the structure gradually
emerges. This simulation is described in \textcite{virgo02}.

\subsubsection{Confronting with reality}

Sometimes we might get the impression that $N$-Body simulations
are better than the real thing, as in the game of `Better Than
Life' played by some of the characters in the BBC TV program {\it
Red Dwarf}.  In the early 1970's people were enthusiastic about a
mere 1000 particles (which reproduced the correct two-point
correlation function so ``it had to be right").  They got even
more enthusiastic with a million particles in the 1990's and now
it is indeed better than life, especially with reality enhancing
graphics, and ready-to-play in your PowerPoint presentation
movies.

Is this enthusiasm justified?  $N$-Body simulations are certainly a
success story, and they certainly make a huge contribution to our
understanding of cosmology.  The models are nevertheless extremely
limited simply because they lack any real gasdynamics, and star
formation which must be important or other things that we know
little about (such as magnetic fields, which one hopes are not
important). There are some salutary lessons, such as the effects of
discreteness in pure $N$-Body models (\textcite{splinter98}), but
there is little or no response to such points from the $N$-Body
community at large.  So maybe we should not worry and just bask in
what is after all better than life.

Up until now, most comparisons between the results of numerical
experiments and the data have been made simply in terms of the
galaxy clustering correlation function.  Even this is fraught with
difficulty since the observed data concerns the distribution of
light whereas the numerical models most readily yield the
clustering properties of the gravitating matter, most of which may
well be dark and invisible. The key ingredient that has to be
added is star formation, and it is perhaps true to say that
attempts at doing this have so far been simple heuristic first
steps.

Another popular model result, the mass function (distribution of
masses) of rich galaxy clusters, depends less on star formation
problems, but knowledge of formation of galaxies and clusters is
certainly necessary to compare the simulated and observed mass
functions.

Some measures, such as the distribution of velocity dispersion of
galaxies and the distribution of halo masses are independent of
the mass-to-light problem, but it is only recently that the large
scale redshift surveys and surveys of real gravitational lenses
have begun to yield the kind of data that is required.

\subsubsection{Scaling in dark matter halos}

$N$-body simulations have revealed fascinating scaling
problems of their own, mostly for smaller scales than
those described in this review. As the initial power
spectrum of perturbations is almost a power law for
comoving scales less than $10 h^{-1}$Mpc, and
cold dark matter and gravitation do not bring in additional
scales, the evolution of structure on these scales, and
the final structure of objects should be similar.

As a proof of this conjecture, $N$-body simulations
show that dark matter halos have well-defined
universal density profiles. There is slight disagreement
between the practitioners on the exact form of this
profile, but the most popular density profile by far
is that found by \textcite{navarro96} (the NFW profile):
\begin{equation}
\label{NFW}
\rho(r)/\rho_c=\frac{\delta_c}{y(1+y)^2},\quad y=r/r_s,
\end{equation}
where $\rho_c$ is the critical cosmological density, $\delta_c$ a
characteristic density contrast, and $r_s$ is a scale radius. The
masses of $N$-body halos are usually defined as that contained
within the ``virial radius'' $r_{200}$, the radius of a sphere of
mean density contrast 200. Then the only parameter describing the
NFW profile for a halo of given mass is the concentration ratio
$c=r_{200}/r_s$.

There have been many studies with differing conclusions on the exact
properties of dark halo profiles; we shall refer the reader to the
latest accurate analysis \cite{navarro04}. The main difficulty is in
eliminating a multitude of possible numerical artifacts, but nobody
seems to doubt that universal profiles exist. Concentration ratios
depend on the mass of a halo, but this seems to be the main
difference.

In connection with observations, the main problem has been the
existence of a density cusp in the center of a halo, and the value
for the logarithmic slope. As this demands probing the very central
regions of galaxy clusters and galaxies, the problem is still open.

\subsubsection{Scaling in galaxy properties}

While the notion of the universal density profile arose from
$N$-body simulations, other scaling laws for cluster- and
galaxy-sized objects have observational origin. The best established
law is called the Fundamental Plane (FP). This scaling law was
discovered simultaneously by \textcite{djorgovski87} and
\textcite{dressler87}. It is rather complex, meaning that elliptical
and S0 (early-type) galaxies form a plane in the 3-space of $(\log
L, \log r_c, \log\sigma)$, where $L$ is the total luminosity of the
galaxy, $r_c$ is its characteristic radius and $\sigma^2$ its
stellar velocity dispersion. (As $L$ and $r_c$ can be combined to
give $\langle I\rangle_c$, the mean surface brightness of the
galaxy, the latter is frequently chosen as one of the three
variables.) These properties of elliptical galaxies are tightly
correlated, and are thought to describe the process of their
formation. Similar correlations have been discovered for galaxy
clusters \cite{lanzoni04}. Their existence demands special scaling
for the mass-luminosity ratio of cluster galaxies with the mass of
the cluster.

As the fundamental plane relation contains the size of a galaxy, it
can be used for estimating the distance to a galaxy. Having a
distance estimate, we can disentangle the proper velocity of a
galaxy from that of the Hubble flow. \textcite{dressler87} (``the
Seven Samurai'') used the newly discovered fundamental plane
relation to derive for the first time the nearby large-scale galaxy
velocity field. In this way the ``Great Attractor'', a large
supercluster complex partly hidden by the Milky Way, was predicted
by \textcite{lilje} from a relatively local sample of galaxies and
discovered by \textcite{ga} using a larger sample of elliptical
galaxies. A recent example of a similar project is the NFP Survey
(NOAO Fundamental Plane Survey), a survey of 100 rich X-ray selected
clusters within $200 h^{-1}$, where the fundamental plane of
early-type cluster galaxies is used to determine cluster distances
and, therefore, large scale cluster flows (\textcite{nfp}).

When talking about scaling laws at galaxy and cluster scales,
one cannot bypass the well-known Tully-Fisher \cite{tully77}
and Faber-Jackson \cite{faber76}
scalings, which declare that the luminosities (or masses)
of galaxies are tightly correlated with their velocity spread.
These scalings can be written as:
\begin{eqnarray*}
L\sim V_{\max}^a,&\qquad& \mbox{Tully-Fisher, spiral galaxies},\\
L\sim \sigma^a,&\qquad& \mbox{Faber-Jackson, elliptical galaxies},
\end{eqnarray*}
where $V_{\max}$ is the maximum rotation velocity of a spiral galaxy,
and $\sigma^2$ is the stellar velocity dispersion of an elliptical galaxy
(in fact, the fundamental plane relation previously explained
is a refinement of the Faber-Jackson relation).
The power-law exponent $a\approx 4$, which can be easily explained,
if there are no dark matter halos around galaxies, and is
difficult to explain for the CDM paradigm. This difficulty has
been of strong support for the MOND theory \cite{milgrom83}.
This theory substitutes the Newtonian theory in the limit of small
accelerations by an empirical formula, which explains the flat
rotating curves of galaxies without invoking the notion of
dark matter, and explains naturally the Tully-Fisher scaling.
MOND does not fit into the present picture of
fundamental physics, as the CDM assumption does,
but it has found a number of followers. A critical (but well-meant)
assessment of MOND can be found in a recent review by
\textcite{binney03}.

\subsection{Statistical models}

The earliest models of galaxy clustering were based on Charlier's
simple notion that the system of galaxies formed some kind of
simple hierarchy.  There was little or no observational basis for
such models.  Later on, in the 1950's when galaxy clusters were
seen as objects in their own right, the clustering process was
seen as aggregates of points (the clusters) scattered randomly in
an otherwise uniform background.

It was not until the systematic analysis of galaxy catalogs and
the discovery of that the two-point clustering correlation
function is a power law that the distribution of galaxies was seen
as being a consequence of gravitational aggregation on all scales.
Galaxy clustering was a general phenomenon and rich galaxy
clusters were seen as something rather rare and special, but
nevertheless a part of the overall clustering process.

\subsubsection{Neyman-Scott processes}

One of the most important of the early attempts to model the
galaxy clustering process came from the Berkeley statisticians
\textcite{neyman52}. They sought to model the distribution of
galaxy clusters as a random spatial superposition of groups of
galaxies of varying size.  The individual groups were to have
their galaxies distributed in a Gaussian density distribution and
they found evidence of superclusters \cite{neyman53}.

Whereas the model in that early form had limited application for
cosmology, the Neyman-Scott process became a discipline in its own
right.  It remains to be seen whether a generalization of these
processes might be resurrected for present day clustering studies.
A recent program in a similar vein is called the halo model;
we shall describe it below.

\subsubsection{Simple fractal models}

There has for a long time been a strong interest in the theory of
random processes which has had a strong impact on many fields of
physics (see for example the collection
of classic papers by \textcite{wax}).
Among the simplest of random processes is the so-called
``Random Walk" in which a particle continually moves a random
distance in a random direction subject to a set of simple rules.
The collection of points at which the particle stops before moving
on has a distribution that can often be calculated.

Many random walks result in distributions of points that are
clustered.  The character of the clustering depends on the
conditions of the walk. It did not take long before someone
suggested that the galaxy distribution could be modeled by a
random walk \cite{four07, mand75}.

What was of interest in these random walk models is that they
could be characterized by a single parameter: a power law index
that related to the mean density profile of the point
distribution.

It should be noted that these simple fractal models have little
direct interest in cosmology: they are merely particularly simple
examples of clustering processes among many. In particular they do
not show the transition to cosmic homogeneity on large scales and
have no relevant dynamical content.

That is not to say that one cannot construct relevant fractal
models.  By 'relevant' we mean that the model should at least be
consistent with or derived from some dynamical theory for the
clustering: anything else is merely descriptive.  Some relevant
ones are described below.

\subsubsection{More complex clustering models}

It was clear at an early stage that the two-point correlation
function for galaxy clustering was by itself an incomplete
descriptor of the galaxy distribution: quite different point
distributions can have the same two-point correlation function.

The obvious step was to compute 3-point and higher order
correlation functions and to seek a more complete description of
the clustering that way.  The key discovery was that the higher
order functions could all be expressed as sums of products of
two-point correlation functions alone.  This lead to a quest to
build clustering hierarchies that embodied these important scaling
properties.

It was evident at the outset that such models would have to be
more sophisticated than the simple fractal hierarchy of
Mandelbrot. The first such model was the clustering hierarchy (a
bounded fractal) of \textcite{sonpee} . This model produced a
galaxy distribution looking remarkably like the observed galaxy
distribution.

The observation that the galaxy distribution was a clustering
hierarchy in which all orders of correlation function could be
related to the basic two-point function could be described in
another way.  Instead of using just one power law index, as in a
simple fractal, to describe the clustering process, it might be
possible to use a distribution of power laws.  This gave rise to
the multifractal model of \textcite{Jones88} in which the
distribution could be generated by a set of simple scaling laws.

\subsubsection{Voronoi tesselations}

The Voronoi tessellation, and the related Delaunay tessellation,
provide well-known tools for investigations into clustering in point
processes.  The construction of such a tessellation starts from a
set of seed points distributed randomly according to some rule
(often Poisson distributed).  A set of walls is constructed around
each point defining a closed polyhedron.  Every point in the
polyhedron has the original seed point as its nearest point among
the set of all seeds.

The polyhedron effectively defines a volume of influence for each
seed point.  The vertices of these polyhedra define a set of
points that is also randomly distributed, but in a way that is
quite different from the distribution of the original seeds.

These tessellations were introduced into astronomy by
\textcite{Icke87} as a model for the galaxy clustering process.
Detailed description of two- dimensional Voronoi tessellations can
be found in \textcite{RipleySS}. The best sources of information on
3-dimensional tessellations in general and in cosmology are
\textcite{rienthesis, rien1}.

\begin{figure}
\includegraphics*[width=6.5 cm]{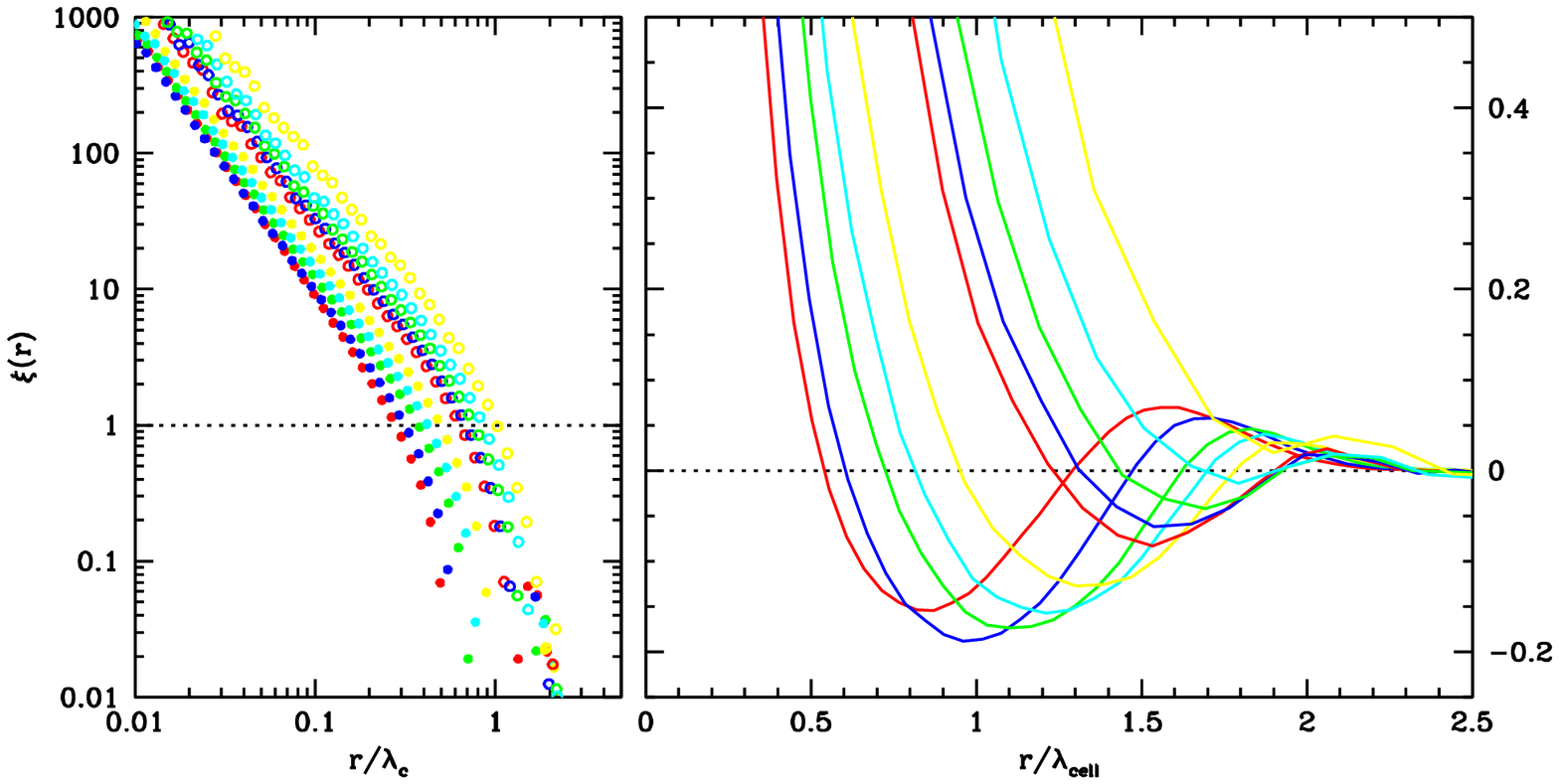}
\caption{\label{fig:voronoi} The scaling of the two-point correlation function
is shown for different subsamples of a Voronoi vertices model. The subsamples
have been selected according to given ``richness'' criterium that mimic
that of the real galaxy cluters, from \textcite{rien2}.}
\end{figure}

What is remarkable is that the two point correlation function for
the Voronoi Vertices generated from Poisson distributed seeds is a
power law that is close to the observed power law of the two-point
galaxy correlation function (see
Fig.~\ref{fig:voronoi}). This tessellation thereby provides a
possible model for the observed galaxy distribution.

Galaxies appear to form on filaments and sheets that surround void
regions. If in the Voronoi model we regard the original seeds as
the centers of expansion of cosmic voids, this model becomes a
dynamically plausible nonlinear model for the formation
large-scale structure formation \cite{Weygaert89}.
The resulting galaxy distribution has many interesting features
that seem to accord with the distribution of galaxies in redshift
surveys \cite{Costarien}.

\subsubsection{Lognormal models and the like}

A rather simplistic yet effective model was presented by
\textcite{coljones} who postulated that the originally Gaussian
density field would evolve into a log-normally distributed density
field. The motivation for this was simply that the hydrodynamic
continuity equation implied that $\log \rho$ would be normally
distributed if the velocity field remained Gaussian. The counts in
cells of various size for $N$-body models and for catalogs of
galaxies are indeed approximately log-normal for a variety of cell
volumes.

Clearly, the contours where the density equaled the mean would
remain fixed: there is no dynamics in such a model.  Such a naive
approach could never reproduce the structure we see today.

There are several relatively simple generalizations of the
lognormal distribution, notably the Poisson lognormal
\cite{Borgani93} and the negative binomial distribution
\cite{Elizalde, betannb}.

\subsubsection{Saslaw-Sheth models}

A novel set of distribution functions was introduced by
\textcite{Saslaw93} and \textcite{Sheth96} derived from a
thermodynamic description of the clustering process. The
distribution functions describe the probability that a randomly
chosen sample volume contains precisely $N$ galaxies. There is
only one free parameter in terms of which the count distribution
for arbitrary values of the volume can be fitted. The resulting
fit is quite remarkable for both $N$-body experiments and for the
data sets that have been analysed \cite{SasCrane}.

The distribution function has some interesting scaling properties
that are discussed in \textcite{saslaw}.

Given the quality of the fit to the data, this is clearly a model
in which the underlying physical motivation deserves more
attention.

\subsubsection{Balian and Shaeffer}
An alternative approach is to create a model for the evolution
some statistically important quantities. \textcite{Balian1}
selected the Void Probability Function: the probability that a
volume $V$ chosen at random would contain no points (galaxies).
This can be generalized to discuss the probability distributions
of volumes containing 1, 2 or $N$ galaxies.

Balian and Shaeffer were able to express many of the details of
the clustering hierarchy in terms of the Void Probability
function, in particular they found a bifractal behavior for the
galaxy distribution \cite{Balian2}. Scaling of voids as a test of
fractality has been studied by \textcite{gaite}.

The mass (luminosity) function was also derived
from similar scaling arguments by \textcite{Bernardeau},
who found the scaling between the galaxy and
cluster luminosity functions to support the theory
of \textcite{Balian2}.

\textcite{Dubrulle} attacked the
problem of gravitational evolution of hierarchical
(fractal) initial conditions. They choose the adhesion
approximation to describe the gravitational dynamics and
demonstrated (with much greater rigour than usual
in cosmological papers) that the mass function has two scaling
regimes, defined by the scaling exponent of the initial
velocity field. This is the only paper that explicitly describes
the evolution of structure on all, even infinitesimally
small scales.

\subsection{Dynamical models }

\subsubsection{Stable clustering models}
The earliest attempt to explain the apparent power law nature of the
two point galaxy correlation function was due to \textcite{peeb74,
peeb74b} and to \textcite{gottrees}. These models were based on the
simple idea that a succession of scales would collapse out of the
expanding background and then settle into some kind of virial
equilibrium.  The input data for the model was a power law spectrum
of primordial imhomogeneities and the output was a power law
correlation function on those scales that had achieved virial
equilibrium.  There would, according to this model, be another power
law on larger scales that had not yet achieved virial equilibrium.

For a primordial spectrum of the form ${\cal{P}}(k) \propto k^{-n}$
the slope of the two-point galaxy correlation function would be
$\gamma = (3n+9) / (n+5)$, which for $n = 0$ gave a respectable
$\gamma = 1.8$, while $n = 1$ gave an almost respectable $\gamma =
2$.

The apparent success of such an elementary model gave great impetus
to the field: we saw something we had some hope of understanding.
However, there were several fundamental flaws in the underlying
assumptions, not the least of which was that the observed clustering
power law extended to such large scales that virial equilibrium was
out of the question.  There were also complications arising out of
the use of spherical collapse models for calculating densities.

Addressing these problems gave rise to a plethora of papers on this
subject, too numerous to detail here. A fine modern attempt at this
is \textcite{shethtor}.  The subject has since evolved into some of
the more sophisticated models for the evolution of large scale
structure that are discussed later (e.g. \textcite{ShethWey}).

\subsubsection{BBGKY hierarchy}

Cosmic structure grows by the action of gravitational forces on
finite amplitude initial density fluctuations with a given power
spectrum.  We see these fluctuations in the COBE anisotropy maps
and we believe they are Gaussian.  This means that the initial
conditions are described as a random process with a given
two-point correlation function. There are no other higher order
correlations: these must grow as a consequence of dynamical
processes.

Given that, it is natural to try to model the initial growth of the
clustering via a BBGKY hierarchy of equations which describe the
growth of the higher order correlation functions. The first attempt
in this direction was made by \textcite{fallsev} though the paper by
\textcite{davpeeb} has certainly been more influential. The full
BBGKY theory of structure formation in cosmology is described in
\textcite{peeb80} and in a series of papers by Fry
\cite{fry82,Frycount}. \textcite{fry85} predicted the 1-point
density distribution function in the BBGKY theory. He also developed
the perturbation theory of structure formation \cite{fry84}, which
has become popular again (see the recent review by
\textcite{bcgs}).

In the perturbative approach, the main question is how many orders
of perturbation theory are required to give sensible results in
the nonlinear regime.

\subsubsection{Pancake and adhesion models}

Very early in the study of clustering, \textcite{zeld} presented a
remarkably simple, yet effective, model for the evolution of
galaxy clustering.  In that model, the gravitational potential in
which the galaxies moved was considered to be known at all times
in terms of the initial conditions.  The particles (galaxies) then
moved kinematically in this field without modifying it. They were
in effect test particles with no self-gravity.  The equations of
motion were arranged so as to give the correct initial, small
amplitude, linear approximation result.

The Zel'dovich model provided a first glimpse of the possible
growth of large scale cosmic structure and led to the prediction
that the galaxy distribution would consist of narrow filaments of
galaxies surrounding large voids.  Nothing of the sort had been
observed at the time, but striking confirmation was later achieved
by the CfA-II Slice sample of \textcite{lapparent} whose redshift
survey revealed for the first time remarkable structures of the
kind predicted by Zel'dovich.

In order to make further progress it was necessary to cure one
problem with the Zel'dovich model: the filaments formed at one
specific instance and then dissolved. The dissolution of the
filaments happened because there was nothing to bind the particles
to the filaments: after the particles entered a filament, they
left.  The cure was simple: make the particles sticky.  This gave
rise to a new series of models, referred to a ``adhesion models"
\cite{gurba89,adhesion}. They were based around the three
dimensional Burgers equation.  In these models structure formed
and once it formed it stayed put: the lack of self gravity within
these models prevented taking them any further.

It was, however, possible to compute the scaling indices for
various physical quantities in the adhesion model. This was
achieved by \textcite{Jones99} using path integrals to solve the
relevant version of the Burgers equation.

\subsubsection{Renormalization group}

\textcite{Peebles85} first recognized that power law clustering
might be described by a renormalization group approach in which
each part of the Universe behaves as a rescaled version of the
large part of the Universe in which it is embedded. This allows
for a recursive method of generating cosmic structure, the outcome
of which is a power law correlation function that is consistent
with the dynamics of the clustering process.

\textcite{Peebles85} used this approach for numerical simulations
of the evolution of structure, hoping that the renormalization
approach would complement the usual $N$-body methods, improving
the usually insufficient spatial resolution and helping to get rid
of the transients caused by imperfect initial conditions. The
first numerical renormalization model had only 1000 particles and
suffered from serious shot noise.

This was later repeated on a much larger scale by
\textcite{Couchman98}. As before, they found that the
renormalization solution produces a stable correlation function.
However, the spatial structures generated by the renormalization
algorithm differed from those obtained by the conventional test
simulation. The relative velocity dispersion was smaller, and the
mass distribution of groups was different. As a rule, the
renormalization solution described small scales better, and the
conventional solution was a better description of the large-scale
structure. As both approaches, the conventional and the
renormalization procedures, suffer from numerical difficulties,
the question of a true simulation remains open.

\subsubsection{The halo model and PThalo model}

\begin{figure}

\resizebox{8.5cm}{!}{\includegraphics*{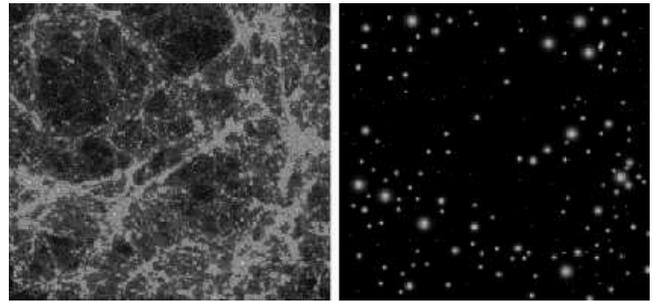}}
\caption{\label{fig:halos} The halo model. The simulated dark
matter distribution (left panel) and its halo model (right panel),
from \textcite{cooray02}.}
\end{figure}

The early statistical model \cite{neyman52} for the galaxy
distribution assumed Poissonian distribution of clusters of
galaxies. This model was resurrected by \textcite{scherr91} and
has found wide popularity in recent years (see the review by
\textcite{cooray02}). In its present incarnation, the halo model
describes nonlinear structures as virialized dark matter halos of
different mass, placing them in space according to the linear
large-scale density field that is completely described by the
initial power spectrum. Such substitution is shown in
Fig.~\ref{fig:halos}, where the exact nonlinear model matter
distribution is compared with its halo model representation.

Once the model for dark matter distribution has been created, the
halos can be populated by galaxies, following different recipes.
This approach has been surprisingly fruitful, allowing calculation
of the correlation functions and power spectra, prediction of
gravitational lensing effects, etc. This also tells us that
low-order (or any-order) correlations cannot be the final truth,
as the two panels in Fig.~\ref{fig:halos} are manifestly
different.

The success of the (statistical) halo model motivated a new
dynamical model to describe the evolution of structure
\cite{pthalos}. The PTHalos formalism, as it is called, creates
the large-scale structure using a second order Lagrangian
perturbation theory (PT) to derive the positions and velocities of
particles, and collects then particles into virialized halos, just
as in the halo model. As this approach is much faster than the
conventional $N$-body simulations, it can be used to sample large
parameter spaces -- a necessary requirement for application of
maximum likelihood methods.

\subsubsection{More advanced models}

Two analytic models in the spirit of the Press-Schechter density
patch model are particuarly noteworthy: the ``Peak Patch" model of
Bond and Myers \cite{BM1,BM2,BM3} and the very recent ``Void
Hierarchy" model of \textcite{ShethWey}.

Both of these models attempt to model the evolution of structure by
breaking down the structure into elements whose individual evolution
is understood in terms of a relatively simple model.  The overall
picture is then synthesized by combining these elements according to
some recipe.  This last synthesis step is in both cases highly
complex, but it is this last step that extends these works far
beyond other like-minded approaches and that lends these models
their high degree of credibility.

The Peak Patch approach is to look at density enhancements, while
the Void Hierarchy approach focusses on the density deficits that
are likely to become voids or are embedded in regions that will
become overdensities.  It somewhat surprising that Peak Patch did
not stimulate further work since, despite its complexity, it is
obviously a good way to go if one wishes to understand the evolution
of denser structures.

The Void Hierarchy approach seems to be particularly strong when it
comes to explaining how large scale structure has evolved: it views
the evolution of large scale structure as being dominated by a
complex hierarchy of voids expanding to push matter around and so
organize it into the observe large scale structures.  At any cosmic
epoch the voids have a size distribution which is well-peaked about
a characteristic void size which evolves self-similarly in time.

\subsection{Hydrodynamic models for clustering}

Let the physical position of a particle at some (Newtonian) time
$t$ be ${\rm \bf r}$.  It is useful to rescale this by the
background scale factor $a(t)$ and label the particle with its
comoving coordinate
\begin{equation}
{\rm \bf x} = {1 \over {a(t)}}{\rm \bf r}
\label{eq6}
\end{equation}
relative to the uniform background. Formation of structure means
that viewed from a frame that is co-expanding with the background,
particles are moving and the values of their coordinates ${\rm \bf
x}$ are changing in time.

There is another coordinate system that can be used: the
Lagrangian coordinate ${\rm \bf q}$ of each particle.  ${\rm \bf
q}$ can be taken to be the value of the comoving coordinate ${\rm
\bf x}$ at some fiducial time, usually at $t = 0$ (the Big Bang)
or a little later, and so remains fixed for each particle.  The
transformation between the Lagrangian coordinate ${\rm \bf q}$ and
the proper (Eulerian) coordinate ${\rm \bf x}$ is achieved via the
equations of motion (see for example Buchert (1992)).

In a homogeneous universe, the particle velocity in physical
coordinates is $\dot{\rm \bf r} = H{\rm \bf r}$, where $H =
\dot{a} / a$ is the Hubble expansion rate.  In this situation the
comoving coordinate ${\rm \bf x}$ of a particle is fixed and there
is no peculiar velocity relative to the co-expanding background
coordinate system.

In an inhomogeneous universe, the displacement of the particles
relative to the co-expanding background coordinate system, ${\rm
\bf x}$ is time dependent.  The velocity relative to these
coordinates is just $\dot {\rm \bf x}$, and this translates back
to a physical ``peculiar'' velocity ${\rm \bf v} = a \dot{\rm \bf
x}$.  We can therefore write the total physical velocity of the
particle (including the cosmic expansion) as
\[
{\rm \bf V} = {\rm \bf v} + H {\rm \bf r}, \qquad
{\rm \bf v} = a \dot{\rm \bf x},
\]
where here the dot derivative is the simple time derivative taken
at a fixed place in the co-expanding frame.

\subsubsection{Cosmological gas dynamics}
As usual, we work in the standard comoving coordinates $\{\rm \bf
x \}$ defined by rescaling the physical coordinates $\{\rm \bf r
\}$ by the cosmic scale factor $a(t)$, as described above.

The motion of a particle is governed by the equations of momentum
conservation, the continuity equation and the Poisson equation.
Expressed relative to the comoving coordinate frame and in terms
of density fluctuation $\delta$ relative to the mean density
$\rho_0(t)$:
\begin{equation}
\delta({\rm \bf x}, t) = {{\rho({\rm \bf x}, t) - \rho_0(t)}
\over {\rho_0(t)}},
\label{eqno8}
\end{equation}
these equations are (\textcite{peeb80, munshi}):
\begin{equation}
{\pd{}{t}} (a {\rm \bf v}) + {\rm \bf {(v.\nabla) v}}
      = - {\pd{\phi}{\rm \bf x}}, \qquad \mbox{\rm momentum
      conservation,}
\label{eqno9}
\end{equation}
\begin{equation}
{\pd{\delta}{t}} +  {1 \over a} {\rm \bf \nabla .}[(1 + \delta)
{\rm \bf v}] = 0, \label{eqno10} \qquad \mbox{\rm continuity,}
\end{equation}
\begin{equation}
{\p2d{\phi}{\rm \bf x}} = 4 \pi G \rho_0 a^2 \delta({\rm \bf x},
t), \label{eqno11} \qquad \mbox{\rm Poisson}.
\end{equation}
Here $\phi({\rm \bf x}, t)$ is the part of the gravitational
potential field induced by the fluctuating part of the matter
density $\rho({\rm \bf x}, t)$ relative to the mean cosmic density
$\bar\rho (t)$.  $G$ is the Newtonian gravitational constant.

Note that here the source of the gravitational potential is the
same density fluctuations that drive the motion of the material
with velocity ${\rm \bf u (x)}$.

\subsubsection{The cosmic Bernoulli equation}
It can be assumed throughout that the cosmic flow is initially
irrotational;  this is justified by the fact that rotational modes
decay during the initial growth of structure or from CMB data.
This assumption makes it possible to take the next step of
introducing a velocity potential that completely describes the
fluid flow and then going on to get the first integral of the
momentum equation: the Bernoulli equation.

Introduce a velocity potential ${\cal V}$ such that
\begin{equation}
{\rm \bf v} = - {\rm \bf \nabla} {\cal V} / a,
\label{eqno12}
\end{equation}
Recalling that the gradient operator is taken with respect to the
comoving ${\rm \bf x}$ coordinates, we see that ${\cal V}$ is the
usual velocity potential for the real flow field ${\rm \bf v}$.
The first integral of the momentum equation becomes
\begin{equation}
{\pd{{\cal V}}{t}} - {1 \over {2a^2}} ({\rm \bf \nabla} {\cal V})^2 = \phi,
\label{eqno13}
\end{equation}
This is referred to as the Bernoulli equation, though in fluid
mechanics we usually find an additional term: the enthalpy $w$
defined by ${\rm \bf \nabla} w = ({\rm \bf \nabla} p) / \rho$.
This vanishes in the post-recombination cosmological context by
virtue of neglecting pressure gradients.

As a matter of interest, for a general (non-potential) flow we
have an integral of the momentum equation that is a constant only
along flow streamlines. Different streamlines can have different
values for this constant. It is only in the case of potential flow
such as is supposed here that the constant must be the same on all
streamlines.

The Bernoulli equation (\ref{eqno13}) is a simple expression of
the way in which the velocity potential (described by $\cal{V}$)
is driven by a gravitational potential $\phi$ in a uniform
expanding background (described by the expansion scale factor
$a(t)$).  Despite its simplicity it has several drawbacks, the
most serious of which is the fact that an additional equation (the
Poisson equation in the form (\ref{eqno11})) or simplifying
assumption is needed to determine the spatially fluctuating
gravitational potential $\phi (\rm \bf x)$.

Another drawback of the Bernoulli equation as presented here is
that it describes a dissipationless flow: there is no viscosity.
Dissipation, be it viscosity or thermal energy transfer, is an
essential ingredient of any theory of galaxy formation since there
has to be a mechanism for allowing the growth of extreme density
contrasts.  Galaxy formation is not an adiabatic process!

A difficulty that presents itself with Eq.~(\ref{eqno13}) is
that the term involving the spatial derivative of the velocity
potential, ${\rm \bf \nabla}{\cal V}$ is multiplied by a function
of time $a(t)$. This can be removed by a further transformation of
the velocity potential:
\begin{equation}
{\cal U} = {{\cal V}  \over {a^2 \dot a}}
\label{eqno14}
\end{equation}
Now, the potential ${\cal U}$ is related to the comoving peculiar
velocity field ${\rm \bf u}$ by ${\rm \bf u} = -a \dot a {\rm \bf
\nabla}{\cal U}$.  In terms of this rescaled potential the
Bernoulli equation takes on a form that is more familiar in
hydrodynamics:
\begin{equation}
{\pd{{\cal U}}{a}} - {1 \over 2}({\rm \bf \nabla} {\cal U})^2 = {3 \over
{2a}}(A \phi - {\cal U}).
\label{eqno15}
\end{equation}
Here we have used the scale factor $a \propto t^{2/3}$ as the time
variable, and noted that $A = -(3 \dot a a^2)^{-1} = $ constant in
an Einstein--de Sitter Universe \cite {kofshan88, kofshan90} (NB.:
in these papers the velocity potential has the opposite sign from
ours).

\subsubsection{Zel'dovich approximation}
The Zel'dovich approximation (\textcite{zeld, shanzel}) to the
cosmic fluid flow was a remarkable first try at describing the
appearance of the large scale structure of the Universe in terms
of structures referred to as ``pancakes'' and ``filaments'' that
surround ``voids''.  Indeed, one might say that through this
approximation Zel'dovich predicted the existence of the structures
mapped later by \textcite{lapparent}.

The Zel'dovich approximation is recovered from the last variant of
the Bernoulli equation above (\ref{eqno15}) by setting $A \phi = -
{\cal U}$.  This latter relationship replaces the Poisson equation
in that approximation.

While predicting the qualitative features of large scale
structure, the Zel'dovich approximation had a number of
shortcomings, notable among which was the fact that particles
passed through the pancakes rather than getting stopped there and
accumulating into substructures (galaxies and groups).

The last decade has seen a host of improvements to the basic
prescription which are nicely reviewed by \textcite{buchert96,
susper} and by \textcite{sahni}. These improvements largely fall
into three categories: ``adhesion'' schemes in which particle
orbits are prevented from crossing by introducing an artificial
viscosity, various ``fixup'' schemes in which simplifying
assumptions are made about the gravitational potential or the
power spectrum and ``nonlinear'' schemes in which the basic
Zel'dovich approximation is taken to a higher order.  We defer the
discussion of the ``adhesion approach'' to the next section.

\subsubsection{Super-Zel'dovich approximations}
Several recipes have been given for improving on the Zel'dovich
approximation in its original nondissipative form without
introducing an {\it ad hoc} artificial viscosity. In these
approximations, the Poisson equation is replaced with some {\it
ansatz} regarding the gravitational potential: it can be set, for
example, equal to a constant, or equal to the velocity potential.

\textcite{matar1, melott1} introduced a variant called the
``Frozen Flow Approximation" (FFA) in which the peculiar velocity
field at any point fixed in the background is frozen at its
original value: the flow is ``steady'' in the comoving frame. (The
initial peculiar velocity field is chosen self-consistently with
the fluctuating potential and the initial density field).

In another approach \textcite{bagla94, bagla95} and
\textcite{brain93}  assume that the fluctuating part of the
gravitational potential at a point expanding with the background
remains constant (as it does in linear theory).  This is referred
to as the ``Frozen Potential Approximation'' (FPA) or ``Linearly
Evolving Potential'' (LEP).  The motivation for this as a
nonlinear extension arises from some special cases where nonlinear
calculations have been done and from $N$-body simulations in which
the potential is seen not to change much in comparison with other
quantities.

\textcite{munshi} point out that the standard Zel'dovich
approximation is equivalent to the assumption that ${\cal V} =
\phi t$, while the Frozen Flow approximation is ${\cal V} = \phi_0
t$ and the Frozen (or Linear) Potential approximation is $\Phi =
\phi_0$.  In any case, this last equation provides an equation for
the velocity potential given a model for the gravitational
potential.

More recently, we have seen the ``Truncated'' Zel'dovich
Approximation \cite{cms93, mps94}, the  ``Optimized'' Zel'dovich
Approximation \cite{msw94} and the ``Completed"  Zel'dovich
Approximation  \cite{betanzel} . These correlate remarkably well
with full $N$-body simulations.

\subsubsection{Nonlinear enhancements}
Various authors have presented nonlinear versions of the
Zel'dovich approximation.  \textcite{gramann93, susper} used a
second order extension, while Buchert (1994) presented a
perturbation scheme that is correct to third order in small
quantities.

\subsection{Nonlinear dynamic models}

The Zel'dovich approximation and its fixups are Lagrangian
descriptions of the cosmic fluid flow.  Their importance lies in
the fact that they capture the gross elements of the nonlinear
clustering while their weakness lies in their side-stepping any
real gravitational forces.  Consequently, they have been used
mainly as short-cut simulators of the evolution of large scale
structure.  Little analytic work has been done using these
approaches.

It is the Lagrangian nature of those equations that makes it
difficult to perform analytic calculations that might lead to an
understanding of how, say, the two-point correlation function
evolves with time.  In order to make progress on an analytic front
it is necessary to cast the equations into analytically tractable
Eulerian forms.  The basis for this was provided by the important
``adhesion approximation'' of \textcite{gss89}, though in the form
presented there it was only ever used for numerical simulations.

\subsubsection{Adhesion Approximations}

The paper by \textcite{gss89} provided a version of the Zel'dovich
approximation in which particle shell-crossing was inhibited: the
material was stopped as it approached the pancakes by an
artificial viscosity introduced on a fairly ad hoc basis into the
equations.  The underlying equation in this approximation turns
out to be the three dimensional Burgers Equation, and so the
approach has the virtues of being simple to use and very easy to
compute (see for example \textcite{weingunn}).

The adhesion approximation is in a sense a linear approximation:
it is allowed to evolve into the nonlinear regime in the
expectation that its behavior will mimic the nonlinear behavior.
This shortcoming has recently been tackled by \textcite{menci}.

Just as the simple Zel'dovich approximation tends to diffuse the
pancakes, the adhesion approximation ensures that asymptotically
they are infinitely thin, and that the particle velocity
perpendicular to these surfaces is zero.  The slowing down of the
particles as they approach the pancakes, the notion of
``viscosity'' in dark matter,  and the lack of a full treatment of
the gravitational field fluctuations leaves open some questions as
to just how good the approximation is for studying, say, large
scale cosmic flow fields.

It is remarkable how much can be done within the framework of the
adhesion model.  \textcite{BabStark} had introduced structure
functions based on the moments of inertia of the local particle
distribution, to describe the local shape of the matter
distribution.  They showed this to be a useful descriptor of the
topology of the galaxy distribution. The evolution of these
structure functions was studied analytically by \textcite{Sathya}.
They analyzed the emergence of large scale filamentary and
pancake-like structures and showed how this might lead to a large
scale coherence in the galaxy distribution. \textcite{SahniVoid}
discussed the evolution of voids using the adhesion approximation.
In their model, ever larger voids emerge at successive epochs,
eventually leaving the largest voids. According to this model, voids
contain some internal filamentary and pancake-like substructures
that dissolve as the voids get older.

\subsubsection{The Random Heat Equation}

The random heat equation was introduced into the subject of cosmic
structure evolution by \textcite{Jones99}.  The Bernoulli
equation (\ref{eqno13}), modified by introducing viscosity (see
\textcite{Jones99}), can be linearised by means of the
Hopf-Cole transformation of variables in which we replace the
velocity potential ${\cal V}$ with a logarithmic velocity
potential $\psi$:
\begin{equation}
{\cal V} = -2 \nu \ln \psi
\label{eqno17}
\end{equation}
If the gravitational potential is rescaled with the viscosity:
\begin{equation}
\phi({\rm \bf x}) = 2 \nu \epsilon({\rm \bf x}),
\label{eqno18}
\end{equation}
Equation (\ref{eqno13}) with the viscosity term
reduces to
\begin{equation}
{\pd{\psi}{t}} = {1 \over a^2} \nu {\p2d{\psi}{\rm \bf x}} + \epsilon ({\rm \bf x})
\psi.
\label{eqno19}
\end{equation}
Again, it is worth stressing that $\nu$ can depend on time, but we
see that invoking a time dependence in $\nu$ means that the new
potential $\epsilon({\rm \bf x})$ gains an explicit time
dependence.

This time dependence can be masked so as to give the random heat
equation in its standard form:
\begin{equation}
{\pd{\psi}{t}} = \nu {\p2d{\psi}{\rm \bf x}} + \epsilon ({\rm \bf x}) \psi.
\label{eqno20}
\end{equation}
It is now to be understood that either $\nu$ or $\epsilon$ (or both)
may contain an explicit time dependence through a multiplying factor.

The renormalised potential field $\epsilon({\rm \bf x})$ is
considered as given and the task is to find the potential $\psi$.
This equation is familiar in slightly different forms in a variety
of fields of physics where it has a variety of names: the Anderson
Model, the Landau-Ginzburg equation, and with a complex time it is
simply the Schrodinger Equation of quantum mechanics. We may hope
to benefit from the vast knowledge that already exists about this
equation.

If we take the limit $\nu \to 0$ and use the definition ${\cal V}
= -2 \nu \ln \psi$, we are led straight back to the familiar
looking dynamical equation
\[
\frac{\partial(a{\rm\bf v})}{\partial t}=\nabla\phi,
\]
telling us that the gravitational potential drives the fluctuating
velocity field. Despite the circuitous route used in deriving the
random heat equation, it still remains very close to the
fundamental physical process that drives the growth of the large
scale structure.

\subsubsection{The Solution of the RH equation}

We can formally solve random heat equation  following the
discussion of \textcite{brax} (but see also \textcite{zmrs1,
zmrs2}).  The solution is expressed in terms of path integrals as
was first given by Feynman and Kac:
\[
\psi({\rm \bf x}, t) = \int K({\rm \bf x}, t, {\rm \bf x}_0, 0) \psi({\rm \bf x_0}, 0)
d{\rm \bf x_0},
\]
where the propagator $K$ is
\begin{equation}
K({\rm \bf x}_1, t, {\rm \bf x}_0, 0) = \int_{{\rm \bf x}(0) = {\rm \bf x}_0}^{{\rm \bf x}(t) =
{\rm \bf x}_1} e^{S({\rm \bf x}(\tau), \tau)} {\cal D}[{\rm \bf x}(\tau)]
\label{eqno22}
\end{equation}
and
\[
S({\rm \bf x}(\tau), \tau) =  - \int_0^\tau  \left[ {1 \over {4\nu}}
\left\vert {{d{\rm \bf x}(\tau')} \over {d \tau'}} \right\vert^2 -
\epsilon({\rm \bf x}(\tau'), \tau') \right] d \tau'
\]
is the action.  This is just the ``free particle'' action with an
additional contribution to the action from the potential
$\epsilon({\rm \bf x}, t)$ evaluated at appropriate places along
the various paths that contribute to the solution
(\textcite{brax}).  The integrand is just the Lagrangian for a
particle moving in a potential $\epsilon({\rm \bf x}, t)$.

What is important here is that the potential $\epsilon({\rm \bf
x}, t)$ contributes to the sum over all paths through an
exponential.  Thus the additive contributions from each part of
the relevant paths results in a multiplicative contribution to the
final solution.  It is this which creates the lognormal
distribution in $\psi({\rm \bf x}, t)$ if the potential
$\epsilon({\rm \bf x}, t)$ in normally distributed.

\subsubsection{Statistical Moments}
\textcite{zmrs1, zmrs2} explain the solution $\psi({\rm \bf x},
t)$ in straightforward terms.  They point out that, of all the
paths that contribute to the integral, one might expect the
dominant contribution to come from those paths that pass rapidly
through high maxima of this potential.  However, there are rarer
paths (optimal trajectories) that are traversed more quickly and
so probe a greater volume that can encounter still larger (and
rarer) maxima of the potential.  These latter paths in fact make
the main contribution to the integral. This is presented
rigorously by \textcite{garmol}.

The outcome of the discussion is that the moments of the
distribution of $\psi$ scale as
\begin{equation}
\left< \psi^q \right> \propto \exp((q \bar\epsilon + {1 \over 2}
q^2 \sigma^2) t) \label{eqno24}
\end{equation}
where $\bar\epsilon$ and $\sigma$ are the mean and variance of the process
$\epsilon$.  This gives intermittency indices
\begin{equation}
\mu_q \propto (q^2 - q)
\label{eqno25}
\end{equation}
\cite{brax}, where the constant of proportionality is determined
by the dimensional characteristics of the random process
$\epsilon({\rm \bf x})$. Thus the solution of the random heat
equation is lognormally distributed for a Gaussian fluctuating
gravitational potential.

In view of the Hopf-Cole transformation, the velocity potential is
in fact the logarithm of the pseudo-potential $\psi$: ${\cal V} =
- 2 \nu \ln \psi$.  Since $\psi$ is lognormally distributed, it
follows that ${\cal V}$ is normally distributed and we can compute
its rms error as
\begin{equation}
\sigma_{\cal V} \propto \sigma_{\phi} t^{1 \over 2}
\label{eqno26}
\end{equation}
Remember that the variance of the gravitational potential
fluctuations $\sigma^2_\phi$ may itself have a time dependence.
This is one of the things that was assumed as given and which in
the single-component model is given by the approximation used to
eliminate the Poisson equation.

\subsubsection{The Schrodinger Equation}

Starting with the coupled Klein-Gordon and Einstein field
equations, \textcite{widkai} produced an ansatz for replacing the
Euler and continuity equations of hydrodynamics with a Schrodinger
equation in the form
\begin{equation}
i{\cal H}{\pd{\Psi}{t}} = -{{\cal H}^2 \over 2m} {\p2d{\Psi}{\rm
\bf x}} + m \phi({\rm \bf x})\Psi. \label{eqno27}
\end{equation}
(see also \textcite{speigel}). ${\cal H}$ here is taken to be an
adjustable parameter controlling spatial resolution.  In this
model the gravitational potential and density fields are given by
\begin{equation}
{\p2d{\phi}{\rm \bf x}} = 4 \pi G \Psi \Psi^*, \qquad \rho = |\Psi^2|
\label{eqno28}
\end{equation}
\textcite{widkai} see this as a means for doing numerical
simulations of the evolution of large scale structure (they use a
Schrodinger solver based on an implicit finite differencing method
called Cayley's Scheme).

The Schrodinger equation for $\Psi$ can be solved analytically by
identical procedures to those described above for solving the random
heat equation, the difference being that the potential $\Psi$ being
solved is complex.   $\Psi$ is directly related to the density
field. This route is advocated by \textcite{coles02} in his very
clear discussion of models for the origin of spatial intermittency.
\textcite{colespen} have taken this further and shown how to add
effects of gas pressure corresponding to a polytropic equation of
state. They present this as a useful approach for modeling the
growth of fluctuations in the mildly nonlinear regime, which is
somewhat short of the ambition of the original Jones (1999) program.

\subsubsection{General Comments}

The relative merits of the random heat equation and the
Schrodinger equation approach are yet to be assessed.  They are
derived from quite different premises: one pretends to be a
derivation from the basic equations while the other is an ansatz
based on interpreting quantum mechanics as a fluid process.  Each
has a level of arbitrariness: one involves an unknown (unphysical)
viscosity that is allowed to tend to zero, while the other
involves a tuning parameter, the effective Planck Constant ${\cal
H}$ that can probably be allowed to become vanishingly small
without changing any results.

In condensed matter physics generalizations of both equations have
played important roles as the basis of analytic models for a
diversity of physical phenomena.  They appear to offer an
important jumping off point for further research based on well
established techniques.

More recently, \textcite{matmoh} have presented a modification of
the adhesion model that they call the {\it forced adhesion model}.
This is based on the {\it forced Burgers equation}, which they
transform into a random heat equation and solve using path
integrals.  It should be noted that this approach is in fact quite
different from that of \textcite{Jones99}:  Matarrese and Moyahaee
use different variables and they claim to model the self-gravity
of the system, thereby avoiding Jones' external field
approximation.

\textcite{menci}, in an approach rather similar to Matarrese and
Moyahaee, also avoids the external field assumption. This is done
by generalizing the simplistic gravitational terms of the
classical adhesion model to a form that, it is claimed, extends
the validity of the gravitational field terms.  Despite the
greater complexity, a solution can be achieved via path integrals.

The main shortcoming of the \textcite{Jones99} model is indeed the
assumption of an externally specified random gravitational
potential field, though it is not clear that the proposed
alternatives are much better.  In the Jones model the intention
had been to write two equations: one collisionless representing
dark matter and providing the main contribution to the
gravitational potential and the other collisional, representing
the baryonic (dissipative) component.  That program was never
completed.

\section{CONCLUDING REMARKS}

\subsection{About scaling}

As we have demonstrated above, there are many scaling laws, which
connect cosmological observables. The main reasons for that are the
scale-free nature of gravitation and the (hopefully) scale-free
initial perturbations.

The gravity scaling could, in principle, extend into very small
scales, if we had only dark matter in the universe. In the real
world the existence of baryons limits the scaling range from below
by typical galaxy masses.

The scaling range starts from satellite galaxy distances, several
tens of kpc, and it may extend up to cluster sizes, 10 Mpc;
two-three decades is a considerable range. The scaling laws at
supercluster distances and larger are determined by the physics of
initial fluctuations.

The first scaling law characterizing the distribution of galaxies is
the power-law behavior of the two-point correlation function at
small scales: $\xi(r) \propto r^{-\gamma}$. Other authors try to fit
the quantity $1+\xi(r)$ to a power law $ \propto r^{D_2-3}$.
Obviously the previous two power laws can only hold simultaneously
within the strong clustering regime, where $\xi(r) \gg 1$ and,
therefore --only at those scales-- the equality $\gamma=3-D_2$
holds. At intermediate scales ($3 < r < 20 \,h^{-1}$ Mpc) the
correlation dimension $D_2$ is $\sim 2$, increasing at larger scales
up to $D_2 \simeq 3$, indicating an unambiguous transition to
homogeneity. Moreover the statistical analysis of the galaxy
catalogs permits to conclude that, within the fractal regime, the
scaling is better described in terms of multifractal inhomogeneous
measures rather than using homogeneous self-similar scaling laws.

Scaling of the galaxy correlation length $r_0$ with the sample size,
$r_0 \propto R_s$, is a strong prediction for a fractal
distribution. Nevertheless, this behavior is clearly ruled out by
the present available redshift catalogs of galaxies. The scaling of
$r_0$ for different kind of objects --from galaxies to clusters
including clusters with different richness-- has been expressed as a
linear dependence of $r_0$ with the intercluster distance $d_c$.
This law, however, does not hold for large values of $d_c$.

One successful scaling law found in the distribution of galaxies is
the scaling of the angular two-point correlation function with the
sample depth. In this case however, the scaling argues against an
unbounded fractal view of the distribution of galaxies, supporting
large-scale homogeneity.

Finally, the hierarchical scaling hypothesis of the $q$-order
correlation function needs further confirmation from the still under
construction deep and wide redshift surveys.

     We have attempted here to provide an overview of the mathematical and
statistical techniques that might be used to characterize the large scale
structure of the universe in coordinate space, velocity space, or both, with,
we hope, enough reference to actual applications and results to indicate
the power of the various techniques and where they are likely to fail.
Of these methods, the ones that have been used most often and so are needed
for reading the current literature are the two-point correlation function
(Sect.  V.B), the power spectrum (Sect. VI.C), counts-in-cells and
the void probability function (Sect. VI.E.3), and fractal and multifractal
measures (Sect. VI.E.4). Those that we
believe have the most potential for the future analysis of the very large
redshift data bases currently becoming available are  the Fourier
methods (Sect. VI.C and Sect. VI.D), although surely the reliable determination
of the two-point correlation function at large scales is still very important
for understanding the large-scale structute \cite{durrer03}.

     Most of the techniques can be applied equally well to real data (in two
or three dimensions) or to the results of numerical simulations of how structure
ought to form in universes with various cosmological parameters, kinds of dark
matter, and so forth (also in three dimensions or two-dimensional projections).

\subsection{Future data gathering}
It may well be that the 2dF and SDSS surveys are the last great
redshift surveys for some time to come.  They have yielded a
phenomenal amount of new information which we have hardly had time
to fully digest.   It is not clear what extra information another
million redshifts might yield: long term funding issues may prevent
us from ever seeing that. However, the future may well lie in the
direction of deeper surveys probing those times when the galaxies
themselves were forming and the large scale structure was coming
into existence.

A number of such surveys are currently under way: 2MASS, COMBO17,
GOODS, DEEP2, CADIS, and the recently funded ALHAMBRA.  With these
we will be able to confront our models with real data, but only
provided we can filter out the effects of galaxy evolution which
will affect sample selection and data interpretation (particularly
if there are luminosity dependent effects).

\subsection{Understanding structure}
We have tried and tested a number of descriptors of the galaxy
distribution with varying amounts of success.  The task has been
helped by ever-growing data sets, but it is nevertheless becoming
clear that a somewhat different approach may be required if we are
to improve substantially on what we understand now.

What different approaches might we take?  Our visual impression of
large scale structure is that it is dominated by voids, filaments
and clusters.  This suggests that instead of looking at sample-wide
statistical measures such as correlation functions, we might try to
isolate the very features that strike us visually and examine them
as individual structures.  Much effort has already been devoted to
isolating ``clusters" of galaxies, but there are currently few, if
any, methods available for isolating either voids or filaments.

Wavelet analysis and its generalizations such as Beamlets and
Ridgelets may prove useful in identifying these structures 
\cite{donoho}.  Other
nonlinear analysis methodologies exist but have not been tried in
this context.  The fact that galaxies (or points in a simulation)
provide a sparse Poisson sample of the underlying data complicates
the application of what might otherwise be standard methods.

The power of having a clear mathematical descriptor lies in being
able to unambiguously identify and study specific objects.  This in
turn provides a tools for confronting simulations with data.

\subsection{About simulations}
Ever since the first simulations by Aarseth, Gott and Turner we have
gazed upon and admired simulations looking ``as good as the real
thing". We were impressed by the gravitational growth of clustering
and we were impressed by the fact that the two-point correlation
function exhibited a power law of approximately the right slope.

Subsequent developments explored the dependency of the results on
initial conditions and extended significantly the range of length
scales over which we could apply our value-judgements. There has
also been a clearer discrimination between dark matter (the stuff of
simulations) and the luminous matter (the stuff we observe).  To
this has been added exceptional computer graphics to render the
simulations as ``observed samples".  They look as good as the real
thing.

Several caveats apply.  First, simulations provide three space and
three velocity coordinates for each mass point at each time.  Data
provide two (angular) space coordinates and a redshift, which is
made up of two terms, one proportional to the third spatial
coordinate (distance) and one representing motion of the point
(galaxy or cluster) relative to uniform cosmic expansion.  These can
be separated only within some model of what real (rather than
$N$-body) clusters ought to be doing in the way of a Virial theorem
or some other way of parcelling out potential and kinetic energy
among the mass points.

     Second, between the simulations of what the (mostly dark) matter is doing
and data on what luminous galaxies are doing lies all of what one might call
gaseous astrophysics (or even gastrophysics).  The intermediate territory
includes inflow of baryons into the potential wells, star formation and wind
energy input, supernovae (which add both kinetic energy and heavy elements,
which change how gas cools and condenses), galactic winds, on-going infall
into the wells, systematic gas flow within galaxies, shocking of baryons
plus heating and/or triggered star formation when halos interact, collide,
and merge, energy input from black hole accretion, and so forth.  Most of these
currently defy real calculation and are represented by parameters and
proportionalities.  Thus the statement that some particular set of cosmological
parameters, initial conditions, and prescriptions for star formation evolve
forward in time to ``fit the data" is not equivalent to being able to say that
this is the way nature did it.

\subsection{Where we stand on theory}
The evolution of cosmic structure is a complex nonlinear process
driven mainly by the force of gravity.  The simplicity of the
underlying driving mechanism, Newtonian attraction, and the fact
that we see simple power law scaling, leads us to believe that the
process of how large scale cosmic structure is organized can be
understood. What is missing is a clear methodology for this, and it
is certain that we shall to borrow tools and methods from other
branches of physics.  This is of course easier said than done since
the driving force, gravity, has infinite range and is always
attractive.

Two approaches look promising at this time.  There is the numerical
Renormalization group simulations of Peebles and Couchman.  Then
there are the analytic models: the Void Hierarchy models of Sheth
and van de Weygaert and the Peak Patch model of Bond and Myers.  The
Random Heat Equation model of Jones and the Schrodinger Equation
approach of Widrow and Kaiser remain to be fully evaluated.

\subsection{And finally ...}
We have good reason to believe that our data samples are now good
enough to unequivocally allow an unambiguous description of the
clustering of galaxies in the Universe.  This description is
entirely consistent with the view of the Universe as a whole that
has emerged from the theoretical and observational research of the
20th. century.  There are many details to fill in and there is much
left to understand.  The details will come with future observational
projects and the understanding will come with further exploitation
of cross-disciplinary physics.  It is the existence of scaling laws
in the galaxy distribution that provides us with a ray of hope that
it is possible to do more than merely models the growth of cosmic
structure: we may be able to understand it.

Arguably the single greatest surprise is how relatively well even
rather simple models appear to reproduce the hard-won data.

\section*{Acknowledgments}

Bernard Jones and Enn Saar spent several weeks at the Valencia
University Observatory, during which time large portions of this
article were written. They want to thank for its kind hospitality as
well as the facilities provided by the Department of Astronomy and
Astrophysics. We are grateful to S. Paredes for providing us with
some figures.  Rien van de Weygaert made extensive comments on the
manuscript for which we are most grateful.  The editor, Julian
Krolik, also provided much appreciated guidance.  Thanks are due to 
all authors, editors and publishers who granted permission to us
to include in this review previously published illustrations, 
images and figures. This work has been
supported by Valencia University through a visiting professorship
for Enn Saar, by the Spanish MCyT projects AYA2000-2045 and
AYA2003-08739-C02-01 (including FEDER), by the Generalitat
Valenciana ACyT project CTIDIB/2002/257 and GRUPOS03/170, and by the
Estonian Science Foundation grant 4695.

\bibliography{scaling}

\end{document}